%% file: main.tex
\documentclass[a4paper,11pt]{article}
\pdfoutput=1

\usepackage{jheppub}
\usepackage[T1]{fontenc}
\usepackage{graphicx}
\usepackage{bm}
\usepackage{amsmath,amssymb}
\usepackage{xcolor}
\usepackage{mathtools}
\usepackage{booktabs}
\usepackage{tabularx}
\newcolumntype{L}[1]{>{\raggedright\arraybackslash}p{#1}}
\newcolumntype{Y}{>{\raggedright\arraybackslash}X}
\allowdisplaybreaks

\newcommand{\as}{\alpha_s}
\newcommand{\Nc}{N_c}
\newcommand{\CA}{C_A}
\newcommand{\CF}{C_F}

\newcommand{\Qs}{Q_s}
\newcommand{\QsF}{Q_{s,F}}
\newcommand{\Sperp}{S_\perp}
\newcommand{\Lamb}{\Lambda}

\newcommand{\dd}{\mathrm{d}}
\newcommand{\ii}{\mathrm{i}}
\newcommand{\SiJF}{\mathcal{J}}
\newcommand{\Mcal}{\mathcal{M}}
\newcommand{\Scal}{\mathcal{S}}

\newcommand{\Hcal}{\mathcal{H}}
\newcommand{\Usud}{\mathcal{U}}
\newcommand{\K}{\mathcal{K}}
\newcommand{\Khard}{\mathcal{K}}
\newcommand{\Tqq}{T_{qq}^{(1)}}
\newcommand{\Tgg}{T_{gg}^{(1)}}
\newcommand{\Tgq}{T_{gq}^{(1)}}
\newcommand{\Tqg}{T_{qg}^{(1)}}
\newcommand{\Mmeas}{\mathcal{M}}
\newcommand{\Wcoh}{\mathcal{W}^{\rm coh}}
\newcommand{\Ufund}{U}
\newcommand{\Uadj}{\mathcal U}
\newcommand{\Sdip}{S}
\newcommand{\Gvac}{\Gamma^{\rm vac}}

\newcommand{\Dg}{\Delta\Gamma^{\rm CGC}}

\begin{document}

\preprint{}

\title{Multiplicity-dependent forward jet production in proton--nucleus collisions}

\author[a]{Lei Wang}
\affiliation[a]{Key Laboratory of Quark and Lepton Physics (MOE)
  and Institute of Particle Physics,
  Central China Normal University,
  Wuhan, Hubei 430079, China}
\emailAdd{leiwang@ccnu.edu.cn}

\date{}

\abstract{
Forward jet production in proton--nucleus collisions probes a dilute projectile
scattering from a dense small-\(x\) nuclear gluon field, while the charged-particle
multiplicity inside the jet probes the final-state cascade in the jet cone.  We develop
a factorization framework for single-inclusive forward jets with an internal
multiplicity measurement by combining the Color Glass Condensate (CGC) description of
the production process with soft-collinear effective theory (SCET) semi-inclusive jet
functions.  At the formal matching level, the construction enforces the inclusive
limit through the stated \(O(\alpha_s)\) matching order.  When the multiplicity weight
vanishes, the measured fixed-order terms reduce to the known inclusive next-to-leading
order (NLO) CGC forward-jet cross section.  The all-order resummed form is retained as
a factorization ansatz.  We define the multiplicity-measured jet operator in a CGC
background.  We then formulate a matching prescription that separates the PDF and
BK/JIMWLK sectors from the measured jet evolution.  Sudakov, SiJF, and multiplicity
radiation are assigned consistently within this prescription.  This separation
identifies the Wilson-line resolution mechanism through which saturation can modify
the internal multiplicity evolution.  The resulting framework shows that
high-multiplicity forward jets are
sensitive both to the production-level quark/gluon mixture and to finite-\(Q_s\)
corrections when an early in-cone splitting is resolved by the target.  A baseline
numerical implementation tests this construction in a phenomenological
nPDF-reweighted per-nucleon setup and illustrates a multiplicity-conditioned
nuclear-side diagnostic.  The displayed baseline uses LO-like production weights,
the accepted diagonal-quark denominator factor, and the local Wilson-line correction
to multiplicity evolution.  Finite measured hard numerators are defined by the
matching construction but are not included in the displayed curves.  The numerical
diagnostic is not a completed calculation of
\(\Delta\Gamma^{A}-\Delta\Gamma^{p}\) with separately normalized proton and nuclear
dipoles.
}


\maketitle
\flushbottom

\section{Introduction}
\label{sec:intro}

Forward jet production in proton--nucleus collisions probes a highly asymmetric
partonic configuration.  The projectile parton carries a moderate or large momentum
fraction \(x_p\) and can be described by ordinary collinear parton distribution
functions, whereas the nuclear target is probed at small \(x_A\).  In this region the
growth of the small-\(x\) gluon density, first described by linear BFKL evolution, is
tamed by nonlinear recombination effects that lead to gluon saturation.  The CGC
effective theory provides the corresponding weak-coupling but high-occupancy
description in terms of Wilson lines averaged over color sources
\cite{BFKL:1977,GLR:1983,MuellerQiu:1986,McLerran:1993ni,McLerran:1993ka,
Balitsky:1995ub,Kovchegov:1999yj,JIMWLK:1997,JIMWLK:2001,Iancu:2001ad,
Gelis:2010nm,KovchegovLevin:2012}.  Forward single-inclusive hadron and jet production
are among the most direct dilute--dense observables because the projectile supplies a
controlled large-\(x\) probe while the transverse momentum of the produced particle or
jet is generated by multiple scattering from the target.  The hybrid factorization
formalism and its NLO extensions have therefore played a central role in quantitative
small-\(x\) phenomenology
\cite{Dumitru:2002qt,Dominguez:2011wm,Dominguez:2011br,Chirilli:2011km,
Chirilli:2012jd,Stasto:2013cha,Watanabe:2015tja,Albacete:2010bs,
Albacete:2012xq,Lappi:2013zma,vanHameren:2014lna,Stasto:2014sea,
Altinoluk:2014eka,Stasto:2016wrf,Iancu:2016vyg,Ducloue:2016shw,
Ducloue:2017dit,Liu:2020mpy,Shi:2021hwx,Liu:2022ijp,Wang:2022zla}.
In particular, the process dependence and universality structure of small-\(x\)
unintegrated gluon distributions were clarified in Ref.~\cite{Dominguez:2011br}.
The complete one-loop hybrid-factorization treatment of inclusive hadron production was
given in Refs.~\cite{Chirilli:2011km,Chirilli:2012jd}.  These results provide the
subtraction logic that later forward-jet calculations must respect.  The subsequent
literature addressed several ingredients needed for stable phenomenology.  These include
matching to collinear factorization and imposing exact kinematic constraints
\cite{Stasto:2014sea,Watanabe:2015tja}.  Other studies clarified the interpretation and
subtraction of NLO rapidity logarithms
\cite{Altinoluk:2014eka,Iancu:2016vyg,Ducloue:2016shw,Ducloue:2017dit}.  The
threshold/Sudakov logarithms relevant at large projectile momentum fraction were also
identified \cite{Mueller:2013wwa,Liu:2020mpy,Shi:2021hwx}.
Closely related small-\(x\) NLO calculations for photons, SIDIS, and dijets further
clarify how finite Wilson-line correlators and real--virtual subtractions appear in
differential observables
\cite{Benic:2016uku,Iancu:2020jch,Hatta:2022lzj,Caucal:2021ent,
Caucal:2022ulg,Taels:2022tza,Bergabo:2022tcu,Bergabo:2022zhe,Tong:2022zwp}.

Jets are particularly attractive in this context.  Compared with identified hadrons,
jets reduce the dependence on fragmentation functions and more directly retain the
transverse momentum flow of the scattered parton.  The forward inclusive jet calculation
in the CGC framework has recently been completed through NLO.  It combines a narrow-jet
treatment with collinear jet functions and threshold/Sudakov resummation
\cite{Liu:2022ijp,Wang:2022zla}.  That calculation also clarifies an important
structural point for the present work.  After the jet algorithm is applied, final-state
collinear singularities are absorbed into jet functions.  Target rapidity divergences
are assigned to BK/JIMWLK evolution.  The remaining finite hard coefficients can then
be identified separately.  A
multiplicity-dependent extension must preserve exactly this organization.

The internal structure of a jet, on the other hand, is naturally described by SCET and
jet functions.  SCET separates hard production, collinear radiation inside jets, and
soft radiation through an expansion in the small jet invariant mass or small cone angle
\cite{Bauer:2000ew,Bauer:2001yt,Bauer:2002nz,Beneke:2002ph,Becher:2008cf,
Becher:2009cu,Becher:2009th,Ellis:2010rwa}.  Semi-inclusive jet
functions (SiJFs) provide a particularly useful language for inclusive jet production.
They describe the probability that a parent parton produces a reconstructed jet carrying
a momentum fraction \(z\).  Their renormalization resums logarithms of the jet radius
\(R\) \cite{Kang:2016mcy,Kang:2016ehg,Dasgupta:2014yra,
Dasgupta:2016bnd,Procura:2009vm,Jain:2011xz}.  Recently, the same jet-function language was extended to the full
charged-particle multiplicity distribution inside inclusive and exclusive jets by
introducing multiplicity generating functions obeying coupled nonlinear,
angular-ordered evolution equations \cite{Duan:2025hmjets}.  This provides a
first-principles way to compute how quark and gluon jets populate the high-multiplicity
tail, including flavor transfer through SiJFs.

High internal jet multiplicity is not merely an extra binning variable.  Measurements of
high-multiplicity jets in pp collisions show that jets selected at large normalized
multiplicity \(\nu=N_{\rm ch}/\langle N_{\rm ch}\rangle\) have substructure and
correlation patterns that differ from inclusive jets \cite{CMS:HighMultiplicityJets,
CMS-PAS-HIN-24-024,CMS:Ridge:2010,CMS:Collectivity:2017,ALICE:2023lyr,
ALICE:IntraJetMultiplicity}.  These measurements connect to a longer line of work on
jet multiplicities, angular ordering, Koba--Nielsen--Olesen scaling, and perturbative
QCD descriptions of multiplicity fluctuations
\cite{Koba:1972ng,Lam:1983vw,Hinz:1985wq,Bassetto:1982ma,DokshitzerWebber:1991,
Dokshitzer:1991wu,Konishi:1978yx,Catani:1990rr,Vertesi:2020utz,Liu:2022bru,
Germano:2024ier,Dokshitzer:2025owq,Duan:2025hmjets}.  Theoretically, such selections
preferentially enhance gluon-initiated jets and multi-prong configurations, thereby
stressing the nonlinear part of the multiplicity generating functional
\cite{Baty:2021ugw,Zhao:2024wqs,Duan:2025hmjets}.  In
pA forward kinematics this is especially interesting: the same selection can change the
quark/gluon mixture produced by the CGC hard kernel and may also increase sensitivity
to whether the nuclear target resolves early in-cone splittings.  High-multiplicity
forward jets are therefore a differential probe of saturation, not only through the
inclusive production rate but also through the conditional final-state cascade.

Beyond counting, a multiplicity selection biases which regions of the parton cascade
are sampled.  High-multiplicity jets preferentially select longer cascades and larger
gluon fractions.  They also favor more resolved multi-prong configurations.  In this indirect but
operational sense, multiplicity selection biases the ensemble of angular configurations
on which more differential jet-substructure observables are measured.  Energy-energy
correlators (EECs), originally introduced as
energy-weighted angular correlations in \(e^+e^-\) annihilation, have become precision
tools for resolving the angular pattern of QCD energy flow inside jets
\cite{Basham:1977iq,Basham:1978bw,Basham:1978zq,PhysRevD.24.2383,
Moult:2018jzp,Dixon:2019uzg,Chen:2020vvp,Duhr:2022yyp}.  Modern EEC studies now span
hadronic collisions and DIS.  They have also been extended to small-\(x\), nuclear,
and heavy-ion environments, including jet quenching
\cite{Li:2020bub,Li:2021txc,Liu:2022wop,Liu:2023aqb,Yang:2023dwc,
Barata:2023bhh,Bossi:2024qho,Ke:2025ibt,CMS:2024mlf,ATLAS:2023tgo,
ALICE:2024dfl,STAR:2025jut}.  Most relevant for the present work is the recently
introduced multiplicity-conditioned EEC jet function.  For jets selected at fixed
normalized multiplicity \(\nu=N_{\rm ch}/\langle N_{\rm ch}\rangle\), it shows that the
EEC acquires a \(\nu\)-dependent anomalous dimension in the perturbative angular region
\(\Lambda_{\rm QCD}/p_{T,\rm jet}\ll\chi\ll R\)
\cite{Duan:2026eecmultiplicity}.  This observation reinforces the central philosophy
of the present paper: internal multiplicity is not only a counting observable, but also
an organizing variable that biases the scale-dependent parton shower and can therefore
feed into angular energy-flow observables.  In a nuclear environment, several effects
may reshape the jet sample.  They include background fluctuations, saturation, medium
response, and energy loss.  Controlling the resulting multiplicity bias is therefore a
prerequisite for interpreting future EEC measurements.

The observable considered here is therefore more differential than the inclusive
forward-jet cross section: besides the jet momentum, it resolves the charged-particle
multiplicity carried by the reconstructed jet.  This additional measurement makes the
factorization problem sharper.  The CGC hard factor already contains real and virtual
radiation generated in the scattering from the dense target, together with the
subtractions that define the projectile PDF and the small-\(x\) evolution of the
target.  The multiplicity distribution inside the jet, by contrast, is governed by
final-state collinear branching and by flavor transfer through SiJFs.  A consistent
description must therefore decide, locally in phase space, which radiation belongs to
the production kernel and which radiation belongs to the measured jet evolution.

This paper develops such a description by combining the inclusive NLO/resummed CGC
forward-jet calculation of Ref.~\cite{Wang:2022zla} with the multiplicity-dependent
jet-function formalism of Ref.~\cite{Duan:2025hmjets}.  The central organizing
principle is the inclusive limit.  At the explicit fixed-order matching accuracy stated
below, setting the Laplace variable conjugate to the charged-particle multiplicity to
\(s=0\) makes the measurement operator the identity, and the measured terms must
reproduce the corresponding inclusive fixed-order cross section.  We impose the same
relation as a consistency requirement on the resummed ansatz, without claiming an
independent all-order factorization proof.  This condition follows from probability
conservation and provides stringent protection against double counting.  The potentially
overlapping sectors include NLO real radiation and BK/JIMWLK evolution.  They also
include Sudakov factors, SiJF evolution, and target-dependent modifications of the
multiplicity anomalous dimension.

The principal result of this work is the operator-level and matching construction.  The
numerical study is deliberately narrower: it tests the explicitly listed production and
evolution ingredients and illustrates the Wilson-line-resolution mechanism, but it does
not implement all finite measured hard channels or a complete
\(\Delta\Gamma^A-\Delta\Gamma^p\) comparison.

The rest of this paper is organized as follows.  In Sec.~\ref{sec:kinematics} we
summarize the forward kinematics and the inclusive CGC baseline to which the
multiplicity-dependent result must reduce.  Section~\ref{sec:generating}
introduces the multiplicity generating functions, and Sec.~\ref{sec:lo-embedding}
embeds them into the leading-order hybrid cross section, making explicit how a
multiplicity cut changes the quark/gluon composition of the forward jet sample.  In
Secs.~\ref{sec:qs-evolution} and \ref{sec:flavor-transfer} we then discuss,
respectively, how saturation can enter the multiplicity evolution and how flavor
transfer is incorporated through semi-inclusive multiplicity jet functions.
Section~\ref{sec:matching} formulates the matching and double-counting subtraction at
the level of raw real-emission kernels.  The operator definition of a
multiplicity-measured jet function in a CGC background is given in
Sec.~\ref{sec:operator}, and its leading Wilson-line-resolved anomalous dimension is
derived in Sec.~\ref{sec:delta-gamma}.  Sections~\ref{sec:ensemble} and
\ref{sec:hard-coefficients} specialize the construction to Gaussian/MV dipoles, state
the rcBK completion, and map the result onto the NLO hard coefficients.  The baseline
numerical implementation and its uncertainty checks are presented in
Secs.~\ref{sec:numerical-status} and \ref{sec:uncertainties}.  We conclude in
Sec.~\ref{sec:conclusions}.  Additional scheme checks and the complete hard-coefficient
import map are collected in Appendices~\ref{app:scheme} and
\ref{app:hard-coefficients}.

\section{Forward kinematics and inclusive CGC baseline}
\label{sec:kinematics}

Before introducing the multiplicity measurement we state the accuracy and scope of the
calculation.  The production kernel is organized in the hybrid dilute--dense expansion:
the projectile is treated in collinear factorization, while the target color field is
kept to all orders through Wilson-line correlators.  The inclusive production recoil is
expanded at leading power in the hard scale \(P_T\).  We therefore neglect corrections
to hard production that are suppressed by \(Q_s^2/P_T^2\) or by the inverse projectile
energy.  Powers of the jet radius beyond the narrow-jet approximation are also omitted.  This
statement should not be confused with the finite-\(Q_s\) correction to the internal
multiplicity kernel.  That correction is organized instead by the ratio
\(Q_s^2/k_{\perp,{\rm split}}^2\).  Here \(k_{\perp,{\rm split}}\) is the relative
transverse momentum of an in-cone splitting and can be much smaller than \(P_T\).

The generic add--subtract matching structure and its inclusive-limit constraints are
formulated through \(O(\alpha_s)\).  The measured real-emission sector is expanded at
this order.  Each singular region is then assigned once to the PDF, BK/JIMWLK, Sudakov,
SiJF, or multiplicity-evolution sector.  A fully differential, channel-by-channel
implementation of every finite multiplicity-measured hard term remains incomplete and
is not used in the displayed numerical baseline.  The numerical results presented
later should therefore not be described as a complete all-channel NLO-accurate
phenomenological prediction over the full scan.
They are a restricted baseline implementation of the factorized construction.  The
production weights are LO-like.  They are supplemented by the diagonal-quark finite
denominator factor only where it passes the stated stability tests.  The reduced
parent-gluon and off-diagonal finite kernels are excluded from the displayed curves.
All finite measured numerators are excluded as well.  This restriction is displayed
explicitly in Secs.~\ref{sec:numerical-status} and \ref{sec:uncertainties}.

Beyond the explicit \(O(\alpha_s)\) matching, the resummed expression is a factorization
ansatz.  It assumes that the projectile-collinear and target-rapidity logarithms remain
separable under their respective evolutions.  The same assumption is made for the
recoil/threshold, semi-inclusive-jet, and multiplicity logarithms.  We do not provide an independent all-order proof of this
joint factorization.  The \(s=0\) closure test is a necessary consistency condition, but
by itself it does not prove the absence of all higher-order overlap terms at nonzero
\(s\).  Within this stated scope, the baseline resums the leading nonlinear
angular-ordered multiplicity logarithms and uses the Sudakov/threshold resummation
inherited from the inclusive forward-jet calculation of Ref.~\cite{Wang:2022zla}.

\begin{figure}[t]
\centering
\includegraphics[width=0.88\textwidth]{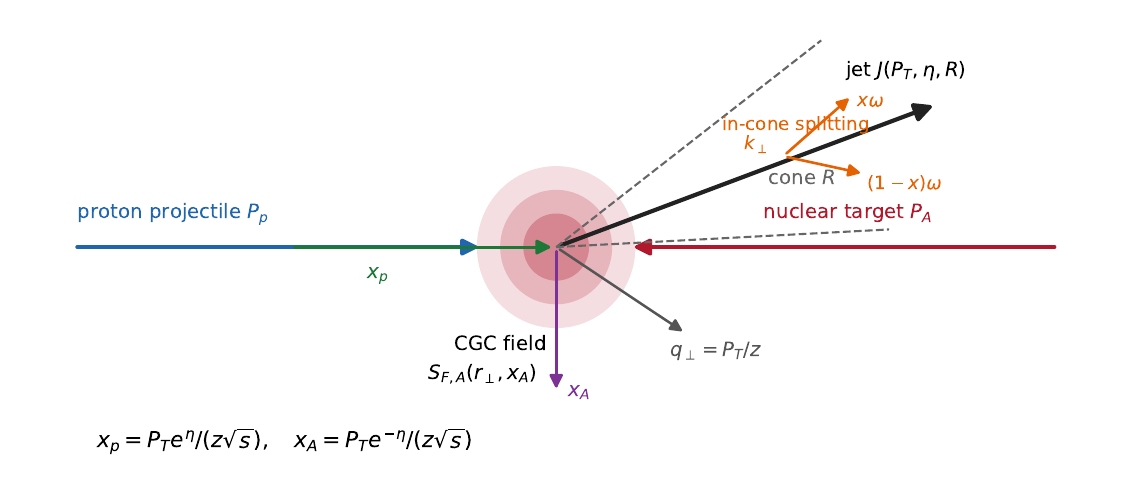}
\caption{Kinematic variables used in the hybrid forward-jet calculation.  The observed
jet carries transverse momentum \(P_T\), rapidity \(\eta\), and radius \(R\).  The
projectile momentum fraction is \(x_p=P_Te^\eta/(z\sqrt{s})\), the target momentum
fraction is \(x_A=P_Te^{-\eta}/(z\sqrt{s})\), and the partonic recoil conjugate to the
target dipole is \(q_\perp=P_T/z\).  The relative transverse momentum
\(k_\perp\) labels an in-cone final-state splitting and is distinct from \(q_\perp\).}
\label{fig:kinematics}
\end{figure}

We consider
\begin{equation}
  p(P_p)+A(P_A)\to J(P_J,R)+X,
\end{equation}
where \(J\) is a forward jet with rapidity \(\eta\), transverse momentum \(P_T\), and
jet radius \(R\).  For a partonic momentum fraction \(z\) entering the jet
measurement, we use the usual hybrid kinematics of forward dilute--dense production
\cite{Dumitru:2002qt,Dominguez:2011wm,Chirilli:2011km,Chirilli:2012jd,
Wang:2022zla},
\begin{equation}
  x_p=\frac{P_T e^\eta}{z\sqrt{s}},
  \qquad
  x_A=\frac{P_T e^{-\eta}}{z\sqrt{s}},
  \qquad
  q_\perp=\frac{P_T}{z}.
  \label{eq:kinematics}
\end{equation}
The lower limit on \(z\) is
\begin{equation}
  \tau=\frac{P_T e^\eta}{\sqrt{s}},
  \qquad
  z\in[\tau,1].
\end{equation}
Throughout the paper we use \(P_T\) or \(P_{J\perp}\) for the observed jet transverse
momentum, \(q_\perp=P_T/z\) for the partonic transverse momentum conjugate to the target
dipole, and \(k_\perp\) for the relative transverse momentum generated inside a
final-state splitting.  In coordinate space the variable \(r_\perp\) is conjugate to the
dipole momentum \(q_\perp\), whereas the daughter separation in the multiplicity kernel
is conjugate to \(k_\perp\).  This distinction is important below: \(q_\perp\) controls
the production recoil from the target, while \(k_\perp\) controls the formation time and
the angular phase space of the in-cone shower.
Figure~\ref{fig:kinematics} summarizes these two transverse-momentum roles and fixes the
kinematic notation used in the subsequent production and multiplicity sectors.

At leading power in the hybrid expansion, the inclusive forward-jet cross section can
be written schematically as
\begin{equation}
  \frac{\dd\sigma_{\rm incl}^{pA\to JX}}{\dd\eta\,\dd^2P_T}
  =
  \sum_i \int_\tau^1 \frac{\dd z}{z^2}\,
  x_p f_i(x_p,\mu)\,
  \Hcal_i(q_\perp,x_A,\mu)\,
  \Usud_{i}^{\rm Sud}(P_T,R,\mu)\,
  J_i(z,\mu),
  \label{eq:inclusive-factorized}
\end{equation}
Here \(i=q,g\).  The function \(J_i\) is the ordinary semi-inclusive jet function or
its LO limit, and \(\Usud_{i}^{\rm Sud}\) denotes the recoil/threshold Sudakov factor.
The inclusive CGC production kernel is denoted by \(\Hcal_i\).  It contains the
target-field information: the dipole amplitude, its BK/JIMWLK evolution, and its
saturation-scale dependence.  At fixed order the Sudakov
factor may be expanded and combined with the finite hard coefficients; in the resummed
baseline it is kept as the exponentiated factor of Ref.~\cite{Wang:2022zla}.
Schematically,
\begin{align}
  \Usud_{i}^{\rm Sud}(P_T,R,\mu)
  &=
  \exp\left[-S_i^{\rm Sud}(P_T,R,\mu)\right],
  \nonumber\\
  S_i^{\rm Sud}
  &=
  \int_{\mu_b}^{\mu}
  \frac{\dd\mu'}{\mu'}\,
  \left[
  A_i(\alpha_s)\ln\frac{P_T^2}{\mu'^2}
  +
  B_i(\alpha_s,R)
  \right],
  \label{eq:sudakov-schematic}
\end{align}
with the coefficient functions and threshold/recoil convention taken from the inclusive
calculation.  Terms generated by expanding Eq.~\eqref{eq:sudakov-schematic} are not
again included in the finite multiplicity numerator; they belong to the Sudakov
subtraction sector in the matching map below.
For example, following the standard CGC notation for fundamental and adjoint Wilson-line
dipoles \cite{Balitsky:1995ub,Kovchegov:1999yj,Iancu:2001ad,KovchegovLevin:2012,
Wang:2022zla}, the LO quark and gluon hard factors are written in terms of the
corresponding dipole Fourier transforms,
\begin{align}
  \Hcal_q^{(0)}(q_\perp,x_A)
  &=
  {\cal N}_q\,F_F(q_\perp,x_A),
  \\
  \Hcal_g^{(0)}(q_\perp,x_A)
  &=
  {\cal N}_g\,F_A(q_\perp,x_A),
  \label{eq:lo-dipoles}
\end{align}
with
\begin{align}
  F_F(k_\perp,x_A)
  =
  \int\frac{\dd^2r_\perp}{(2\pi)^2}\,
  e^{-\ii k_\perp\cdot r_\perp}\,
  S_F(r_\perp,x_A),
  \nonumber\\
  F_A(k_\perp,x_A)
  =
  \int\frac{\dd^2r_\perp}{(2\pi)^2}\,
  e^{-\ii k_\perp\cdot r_\perp}\,
  S_A(r_\perp,x_A),
  \label{eq:dipole-fourier}
\end{align}
where \(r_\perp=x_\perp-y_\perp\).  The coordinate-space dipoles are
\begin{align}
  S_F(r_\perp,x_A)
  &=
  \frac{1}{\Nc}
  \left\langle{\rm Tr}\,
  U(x_\perp)U^\dagger(y_\perp)\right\rangle_{x_A},
  \nonumber\\
  S_A(r_\perp,x_A)
  &=
  \frac{1}{\Nc^2-1}
  \left\langle{\rm Tr}\,
  \mathcal U(x_\perp)\mathcal U^\dagger(y_\perp)\right\rangle_{x_A}.
  \label{eq:fund-adj-dipoles}
\end{align}
with normalization factors \({\cal N}_{q,g}\) fixed by the convention used for the
hybrid cross section.  The representation in
Eq.~\eqref{eq:dipole-fourier} is used only schematically here; the numerical
implementation keeps track of the specific convention used in Ref.~\cite{Wang:2022zla}
coefficients.
The fundamental dipole \(S_F\) is the eikonal scattering amplitude for a color
fundamental parton at transverse positions \(x_\perp\) and \(y_\perp\), averaged over
the target color fields at rapidity \(Y=\ln(1/x_A)\).  The adjoint dipole \(S_A\)
describes the corresponding scattering of a gluon.  In the forward regime their Fourier
transforms provide the transverse momentum transferred from the dense target to the
projectile parton.  The convention used here normalizes \(F_{F,A}\) as Fourier
transforms of \(S_{F,A}\).  Some references Fourier transform \(1-S\) instead.  Away
from the zero-momentum contact term, and within the hard-factor convention imported
from Ref.~\cite{Wang:2022zla}, the two conventions are related by the corresponding
normalization and subtraction prescription.

Beyond LO the statement is not a naive sum of independent physical cross sections.
Rather, the renormalized NLO/resummed result of Ref.~\cite{Wang:2022zla} is organized by
subtraction sectors.  Projectile-collinear poles are absorbed into the PDF, target
rapidity logarithms into BK/JIMWLK evolution, jet-collinear logarithms into the SiJF,
and soft recoil logarithms into the Sudakov factor.  After these subtractions one is left
with finite hard coefficients, denoted below by \(\dd\sigma_{\rm hard}\) or by the
numbered \(\dd\sigma^i\)'s of Ref.~\cite{Wang:2022zla}.  Only those finite remainders can be imported into a
multiplicity measurement without double-counting logarithmic radiation already assigned
to other evolution factors.

\section{Multiplicity generating functions}
\label{sec:generating}

This section fixes the multiplicity notation used in the rest of the paper.  The
definitions and vacuum evolution equations are not new results of the present work:
they are the jet-function multiplicity formalism developed in
Ref.~\cite{Duan:2025hmjets}.  We reproduce only the ingredients needed to couple that
formalism to the forward-CGC production kernel.  For a jet initiated by a parton of
flavor \(i\), the multiplicity generating function is
\begin{equation}
  Z_i(u,\mu)
  =
  \sum_{n=0}^{\infty}P_i(n,\mu)u^n,
  \qquad
  Z_i(1,\mu)=1.
  \label{eq:u-generating}
\end{equation}
Equivalently, in Laplace space one may use \(u=e^{-s}\),
\begin{equation}
  Z_i(s,\mu)=\sum_n e^{-ns}P_i(n,\mu),
  \qquad
  Z_i(0,\mu)=1.
  \label{eq:s-generating}
\end{equation}
The fixed-multiplicity distribution is recovered by inversion:
\begin{equation}
  P_i(n,\mu)
  =
  \frac{1}{2\pi \ii}
  \int_{\cal C}\dd s\,
  e^{ns}Z_i(s,\mu).
  \label{eq:inverse-laplace}
\end{equation}
Moments are generated by derivatives at \(s=0\):
\begin{equation}
  \langle n\rangle_i
  =
  -\left.\frac{\partial Z_i(s)}{\partial s}\right|_{s=0},
  \qquad
  \langle n(n-1)\rangle_i
  =
  \left.\frac{\partial^2 Z_i(s)}{\partial s^2}\right|_{s=0}.
  \label{eq:moments}
\end{equation}

The vacuum branching evolution is the nonlinear Markov evolution derived in
Ref.~\cite{Duan:2025hmjets}.  Written with the light-cone energy \(\omega\) of the
parent parton, the daughter generating functions must be evaluated at their daughter
energies:
\begin{align}
  \frac{\dd Z_q(s,\omega,t)}{\dd t}
  &=
  \int_0^1\dd x\,
  \Gamma_{q\to qg}(x,\omega,t)
  \Big[
  Z_q(s,x\omega,t)Z_g(s,(1-x)\omega,t)
  -
  Z_q(s,\omega,t)
  \Big],
  \label{eq:vac-zq}
  \\
  \frac{\dd Z_g(s,\omega,t)}{\dd t}
  &=
  \frac{1}{2}\int_0^1\dd x\,
  \Gamma_{g\to gg}(x,\omega,t)
  \Big[
  Z_g(s,x\omega,t)Z_g(s,(1-x)\omega,t)
  -
  Z_g(s,\omega,t)
  \Big]
  \nonumber\\
  &\quad+
  \sum_{q_f}\int_0^1\dd x\,
  \Gamma_{g\to q_f\bar q_f}(x,\omega,t)
  \Big[
  Z_{q_f}(s,x\omega,t)Z_{\bar q_f}(s,(1-x)\omega,t)
  -
  Z_g(s,\omega,t)
  \Big],
  \label{eq:vac-zg}
\end{align}
where \(t=\ln(\mu^2/Q_0^2)\), as in the backward evolution of
Ref.~\cite{Duan:2025hmjets}.  The right-hand side vanishes at \(s=0\), since
\(Z_i(0,\omega,t)=1\) for every flavor and every energy.  This ensures probability
conservation at every scale.  The factor \(1/2\) in the \(g\to gg\) term avoids double
counting identical daughter gluons when the \(x\) integration covers the full interval.
In a phenomenological implementation the \(\Gamma\)'s are angularly integrated splitting
kernels with cone and infrared restrictions, following the energy-sharing structure of
Ref.~\cite{Duan:2025hmjets}.

\section{LO CGC embedding of the multiplicity measurement}
\label{sec:lo-embedding}

At LO, we insert the multiplicity generating functions of
Ref.~\cite{Duan:2025hmjets} into the standard hybrid CGC forward-jet factorization
structure used in Ref.~\cite{Wang:2022zla}.  The CGC production kernel fixes the
distribution of parent parton flavors entering the jet, while the internal multiplicity
is described by the conditional generating function.  The multiplicity-dependent cross
section in \(s\)-space is
\begin{equation}
  \frac{\dd\sigma_{\rm LO}(s)}{\dd\eta\,\dd^2P_T}
  =
  \sum_{i=q,g}\int_\tau^1\frac{\dd z}{z^2}\,
  x_p f_i(x_p,\mu)\,\Hcal_i^{(0)}(q_\perp,x_A)\,J_i^{(0)}(z)\,Z_i(s,\mu_J).
  \label{eq:lo-multiplicity}
\end{equation}
Equation~\eqref{eq:lo-multiplicity} immediately satisfies the inclusive limit because
\(Z_i(0)=1\).

The fixed-multiplicity spectrum is
\begin{equation}
  \frac{\dd\sigma_{\rm LO}(n)}{\dd\eta\,\dd^2P_T}
  =
  \sum_i
  \K_i^{(0)}(P_T,\eta)\,
  P_i(n,\mu_J),
  \label{eq:fixed-n-lo}
\end{equation}
where
\begin{equation}
  \K_i^{(0)}(P_T,\eta)
  =
  \int_\tau^1\frac{\dd z}{z^2}\,
  x_p f_i(x_p,\mu)\,
  \Hcal_i^{(0)}(q_\perp,x_A)\,
  J_i^{(0)}(z).
  \label{eq:lo-weight}
\end{equation}
This expression makes clear that a multiplicity bias changes the effective mixture of
quark and gluon jets even if the production kernel is unchanged.  The gluon fraction at
fixed multiplicity is
\begin{equation}
  f_g(n;P_T,\eta)
  =
  \frac{\K_g^{(0)}P_g(n)}
  {\K_q^{(0)}P_q(n)+\K_g^{(0)}P_g(n)}.
  \label{eq:fg-n}
\end{equation}
Since gluon jets typically have larger multiplicities, selecting large \(n\) enhances
the gluon fraction.  This effect is kinematic and flavor-compositional; it is distinct
from a genuine CGC modification of the multiplicity evolution.

The same representation also gives the multiplicity moments of the pA jet sample.  The
mean multiplicity at fixed \(P_T,\eta\) is
\begin{equation}
  \langle n\rangle_{pA}
  =
  \frac{\K_q^{(0)}\langle n\rangle_q+\K_g^{(0)}\langle n\rangle_g}
       {\K_q^{(0)}+\K_g^{(0)}}.
  \label{eq:mean-pa}
\end{equation}
The variance contains both the intrinsic quark/gluon jet variances and a mixture term,
\begin{align}
  {\rm Var}_{pA}(n)
  &=
  f_q\,{\rm Var}_q(n)+f_g\,{\rm Var}_g(n)
  \nonumber\\
  &\quad
  +
  f_q f_g
  \left(\langle n\rangle_g-\langle n\rangle_q\right)^2,
  \label{eq:variance-pa}
\end{align}
with \(f_i=\K_i^{(0)}/(\K_q^{(0)}+\K_g^{(0)})\).  This equation is useful because it
separates two effects that can otherwise be confused: a broad multiplicity distribution
can arise from genuine nonlinear branching inside a jet, or simply from event-by-event
flavor mixing.
In the baseline approximation \(Z_i^{pA}=Z_i^{\rm vac}\); corrections from
\(\Delta Z_i^{\rm CGC}\) to the gluon-jet mean are quantified in
Sec.~\ref{sec:numerical-status}.

For nuclear modification studies one may define a multiplicity-differential ratio
\begin{equation}
  R_{pA}(n;P_T,\eta)
  =
  \frac{1}{A}
  \frac{\dd\sigma_{pA}(n)/\dd\eta\,\dd^2P_T}
       {\dd\sigma_{pp}(n)/\dd\eta\,\dd^2P_T}.
  \label{eq:rpa-n}
\end{equation}
In the baseline where the internal generating functions are vacuum-like and all nuclear
dependence is in the production kernels, this becomes
\begin{equation}
  R_{pA}(n)
  =
  \frac{1}{A}
  \frac{\K_q^{pA}P_q(n)+\K_g^{pA}P_g(n)}
       {\K_q^{pp}P_q(n)+\K_g^{pp}P_g(n)}.
  \label{eq:rpa-baseline}
\end{equation}
Equation~\eqref{eq:rpa-baseline} is the first observable consequence of the framework.
Even without a target-induced modification of \(Z_i\), the multiplicity selection can
change the apparent nuclear modification by reweighting quark and gluon production
channels differently in pA and pp.  A genuine saturation correction to the multiplicity
evolution should therefore be identified only after this flavor-mixture baseline has been
subtracted or independently constrained.
Equation~\eqref{eq:rpa-baseline} holds in the baseline
\(Z_i^{pA}=Z_i^{\rm vac}\).  Corrections from \(\Delta Z_i^{\rm CGC}\) enter at
\({\cal O}(Q_s^2/k_{\perp,{\rm split}}^2)\) and are incorporated in
Sec.~\ref{sec:delta-gamma}.  Here \(k_{\perp,{\rm split}}\) is the relative transverse
momentum of the in-cone branching.

\section{How saturation can enter multiplicity evolution}
\label{sec:qs-evolution}

Before constructing the operator definition, it is useful to state the constraints on any
\(Q_s\)-dependent multiplicity evolution.  A target-dependent generating function may
be written schematically as
\begin{equation}
  Z_i^{pA}(s,\omega,\zeta;Y)
  =
  Z_i^{\rm vac}(s,\omega,\zeta)
  +
  \Delta Z_i^{\rm CGC}(s,\omega,\zeta;Y).
  \label{eq:z-cgc-decomp}
\end{equation}
Here \(\zeta\) denotes a generic angular-ordering coordinate.  In the operator
discussion it may be taken as the physical-angle variable; the forward-jet numerical
implementation uses its boost-invariant hadron-collider counterpart
\(\rho=\Delta R^2/2\), defined in Eq.~\eqref{eq:hadron-collider-angular-map}.
The correction is allowed by factorization only if it satisfies three limits.  First, the
vacuum limit requires
\begin{equation}
  \lim_{Q_s\to0}\Delta Z_i^{\rm CGC}=0.
  \label{eq:qs-vac-limit}
\end{equation}
Second, probability conservation requires
\begin{equation}
  \Delta Z_i^{\rm CGC}(s=0,\omega,\zeta;Y)=0.
  \label{eq:qs-probability}
\end{equation}
Third, perturbative decoupling requires the correction to vanish when the transverse
scale of the measured splitting is much larger than the target resolution scale,
\begin{equation}
  \Delta\Gamma_{i\to ab}^{\rm CGC}
  =
  {\cal O}\!\left(\frac{Q_s^2(Y)}{k_\perp^2}\right),
  \qquad
  k_\perp^2\gg Q_s^2(Y).
  \label{eq:qs-decoupling}
\end{equation}
These constraints are more important than the details of any single model.  They ensure
that a target-dependent multiplicity evolution does not contaminate the inclusive CGC
cross section or double count the production kernel.

At the phenomenological level there are three logically distinct ways in which the
nuclear target can affect the final multiplicity distribution.  They correspond to
different time orderings of the shower relative to the nuclear shockwave and should not
be conflated:
\begin{enumerate}
\item a target-dependent nonperturbative boundary condition for the multiplicity
generating function;
\item a broadening-induced deformation of the perturbative shower phase space;
\item a Wilson-line-resolved branching kernel generated when the target resolves the
daughter color charges of an early in-cone splitting.
\end{enumerate}
The first two mechanisms are model dependent.  They parameterize how nuclear color
fields may affect either the late boundary condition or the available perturbative
phase space.  Similar ideas appear in medium-modified shower and broadening descriptions
\cite{Chien:2015hda,Kang:2017frl,Casalderrey-Solana:2016jvj,
Mehtar-Tani:2016aco,Tachibana:2017syd}.  The third mechanism is different.  It is the
first-principles CGC contribution developed in this work, because it follows from
inserting the multiplicity measurement before the Wilson-line correlator is reduced to
an inclusive dipole.  Mechanism 1 is a \(Q_s\)-dependent nonperturbative boundary
condition:
\begin{equation}
  Z_i(s,\omega,\zeta_0)
  \to
  Z_i(s,\omega,\zeta_0;Q_s),
  \label{eq:model-boundary}
\end{equation}
This boundary condition can represent target-dependent hadronization, color
reconnection, or late soft rescattering below the perturbative shower scale.  It can change the
observed charged multiplicity even if the perturbative parton shower is vacuum-like.
However, it is not predicted by the perturbative CGC calculation itself; it requires a
model for how color flow and hadronization respond to the nuclear environment.

Mechanism 2 is a broadening-modified perturbative phase space, for example
\begin{equation}
  k_{\perp,{\rm eff}}^2
  =
  k_\perp^2+c_i Q_s^2(Y),
  \label{eq:model-broadening}
\end{equation}
which changes where the angular evolution encounters the nonperturbative cutoff.  This
mechanism is closer to the perturbative CGC picture because transverse momentum
broadening is generated by multiple scattering, but it still treats the shower
modification as a kinematic deformation of a vacuum branching kernel.  It can be useful
for uncertainty studies, yet it does not by itself demonstrate that the target has
resolved the individual daughter partons of an in-cone splitting.  Equation
\eqref{eq:model-broadening} is therefore a phenomenological ansatz rather than a result
derived from the CGC operator.  The coefficient \(c_i\) is an order-one,
flavor-dependent parameter.  It may be fixed by data or by a dedicated broadening
calculation.  Alternatively, it may be varied as part of a model uncertainty.  It is
not used as an independent fitted parameter in the baseline calculation below.

Mechanism 3 is the genuine Wilson-line-resolved branching kernel pursued in the
operator analysis below:
\begin{equation}
  \Gamma_{i\to ab}^{pA}
  =
  \Gamma_{i\to ab}^{\rm vac}
  +
  \Delta\Gamma_{i\to ab}^{\rm CGC}.
  \label{eq:model-deltagamma}
\end{equation}
Only the third mechanism is directly tied to a first-principles CGC operator.  It occurs
when the splitting forms sufficiently early for the target to resolve it.  Equivalently,
its transverse size must be large enough that the target does not see only the coherent
parent charge.  The
Wilson-line correlator then changes from a parent dipole to a multi-line correlator
associated with the daughter color charges.  This is the origin of a genuine
\(\Delta\Gamma^{\rm CGC}_{i\to ab}\).  The first two mechanisms are still physically
possible and should be included in a complete phenomenological uncertainty analysis, but
they should not be presented as unique perturbative predictions of the CGC effective
theory.
To realize these three physical scenarios within a QCD-factorized framework, we first
need a final-state object that keeps track of both the reconstructed jet flavor and its
internal multiplicity.  This is the role of the semi-inclusive multiplicity jet
functions introduced in the next section.
The relative size of these mechanisms is kinematics dependent.  In a strict
small-dipole expansion, the Wilson-line correction is power suppressed by
\(Q_s^2/k_{\perp,{\rm split}}^2\).  A finite-dipole implementation can instead produce
a visible effect when the resolved in-cone scale \(k_{\perp,{\rm split}}\) is comparable
to the saturation scale.  For this reason the
phenomenology below separates the flavor-composition baseline from the genuine
Wilson-line correction through the double ratio \(\mathcal R_{\rm mult}\) defined in
Sec.~\ref{sec:numerical-status}, rather than assuming a fixed hierarchy a priori.

\section{Flavor transfer and semi-inclusive multiplicity jet functions}
\label{sec:flavor-transfer}

Beyond the diagonal LO approximation, a parent parton \(i\) can produce a reconstructed
jet of flavor \(j\).  We follow the semi-inclusive jet-function factorization of
Refs.~\cite{Kang:2016mcy,Kang:2016ehg} and the multiplicity extension of
Ref.~\cite{Duan:2025hmjets}.  The appropriate final-state object is then a
semi-inclusive multiplicity jet function,
\begin{equation}
  \widetilde{\SiJF}_{ji}(z,s,\mu)=\SiJF_{ji}(z,\mu)\,Z_j(s,\mu)+\text{nonfactorizing finite terms}.
  \label{eq:sijf-z}
\end{equation}
The factorized form is the leading matching ansatz: \(\SiJF_{ji}\) determines the
inclusive probability for the reconstructed jet flavor and momentum fraction, while
\(Z_j\) gives the conditional multiplicity distribution inside that jet.
This separation follows the logic of SiJF factorization in SCET.  The perturbative
object \(\SiJF_{ji}\) contains the parent flavor, the momentum fraction \(z\), and the
jet-algorithm dependence.  A flavor-dependent generating function encodes the
long-distance charged-particle distribution.  The separation is not a statement that
multiplicity is perturbatively calculable by itself.  Rather, it states that the
scale-dependence and flavor transfer of the semi-inclusive jet can be computed
perturbatively, while the boundary condition for \(Z_j\) remains nonperturbative.
The nonfactorizing finite terms in Eq.~\eqref{eq:sijf-z} are finite matching
contributions beyond the leading multiplicity-SiJF product ansatz.  They are not
included in the present baseline numerical implementation as independent
multiplicity-dependent SiJF corrections.  Instead, the numerical study uses the
restricted production rows specified in
Table~\ref{tab:measured-implementation-map}; any nonfactorizing finite SiJF
contribution is part of the stated matching uncertainty of the baseline calculation.

The corresponding multiplicity-dependent SiJF cross section is
\begin{equation}
  \frac{\dd\sigma_{\rm SiJF}(s)}{\dd\eta\,\dd^2P_T}=
  \sum_{i,j}\int_\tau^1\frac{\dd z}{z^2}\,x_p f_i(x_p,\mu)\,
  \Hcal_i^{\rm CGC}(q_\perp,x_A,\mu)\,\SiJF_{ji}(z,\mu)\,Z_j(s,\mu_J).
  \label{eq:sijf-cross-section}
\end{equation}
Expanding the product through NLO gives
\begin{equation}
  \widetilde{\SiJF}_{ji}^{(1)}
  =
  \SiJF_{ji}^{(1)}Z_j^{(0)}
  +
  \SiJF_{ji}^{(0)}Z_j^{(1)}.
  \label{eq:sijf-expansion}
\end{equation}
This equation is also a double-counting constraint.  The same real splitting cannot be
inserted independently into \(\SiJF_{ji}^{(1)}\) and \(Z_j^{(1)}\) unless it appears
through the controlled expansion in Eq.~\eqref{eq:sijf-expansion}.
The reason is local in the jet algorithm.  At one emission, the jet cone partitions the
real phase space into two mutually exclusive regions.  In the in-cone region, both
daughters contribute to the measured multiplicity.  In the out-of-cone region, the
emission changes the semi-inclusive jet matching coefficient but is not counted as an
internal splitting of the measured jet.  Schematically,
\begin{equation}
  1=\Theta_{\rm in}+\Theta_{\rm out},
  \qquad
  Z^{(1)}\propto \Theta_{\rm in},
  \qquad
  \SiJF^{(1)}\propto \Theta_{\rm out},
  \label{eq:sijf-z-cone-partition}
\end{equation}
up to endpoint and virtual terms required by the same plus-prescription.  The fully
differential version of this cone partition, used for the raw NLO kernels, is written
in Sec.~\ref{sec:matching}.

The flavor-transfer structure is important in pA because the forward CGC production
kernel does not generate the same quark/gluon mixture as a central pp hard process.
For example, a parent quark can produce a reconstructed gluon jet through a finite
\(q\to gq\) configuration, while a parent gluon can produce a quark jet through
\(g\to q\bar q\).  The multiplicity distribution measured for the reconstructed jet is
then the one associated with the observed jet flavor, not the original incoming
projectile flavor.  This is why Eq.~\eqref{eq:sijf-cross-section} carries \(Z_j\) rather
than \(Z_i\).  In the inclusive limit this distinction disappears because all
generating functions become unity, but at large \(n\) it can be numerically important.

\section{Matching and double-counting}
\label{sec:matching}

The full NLO CGC hard factor contains several radiation regions.  Following the
subtraction organization of the inclusive NLO/resummed forward-jet calculation
\cite{Wang:2022zla}, we assign them as follows:
\begin{align}
  \text{target rapidity logarithms} &\to \text{BK/JIMWLK evolution},
  \nonumber\\
  \text{projectile collinear logarithms} &\to \text{PDF evolution},
  \nonumber\\
  \text{threshold or recoil soft logarithms} &\to \text{Sudakov resummation},
  \nonumber\\
  \text{jet-cone collinear logarithms} &\to \text{SiJF evolution},
  \nonumber\\
  \text{resolved in-cone multiplicity branchings} &\to Z_i,
  \nonumber\\
  \text{finite hard leftovers} &\to \dd\sigma^{\rm match}.
  \label{eq:region-assignment}
\end{align}

A safe matching formula is the add-subtract expression
\begin{equation}
  \dd\sigma^{\rm matched}(s)
  =
  \dd\sigma^{\rm resum}(s)
  +
  \left[
  \dd\sigma^{\rm NLO}(s)
  -
  \left.\dd\sigma^{\rm resum}(s)\right|_{\rm NLO}
  \right].
  \label{eq:add-subtract}
\end{equation}
The bracket defines the finite matching term.  The expanded resummed cross section
contains the logarithmic pieces already assigned to PDFs, BK evolution, Sudakov
factors, SiJFs, and \(Z_i\).  Subtracting this expansion from the fixed-order NLO result
leaves only finite hard remainders.
Equation~\eqref{eq:add-subtract} is an \(O(\alpha_s)\) matching identity.  The resummed
expression used in the baseline implementation is obtained by evolving each subtraction
sector with its own anomalous dimension after this fixed-order separation has been
made.  This is a matching prescription, not an independent all-order proof of the joint
CGC--SCET--multiplicity factorization at nonzero \(s\).  The \(s=0\) expansion of the
matched expression must reproduce the inclusive NLO/resummed denominator.  This closure
is necessary for consistency.  It does not, however, prove that all higher-order overlap
terms vanish for a nontrivial multiplicity measurement.  Moreover, the identity does
not imply that every finite measured channel contribution has been derived and
implemented after reducing the fully differential cut kernel.  The restricted
numerical content is listed explicitly in
Table~\ref{tab:measured-implementation-map}.

The numerical implementation uses the following explicit channel decomposition.  For a
given \((P_T,\eta)\) and flavor channel \(i\),
\begin{align}
  D_i(P_T,\eta)
  &=
  \int_\tau^1\frac{\dd z}{z^2}\,
  x_p f_i(x_p,\mu)\,
  \Usud_{i}^{\rm Sud}(P_T,R,\mu)\,
  \left[
  {\cal K}_{i}^{\rm LO}(q_\perp,x_A)
  +
  {\cal K}_{i}^{\rm fin}(q_\perp,x_A,\mu)
  \right],
  \nonumber\\
  N_i^{\rm mult}(s;P_T,\eta)
  &=
  \sum_{i\to ab}
  \int\dd\Pi_{i\to ab}^{\rm fin}\,
  x_p f_i(x_p,\mu)\,
  \Usud_{i}^{\rm Sud}(P_T,R,\mu)\,
  {\cal K}_{i\to ab}^{\rm fin}
  \left[
  {\cal M}_{i\to ab}^{\rm raw}(s)-{\cal M}_{i}^{\rm coh}(s)
  \right].
  \label{eq:numerical-matched-dn}
\end{align}
Here \({\cal K}^{\rm fin}\) denotes the finite hard-coefficient combination that
remains after local subtraction.  The removed sectors are the PDF, BK/JIMWLK, Sudakov,
SiJF, and \(\Delta\Gamma\) kernels.  In the complete matched construction the same
\(D_i\) normalizes the inclusive and multiplicity-weighted terms.  Equation
\eqref{eq:numerical-matched-dn} defines that target formula; it is not a statement that
every term has been inserted in the displayed baseline.  Section
\ref{sec:numerical-status} specifies the restricted subset actually used, with
Table~\ref{tab:measured-implementation-map} serving as the authoritative binary status
map.  In those plots the target dependence of the internal cascade is supplied only by
the local \(\Delta\Gamma\) of Sec.~\ref{sec:delta-gamma}.

Table~\ref{tab:channel-import-map} summarizes the channel-level import map of the formal
matched construction.  It distinguishes the inclusive denominator from the resolved
multiplicity numerator.  It also identifies the subtraction or evolution sector to
which each omitted term is assigned.  Endpoint,
virtual, and pure evolution pieces are retained in the inclusive denominator but are
not multiplied by a resolved daughter measurement.

For compactness, define the finite channel combinations
\begin{align*}
\Sigma_{qq}^{\rm fin}&\equiv
\sigma_{qq}^{3,4,5b,6,7,8,10,11}+\frac12\sigma_{qq}^{5a},
&
\Sigma_{gg}^{\rm fin}&\equiv
\sigma_{gg}^{3,4,5,6,7b,8,9,10}+\frac12\sigma_{gg}^{7a},
\\
\Sigma_{gq}^{\rm fin}&\equiv\sigma_{gq}^{2,3,5},
&
\Sigma_{qg}^{\rm fin}&\equiv\sigma_{qg}^{2,3,5}.
\end{align*}
The individual coefficients follow the convention of
Appendix~\ref{app:hard-coefficients}.

\begin{table}[ht!]
\centering
\footnotesize
\setlength{\tabcolsep}{3pt}
\begin{tabularx}{\textwidth}{L{0.13\textwidth}L{0.27\textwidth}Y L{0.20\textwidth}}
\toprule
Channel & Inclusive content & Resolved weight & Assigned sector \\
\midrule
\(q\to qg\) diagonal &
\(D_{\rm LO}+D_{\rm PDF}+D_{\rm Sudakov}+D_{\rm SiJF}+\Sigma_{qq}^{\rm fin}\) &
\({\cal M}_{q\to qg}^{\rm raw}-{\cal M}_{q}^{\rm coh}\) &
PDF, Sudakov, SiJF, \(\Delta\Gamma\) \\
\(g\to gg\) diagonal &
LO gluon term \(+\Sigma_{gg}^{\rm fin}\) &
\({\cal M}_{g\to gg}^{\rm raw}-{\cal M}_{g}^{\rm coh}\) &
BK/JIMWLK, Sudakov, SiJF, \(\Delta\Gamma\) \\
\(q\to gq\) off diagonal &
\(\Sigma_{gq}^{\rm fin}\) &
\({\cal M}_{q\to gq}^{\rm raw}-{\cal M}_{q}^{\rm coh}\) &
PDF, SiJF \\
\(g\to q\bar q\) off diagonal &
\(\Sigma_{qg}^{\rm fin}\) &
\({\cal M}_{g\to q\bar q}^{\rm raw}-{\cal M}_{g}^{\rm coh}\) &
PDF, SiJF \\
\bottomrule
\end{tabularx}
\caption{Channel import map for the formal matched construction.  The coefficients
are taken from Ref.~\cite{Wang:2022zla}.  The subset used in the displayed baseline is
listed in Table~\ref{tab:measured-implementation-map}.}
\label{tab:channel-import-map}
\end{table}

The coordinate-space terms in Eqs.~(82)--(85) of Ref.~\cite{Wang:2022zla} were also
examined as a possible representation of the parent-gluon finite remainder.  Because
those inclusive kernels have already undergone part of the daughter phase-space
integration, they do not provide a unique resolved multiplicity assignment.  A proper
measured numerator must instead be derived from the fully differential cut kernel by
inserting the cone partition and the resolved-minus-coherent weight before daughter
integration and local subtraction.  An independent quadrature refinement of the
reduced representation was not convergent, so none of the corresponding denominator or
measured terms is used in the displayed baseline.  The momentum-space coefficients in
Appendix~\ref{app:hard-coefficients} remain the formal bookkeeping reference.

For the multiplicity measurement, the finite real-emission contribution has the
generic form
\begin{equation}
  \dd\sigma_{\rm finite}(s)
  =
  \sum_{i\to ab}
  \int\dd\Pi_{i\to ab}\,
  \Khard_{i\to ab}^{\rm finite}\,
  \Mmeas_{i\to ab}(s),
  \label{eq:finite-s}
\end{equation}
where \(\dd\Pi\) denotes the appropriate \(z,\xi\), transverse-momentum, or coordinate
phase space.  If the real splitting \(i\to ab\) shares the parent energy \(\omega\) as
\(x\omega\) and \((1-x)\omega\), the measurement weight depends on which daughters are
inside the jet:
\begin{align}
  \Mmeas_{i\to ab}^{\rm both}
  &=
  Z_a(s,x\omega)Z_b(s,(1-x)\omega),
  \\
  \Mmeas_{i\to ab}^{a{\rm\ only}}
  &=
  Z_a(s,x\omega),
  \\
  \Mmeas_{i\to ab}^{b{\rm\ only}}
  &=
  Z_b(s,(1-x)\omega),
  \\
  \Mmeas_i^{\rm unresolved/virtual}
  &=
  Z_i(s,\omega).
  \label{eq:cone-weights}
\end{align}
The coherent subtraction is obtained by replacing the two resolved daughters by the
parent weight \(\Wcoh_i(s)=Z_i(s,\omega)\).  Therefore a convenient matched numerator is
\begin{equation}
  N_{i\to ab}^{\rm mult}(s)
  =
  \int\dd\Pi\,
  \Khard_{i\to ab}^{{\rm real}}\,
  \left[
  \Mmeas_{i\to ab}(s)-\Wcoh_i(s)
  \right].
  \label{eq:matched-numerator-s}
\end{equation}
At \(s=0\) all \(Z\)'s are unity and the bracket vanishes, so the multiplicity-dependent
piece does not spoil the inclusive cross section.

The preceding equations are compact.  For the actual NLO matching one needs the
fully differential cone decomposition before real--virtual cancellation and before the
Fourier transform to the final momentum-space hard factor.  Let a real splitting
\(i\to a b\) have daughter momenta
\begin{equation}
  p_a^\mu=x\,p_i^\mu+k^\mu,\qquad
  p_b^\mu=(1-x)p_i^\mu-k^\mu,
  \label{eq:raw-daughter-momenta}
\end{equation}
with relative transverse momentum \(k_\perp\).  The jet algorithm separates three
resolved regions:
\begin{align}
  \Theta_{\rm both}
  &=
  \Theta(\Delta R_{ab}<R),
  \\
  \Theta_a
  &=
  \Theta(\Delta R_{aJ}<R)\,
  \Theta(\Delta R_{bJ}>R),
  \\
  \Theta_b
  &=
  \Theta(\Delta R_{bJ}<R)\,
  \Theta(\Delta R_{aJ}>R),
  \label{eq:cone-regions}
\end{align}
plus an unresolved or virtual region carrying the parent weight.  The multiplicity
measurement inserted in the raw real integrand is therefore
\begin{align}
  {\cal M}_{i\to ab}^{\rm raw}(s)
  &=
  \Theta_{\rm both}\,Z_a(s,x\omega)Z_b(s,(1-x)\omega)
  \nonumber\\
  &\quad
  +\Theta_a\,Z_a(s,x\omega)
  +\Theta_b\,Z_b(s,(1-x)\omega)
  \nonumber\\
  &\quad
  +\Theta_{\rm unres}\,Z_i(s,\omega).
  \label{eq:raw-measurement}
\end{align}
The subtraction local in the same variables is obtained by replacing the resolved
daughter measurement by the parent measurement,
\begin{equation}
  {\cal M}_{i}^{\rm coh}(s)=Z_i(s,\omega).
  \label{eq:coherent-measurement}
\end{equation}
Thus the part of a raw real kernel that contributes to the multiplicity anomalous
dimension is proportional to
\begin{equation}
  {\cal M}_{i\to ab}^{\rm raw}(s)-{\cal M}_{i}^{\rm coh}(s).
  \label{eq:raw-measurement-difference}
\end{equation}
This is the formula that should be inserted into the coordinate-space CGC real kernels
before the Wilson-line correlators are reduced to dipoles or Fourier transformed.  It is
also the most transparent way to see the \(s=0\) cancellation: every term in
Eq.~\eqref{eq:raw-measurement-difference} vanishes locally when all \(Z\)'s are set to
unity.

It remains to specify why the target-induced anomalous dimension introduced later does
not count the same radiation twice.  Write the multiplicity evolution kernel as
\(\Gamma=\Gamma^{\rm vac}+\Delta\Gamma^{\rm CGC}\), and expand the corresponding
generating function to first order in the real-emission phase space.  The target-induced
piece contributes
\begin{equation}
  \Delta Z_i^{(1),{\rm CGC}}(s,\omega)=
  \sum_{a b}\int\dd\Pi_{i\to ab}^{\rm sing}\,\Delta\Gamma_{i\to ab}^{\rm CGC}
  \left[{\cal M}_{i\to ab}^{\rm sing}(s)-{\cal M}_{i}^{\rm coh}(s)\right].
  \label{eq:deltagamma-low-order}
\end{equation}
The same singular limit must be subtracted from the measured NLO hard factor.  Thus the
finite multiplicity-sensitive hard kernel is not the full real-emission kernel, but
\begin{equation}
  {\cal K}_{i\to ab}^{\rm mult,fin}
  =
  {\cal K}_{i\to ab}^{\rm real}
  -\left({\cal K}_{i\to ab}^{\rm PDF}+{\cal K}_{i\to ab}^{\rm BK}
  +{\cal K}_{i\to ab}^{\rm Sud}+{\cal K}_{i\to ab}^{\rm SiJF}
  +{\cal K}_{i\to ab}^{\Delta\Gamma}\right).
  \label{eq:local-subtraction-map}
\end{equation}
Equation~\eqref{eq:local-subtraction-map} is the subtraction statement behind our
matching prescription.  The subtraction kernel
\({\cal K}_{i\to ab}^{\Delta\Gamma}\) is chosen to be precisely the singular,
Wilson-line-resolved part of the real-emission kernel that generates
\(\Delta\Gamma_{i\to ab}^{\rm CGC}\).  In the notation of
Ref.~\cite{Wang:2022zla}, this means that the same singular real-emission limits that
are assigned to the inclusive NLO subtraction sectors are removed locally before a
multiplicity weight is assigned.  The \(\Delta\Gamma^{\rm CGC}\) term is allowed only
for the Wilson-line-resolved singular region that is not already removed by
vacuum-like collinear, rapidity, or Sudakov subtractions.  With this definition,
expanding the resummed multiplicity evolution through \(O(\alpha_s)\) removes exactly
the same target-resolved logarithmic contribution from the NLO hard coefficient, while
the remaining finite pieces are kept in the matched numerator.

The multiplicity-sensitive NLO correction can then be written in a subtraction-local
form,
\begin{align}
  \Delta\dd\sigma_{\rm mult}^{(1)}(s)
  &=
  \sum_{i\to ab}
  \int\dd\Pi_{i\to ab}^{\rm raw}
  \left\{
  {\cal K}_{i\to ab}^{\rm real}
  \left[
  {\cal M}_{i\to ab}^{\rm raw}(s)-{\cal M}_{i}^{\rm coh}(s)
  \right]
  \right.
  \nonumber\\
  &\qquad\left.
  -
  {\cal K}_{i\to ab}^{\rm sub}
  \left[
  {\cal M}_{i\to ab}^{\rm sing}(s)-{\cal M}_{i}^{\rm coh}(s)
  \right]
  \right\}
  .
  \label{eq:subtraction-local-mult}
\end{align}
Here \({\cal K}^{\rm sub}\) is the singular approximation assigned to an evolution
sector.  Depending on the limit, this sector is the PDF, BK, Sudakov, SiJF, or \(Z_i\)
evolution.  The quantity \({\cal M}^{\rm sing}\) is the corresponding measurement in
that same singular limit.  The finite remainder is imported only after
these logarithmic pieces have been removed.  Purely inclusive finite remainders and
virtual pieces that are not associated with a resolved daughter configuration are
evaluated with the coherent parent measurement; equivalently, in the isolated
multiplicity numerator they are evaluated at \(s=0\) and do not carry a
\(Z_aZ_b-Z_i\) weight.  This equation is the bridge between the operator-level
multiplicity measurement and the numbered hard coefficients used in the inclusive NLO
forward-jet calculation.

\section{Multiplicity-measured operators in a CGC background}
\label{sec:operator}

The previous sections treat \(Z_i\) as a final-state object multiplied onto the CGC
production kernel.  To understand when \(Z_i\) itself can depend on the target
saturation scale, one must define the multiplicity measurement before factorizing the
target field.  The operator language below combines the SCET jet-operator construction
\cite{Bauer:2000ew,Bauer:2001yt,Bauer:2002nz,Kang:2016mcy,Kang:2016ehg} with the CGC
Wilson-line description of high-energy scattering
\cite{McLerran:1993ni,McLerran:1993ka,Balitsky:1995ub,Kovchegov:1999yj,Iancu:2001ad,
KovchegovLevin:2012}.

For charged multiplicity inside a jet of radius \(R\), define the number operator
\begin{equation}
  \widehat N_{\rm ch}^{J_R}
  =
  \sum_{h\in{\rm charged}}
  \int\frac{\dd^3p}{(2\pi)^3 2E_p}\,
  \Theta_{J_R}(p)\,
  a_h^\dagger(p)a_h(p),
  \label{eq:number-operator}
\end{equation}
and the Laplace measurement operator
\begin{equation}
  \widehat{\Mcal}_s^{J_R}
  =
  e^{-s\widehat N_{\rm ch}^{J_R}},
  \qquad
  \widehat{\Mcal}_{s=0}^{J_R}=1.
  \label{eq:m-measurement}
\end{equation}

Let \(A^-\) be a fixed CGC background.  A fast quark and gluon crossing the target
acquire Wilson lines
\begin{align}
  \Ufund(x_\perp)
  &=
  {\cal P}\exp\left[
  \ii g\int\dd x^+\,t^a A^-_a(x^+,x_\perp)
  \right],
  \\
  \Uadj^{ab}(x_\perp)
  &=
  {\cal P}\exp\left[
  \ii g\int\dd x^+\,T^c_{ab} A^-_c(x^+,x_\perp)
  \right].
  \label{eq:wilson-lines}
\end{align}
The fixed-background quark multiplicity jet correlator can be written schematically as
\begin{equation}
  \widetilde\Mcal_q^{[A]}(z,s,\omega_J,R)
  =
  \frac{z}{2\Nc}
  {\rm Tr}
  \left\langle 0\left|
  \bar\chi_n
  \frac{\bar n\!\!\!/}{2}
  \Scal_A^\dagger
  \widehat{\Mcal}_s^{J_R}
  \widehat{\cal J}_{z,R}
  \Scal_A
  \chi_n
  \right|0\right\rangle,
  \label{eq:fixed-background-operator}
\end{equation}
where \(\Scal_A\) is the eikonal scattering operator generated by the background field
and \(\widehat{\cal J}_{z,R}\) imposes the jet measurement.  The gluon operator is
obtained by replacing \(\chi_n\) by the gauge-invariant collinear gluon building block
and fundamental Wilson lines by adjoint Wilson lines.

The physical CGC average is
\begin{equation}
  \widetilde\Mcal_i^{pA}(z,s,\omega_J,R;Y)
  =
  \int[D\rho]\,
  W_Y[\rho]\,
  \widetilde\Mcal_i^{[A[\rho]]}(z,s,\omega_J,R).
  \label{eq:cgc-average}
\end{equation}
At \(s=0\), this becomes the ordinary semi-inclusive jet object in the CGC background:
\begin{equation}
  \widetilde\Mcal_i^{pA}(z,0,\omega_J,R;Y)
  =
  J_i^{pA}(z,\omega_J,R;Y).
  \label{eq:s0-operator}
\end{equation}
This identity is the operator-level origin of the inclusive-limit constraint stated in
the introduction.

The multiplicity measurement is infrared sensitive but it is not a new ultraviolet
operator insertion.  At fixed partonic external states, the UV poles of the measured
semi-inclusive jet correlator are the same as those of the ordinary SiJF; the measurement
operator changes how real radiation is weighted after the UV region has been separated,
but it does not introduce a new short-distance counterterm.  This is the same
renormalization logic used for semi-inclusive jet functions and their multiplicity
extension \cite{Kang:2016mcy,Kang:2016ehg,Duan:2025hmjets}.  Equivalently,
\begin{equation}
  \widetilde{\SiJF}_{ji}^{\rm bare}(z,s)
  =
  \sum_k\int_z^1\frac{\dd z'}{z'}\,
  Z_{jk}^{\rm SiJF}\!\left(\frac{z}{z'},\mu\right)
  \widetilde{\SiJF}_{ki}(z',s,\mu),
  \label{eq:measured-sijf-renorm}
\end{equation}
with the same renormalization factor \(Z_{jk}^{\rm SiJF}\) as in the unmeasured case.
All \(s\)-dependence resides in the renormalized matrix element and in its
nonperturbative boundary condition.  This separation is essential: perturbative UV
evolution may be evolved with the standard SiJF/DGLAP kernels, while the multiplicity
distribution itself remains an infrared observable.

If the correlator factorizes for a given background, one may define
\begin{equation}
  Z_i^{[A]}(s,z,\omega_J,R)
  =
  \frac{\widetilde\Mcal_i^{[A]}(z,s,\omega_J,R)}
       {\widetilde\Mcal_i^{[A]}(z,0,\omega_J,R)},
  \qquad
  Z_i^{[A]}(0)=1.
  \label{eq:background-z}
\end{equation}
The target-averaged object relevant for inclusive phenomenology is then
\begin{equation}
  Z_i^{pA}(s)
  =
  \frac{\int[D\rho]W_Y[\rho]\widetilde\Mcal_i^{[A]}(s)}
       {\int[D\rho]W_Y[\rho]\widetilde\Mcal_i^{[A]}(0)}.
  \label{eq:averaged-z}
\end{equation}
The phenomenological object in Eq.~\eqref{eq:averaged-z} is defined as the ratio of
ensemble-averaged matrix elements.  In general it must not be identified with
\(\langle Z_i^{[A]}(s)\rangle_Y\), the ensemble average of the fixed-background ratio,
unless an additional factorization or decorrelation approximation between
\(\widetilde\Mcal_i^{[A]}(0)\) and \(Z_i^{[A]}(s)\) is imposed.  Gaussianity of the
target weight alone does not guarantee that decorrelation.

In the thin-shockwave limit, if the resolved in-cone shower forms after the projectile
has crossed the target, \(\Scal_A\) acts only on the coherent parent parton.  Then
\begin{equation}
  \widetilde\Mcal_i^{pA}(z,s)
  =
  {\cal F}_i^{\rm CGC}(Y)\,
  \sum_j \SiJF_{ji}^{\rm vac}(z)Z_j^{\rm vac}(s)
  +
  {\cal O}\!\left(\frac{\Qs^2}{k_{\perp,{\rm split}}^2}\right).
  \label{eq:vac-factorization-limit}
\end{equation}
Here \(k_{\perp,{\rm split}}\) is the relative transverse momentum of the in-cone
splitting, not the production recoil \(q_\perp=P_T/z\).  Its upper perturbative range
is parametrically set by \(k_{\perp,{\rm split}}\sim P_T R\).  The same range may be
expressed through the physical angular variable \(\zeta_{\rm phys}\) or through the
hadron-collider variable \(\rho\) introduced below.  This is the
vacuum-multiplicity baseline.  Corrections
arise when the target resolves the daughter partons in an in-cone splitting.

\section{One-loop CGC correction to the multiplicity anomalous dimension}
\label{sec:delta-gamma}

Consider a parent parton \(i\) with SCET label momentum
\(\omega\equiv\bar n\!\cdot p_i\simeq2E_i\) splitting into daughters \(a,b\) with
energy fractions \(x\) and \(1-x\), separated by a transverse distance
\(r_\perp\).  The leading light-front splitting probability,
written in the usual small-\(x\) light-front wave-function language
\cite{KovchegovLevin:2012,Dominguez:2011wm,Chirilli:2011km,Chirilli:2012jd}, has
the structure
\begin{equation}
  |\psi_{i\to ab}(x,\bm r)|^2
  =
  \frac{\as}{2\pi^2}
  P_{i\to ab}(x)\frac{1}{r^2}
  +
  \text{spin-dependent finite terms}.
  \label{eq:lf-wavefunction}
\end{equation}
Let \(\theta\) denote the physical three-dimensional opening angle and
\(\zeta_{\rm phys}=1-\cos\theta\).  The anti-\(k_T\) two-particle boundary is most
directly expressed through
\(k_\perp=x(1-x)\omega\tan(\theta/2)\)~\cite{Kang:2016mcy,Duan:2025hmjets}.
Consequently, in vacuum the angular branching kernel is
\begin{equation}
  \Gvac_{i\to ab}(x,\omega,\zeta_{\rm phys})
  =
  \frac{\as}{2\pi}P_{i\to ab}(x)
  \Theta(k_\perp^2-Q_0^2),
  \qquad
  k_\perp^2=x^2(1-x)^2\omega^2
  \frac{\zeta_{\rm phys}}{2-\zeta_{\rm phys}}.
  \label{eq:vac-gamma}
\end{equation}
For a forward hadron-collider jet the physical cone angle is
\(\mathcal R\simeq R/\cosh\eta\), not the rapidity--azimuth distance \(R\) itself.
Introducing the transverse label energy
\(\omega_T\equiv\omega/(2\cosh\eta)\simeq P_T\) and
\(\rho\equiv(\Delta R)^2/2\), the same small-angle kinematics becomes
\begin{equation}
  k_\perp^2=2\rho\,x^2(1-x)^2\omega_T^2,
  \qquad \rho_R=R^2/2,
  \label{eq:hadron-collider-angular-map}
\end{equation}
which is the boost-invariant form used in the numerical implementation.

In a CGC background the two daughters scatter with separate Wilson lines.  We first
divide out the coherent parent scattering factor.  We then account for the finite
overlap between the splitting formation time and the target coherence length.  The
real-emission kernel is thereby modified by an effective resolution factor,
\begin{equation}
  {\cal R}_{i\to ab}(k_\perp,Y)
  \equiv
  \frac{
  \left\langle{\cal C}_{ab}^{\rm daughter}
  (x,1-x;r_\perp,Y)\right\rangle
  }{
  \left\langle{\cal C}_{i}^{\rm parent}
  (r_\perp,Y)\right\rangle
  },
  \qquad
  r_\perp\simeq \frac{1}{k_\perp}.
  \label{eq:resolution-factor-general}
\end{equation}
The numerator is the Wilson-line correlator appropriate to the resolved daughter color
charges and the denominator is the coherent parent correlator in the same channel.
This ratio equals unity when the target sees only the parent charge, and differs from
unity when the daughter separation is resolved by the target field.  Its large-\(\Nc\)
realization and its small-dipole expansion are given below.
In the large-\(\Nc\) dipole reduction used in the baseline numerical study, the parent
correlators are
\begin{equation}
  {\cal C}^{\rm parent}_q(r,Y)=S_F(r,Y),
  \qquad
  {\cal C}^{\rm parent}_g(r,Y)=S_A(r,Y),
  \label{eq:parent-correlators}
\end{equation}
and the daughter correlators are
\begin{align}
  {\cal C}^{\rm daughter}_{q\to qg}(x,r,Y)
  &=
  S_F((1-x)r,Y)S_A(xr,Y),
  \nonumber\\
  {\cal C}^{\rm daughter}_{g\to gg}(x,r,Y)
  &=
  S_A(xr,Y)S_A((1-x)r,Y),
  \nonumber\\
  {\cal C}^{\rm daughter}_{g\to q\bar q}(x,r,Y)
  &=
  S_F(r,Y),
  \label{eq:daughter-correlators-explicit}
\end{align}
up to the finite-\(\Nc\) and soft-pair caveats discussed below.  These definitions fix
the normalization of \({\cal R}_{i\to ab}\).  In the small-separation limit
\(r_\perp\to0\), where the target cannot resolve the daughter separation, the
normalized ratios satisfy \({\cal R}_{i\to ab}\to1\).
\begin{equation}
  \Gamma^{h}_{i\to ab}(x,\omega,\zeta_{\rm phys};Y)
  =
  \Gvac_{i\to ab}(x,\omega,\zeta_{\rm phys})
  \left[
  1+
  \left({\cal R}^{h}_{i\to ab}(k_\perp,Y)-1\right)
  {\cal T}_{\rm form}(k_\perp,\omega,x;L_h^+)
  \right].
  \label{eq:cgc-gamma}
\end{equation}
Here \(h=p,A\) labels the target ensemble.  In a complete target-dipole comparison, the
physical nuclear correction to the multiplicity kernel is the difference between the
finite-dipole corrections generated by \(S_A\) and by the corresponding proton dipole
\(S_p\), not the nuclear correction relative to a structureless target.  The
correction relative to the vacuum kernel for a given target ensemble is
\begin{equation}
  \Delta\Gamma^h_{i\to ab}
  =
  \Gvac_{i\to ab}
  \left[
  {\cal R}^h_{i\to ab}(k_\perp,Y)-1
  \right]
  {\cal T}_{\rm form}(k_\perp,\omega,x;L_h^+).
  \label{eq:delta-gamma}
\end{equation}
The formation-time gate accounts for whether the target can resolve the splitting:
\begin{equation}
  {\cal T}_{\rm form}
  =
  1-\exp\left[-\frac{L^+}{t_f}\right],
  \qquad
  t_f=\frac{2x(1-x)\omega}{k_\perp^2}.
  \label{eq:formation-gate}
\end{equation}
This is the light-front \(x^+\) formation time obtained from
\(t_f=1/\Delta p^-\), where
\begin{equation}
  \Delta p^-=\frac{k_\perp^2}{2x(1-x)p_i^+}.
\end{equation}
If one instead uses the SCET label
\(\bar n\!\cdot p\) or the ordinary energy \(E\) as the large variable, the overall
factor in \(t_f\) changes by convention.  Such factors are absorbed into the effective
target length \(L_{\rm eff}^+\); the physical gate depends on the ratio of the
formation length to the target coherence length, not on the notation used for the large
light-cone momentum.
For the boost-invariant numerical variables of
Eq.~\eqref{eq:hadron-collider-angular-map}, we divide both longitudinal quantities by
the same factor \(2\cosh\eta\).  This defines
\(t_{f,T}=2x(1-x)\omega_T/k_\perp^2\) and
\(L_{{\rm eff},T}=L_{\rm eff}^+/(2\cosh\eta)\).  The gate is unchanged,
\(L_{\rm eff}^+/t_f=L_{{\rm eff},T}/t_{f,T}\).  Thus no forward-rapidity boost factor
is hidden in the numerical use of \(R\).
If the jet shower occurs long after crossing a zero-thickness shockwave, then
${\cal T}_{\rm form}\to0\) and the internal multiplicity evolution is vacuum-like.  If
the formation time is comparable to the target coherence length, or if
\(k_\perp^2\sim \Qs^2\), the Wilson-line scattering can resolve the daughters and
\(\Dg\) is nonzero.

There are two competing suppressions.  A highly boosted nucleus is Lorentz contracted,
so the light-cone thickness \(L^+\) is small and late splittings with \(t_f\gg L^+\) are
formed after the shockwave.  Those splittings contribute to the usual vacuum-like jet
evolution.  On the other hand, very early splittings require a large relative transverse
momentum, with \(t_f\simeq 2x(1-x)\omega/k_\perp^2\).  The small-dipole expansion then
suppresses the target resolution by \(Q_s^2/k_\perp^2\).  The genuine
Wilson-line-resolved correction is therefore largest in the intermediate window
\begin{equation}
  k_\perp^2 \sim Q_s^2(Y),
  \qquad
  t_f(k_\perp,x,\omega)\lesssim L^+_{\rm eff},
  \qquad
  k_\perp \lesssim P_T R,
  \label{eq:resolution-window}
\end{equation}
where \(L^+_{\rm eff}\) denotes the effective coherence length over which the splitting
and the target field overlap.  The last inequality is the jet-cone constraint.  At
large \(P_T\) and fixed \(R\), the available perturbative phase space grows, but the
Wilson-line resolution correction remains concentrated near \(k_\perp\) of order
\(Q_s\).  At very small \(P_T R\), by contrast, the jet cone may not contain enough
transverse phase space for a perturbative, target-resolved in-cone splitting.  This is
the kinematic reason why a \(Q_s\)-dependent multiplicity anomalous dimension is
physically allowed but not automatically dominant.

At large \(\Nc\), the channel-dependent resolution correlators may be approximated by
dipoles:
\begin{align}
  {\cal C}_{q\to qg}(\bm r;Y)
  &=
  \Sdip_F((1-x)\bm r;Y)\,
  \Sdip_A(x\bm r;Y),
  \\
  {\cal C}_{g\to gg}(\bm r;Y)
  &=
  \Sdip_A(x\bm r;Y)\,
  \Sdip_A((1-x)\bm r;Y),
  \\
  {\cal C}_{g\to q\bar q}(\bm r;Y)
  &=
  \Sdip_F(\bm r;Y).
  \label{eq:channel-correlators}
\end{align}
The last relation uses the large-\(\Nc\) approximation in the soft-pair limit where the
\(q\bar q\) pair shares the parent transverse coordinate; corrections of order
\(x(1-x)r^2Q_s^2\) are suppressed in the small-dipole expansion.
Away from large \(\Nc\), quadrupoles and higher Wilson-line correlators enter, and the
closed BK approximation must be replaced by JIMWLK or a Gaussian truncation.

For \(k_\perp^2\gg\Qs^2\), the dipole expansion
\begin{equation}
  \Sdip_R(r;Y)
  =
  1-\frac{C_R}{C_F}\frac{r^2\QsF^2(Y)}{4}
  +
  {\cal O}(r^4\Qs^4)
  \label{eq:small-dipole}
\end{equation}
gives
\begin{equation}
  {\cal R}_{i\to ab}(k_\perp,Y)-1
  =
  -\lambda_{i\to ab}(x)
  \frac{\QsF^2(Y)}{k_\perp^2}
  +
  {\cal O}\!\left(\frac{\Qs^4}{k_\perp^4}\right),
  \label{eq:small-dipole-resolution}
\end{equation}
with, in the minimal angular-shell estimate,
\begin{align}
  \lambda_{q\to qg}(x)
  &=
  c_{\rm sh}
  \left[(1-x)^2+\frac{\CA}{\CF}x^2-1\right],
  \\
  \lambda_{g\to gg}(x)
  &=
  -c_{\rm sh}\frac{\CA}{\CF}\,2x(1-x),
  \\
  \lambda_{g\to q\bar q}(x)
  &=
  c_{\rm sh}\left(1-\frac{\CA}{\CF}\right).
  \label{eq:lambdas}
\end{align}
The scheme-dependent constant \(c_{\rm sh}\) must be fixed with the same angular-shell
prescription used in the numerical multiplicity solver.  The expressions above should
be read as the small-dipole limits of the channel correlators in
Eq.~\eqref{eq:channel-correlators}.  In particular,
\(\lambda_{g\to gg}\) and \(\lambda_{g\to q\bar q}\) are negative, while
\(\lambda_{q\to qg}\) changes sign as a function of the daughter energy fraction.
Therefore the sign of the small-dipole correction is channel and phase-space
dependent; the numerical analysis below uses the full dipole ratios rather than
imposing a sign from the expansion.  Combining Eqs.~\eqref{eq:delta-gamma} and
\eqref{eq:small-dipole-resolution} gives the leading parametric correction to the
multiplicity evolution,
\begin{equation}
  \Delta\Gamma^h_{i\to ab}
  \sim
  -\Gvac_{i\to ab}
  \lambda_{i\to ab}(x)
  \frac{\left(Q_s^h\right)^2}{k_\perp^2}
  {\cal T}_{\rm form}.
  \label{eq:parametric-dg}
\end{equation}
When the finite-dipole ratio satisfies \({\cal R}_{i\to ab}>1\), the daughters scatter
more independently than a coherent parent and the in-cone branching rate is enhanced.
When \({\cal R}_{i\to ab}<1\), the same formula describes a depletion.  This equation
therefore explains why a \(\Qs\)-dependent \(Z_i\) is physically allowed but not
guaranteed to dominate: it is power suppressed when the splitting transverse momentum
is much larger than \(\Qs\), and it vanishes if the shower forms after the target.

\section{Gaussian CGC ensemble and small-\(x\) dipole input}
\label{sec:ensemble}

To make the Wilson-line correlators calculable, we use a Gaussian truncation of the
CGC ensemble.  In a first-principles small-\(x\) implementation, its dipole input can be
evolved by the BK/rcBK framework
\cite{Balitsky:1995ub,Kovchegov:1999yj,Iancu:2001ad,Albacete:2010bs,
KovchegovLevin:2012}.  The fundamental dipole is
\begin{equation}
  S_F(r,Y)
  =
  \frac{1}{\Nc}
  \left\langle
  {\rm Tr}\,U(x)U^\dagger(y)
  \right\rangle_Y,
  \qquad r=x-y.
  \label{eq:fund-dipole}
\end{equation}
The Gaussian truncation assumes that all target averages can be expressed in terms of a
two-point color charge correlator.  Equivalently, multi-Wilson-line operators are
obtained by exponentiating a finite color-transition matrix built from pairwise dipole
exponents.  Let \(\{|B_\alpha\rangle\}\) be an orthonormal basis of the overall singlet
subspace of the amplitude--conjugate-amplitude Wilson-line system,
\(\langle B_\alpha|B_\beta\rangle=\delta_{\alpha\beta}\), and let
\(P_{\rm singlet}=\sum_\alpha |B_\alpha\rangle\langle B_\alpha|\).  For a set of
Wilson lines at transverse positions \(x_i\), define
\begin{equation}
  \mathbb E(\{x_i\};Y)
  =
  \exp\left[
  -\sum_{i<j}
  T_i^a T_j^a\,\Gamma(x_i-x_j,Y)
  \right].
  \label{eq:gaussian-evolution-operator}
\end{equation}
We use an all-lines-outgoing color convention.  A Wilson line belonging to the
conjugate amplitude therefore carries the conjugate-representation generator
\(\bar T^a=-(T^a)^*\); with this convention the pairwise products in
Eq.~\eqref{eq:gaussian-evolution-operator} act on a single
amplitude--conjugate-amplitude color space.
The generic singlet correlator is then
\begin{equation}
  {\cal C}^{\rm singlet}_{\alpha\beta}(\{x_i\};Y)
  =
  \langle B_\alpha|\mathbb E(\{x_i\};Y)|B_\beta\rangle
  =
  \left[
  P_{\rm singlet}\,\mathbb E(\{x_i\};Y)\,P_{\rm singlet}
  \right]_{\alpha\beta}.
  \label{eq:gaussian-master}
\end{equation}
where the scalar function \(\Gamma(r,Y)\) is fixed by the dipole,
\begin{equation}
  S_F(r,Y)
  =
  \exp[-C_F\Gamma(r,Y)].
  \label{eq:gamma-from-dipole}
\end{equation}
This representation is more general than the large-\(\Nc\) dipole product used in
Eq.~\eqref{eq:channel-correlators}.  Formally, it keeps finite-\(\Nc\) color
transitions in a reduced singlet basis while still closing the hierarchy in terms of
the dipole input.  More explicitly, the \(q\to qg\) operator belongs to the singlet
sector of the fundamental--adjoint--fundamental-conjugate color space.  The \(g\to gg\)
operator belongs to the singlet sector of adjoint Wilson lines in the amplitude and
conjugate amplitude.  The \(g\to q\bar q\) operator belongs to the singlet
fundamental-pair sector.  In
a fully finite-\(\Nc\) implementation the matrices \(T_i^aT_j^a\) in
Eq.~\eqref{eq:gaussian-master} are evaluated in these channel-dependent singlet bases;
for the three-adjoint singlet subspace one may equivalently use the standard
\(f^{abc}\) and \(d^{abc}\) color tensors as a basis.  The finite-\(\Nc\) singlet-basis
formula is therefore an operator-level definition and a matching target for future
Gaussian-ensemble implementations.  It is not the object evaluated in the baseline
numerical plots.  The baseline numerical results in Sec.~\ref{sec:numerical-status}
use the large-\(\Nc\)/Gaussian dipole reductions displayed in
Eq.~\eqref{eq:channel-correlators}, not the full finite-\(\Nc\) matrix exponential.
In the McLerran--Venugopalan (MV) model \cite{McLerran:1993ni,McLerran:1993ka},
\begin{equation}
  S_F^{\rm MV}(r)
  =
  \exp\left[
  -\frac{r^2 Q_{s0}^2}{4}
  \ln\left(e+\frac{1}{r\Lambda_{\rm IR}}\right)
  \right],
  \label{eq:mv-dipole}
\end{equation}
possibly supplemented by a phenomenological rapidity dependence
\begin{equation}
  Q_s^2(Y)=Q_{s0}^2 e^{\lambda_s Y}.
  \label{eq:qs-y}
\end{equation}
For a genuine rcBK implementation, the tabulated dipole would instead be obtained by
solving the running-coupling BK equation,
\begin{align}
  \frac{\partial S_F(r,Y)}{\partial Y}
  &=
  \int \dd^2 r_1\,
  K_{\rm rc}(r,r_1,r_2)
  \left[
  S_F(r_1,Y)S_F(r_2,Y)-S_F(r,Y)
  \right],
  \nonumber\\
  r_2&=r-r_1,
  \label{eq:rcbk}
\end{align}
with an MV-type initial condition at \(Y=Y_0\).  This equation defines the target
evolution needed for a future first-principles production table; it is not the equation
numerically solved for the baseline curves in Sec.~\ref{sec:numerical-status}.  The
adjoint dipole is approximated by
the Gaussian Casimir relation
\begin{equation}
  S_A(r,Y)=\left[S_F(r,Y)\right]^{\CA/\CF}.
  \label{eq:adjoint-gaussian}
\end{equation}
This relation is exact within a Gaussian two-point exponent and provides a practical
closure for the coordinate-space kernels.

The channel correlators entering \(\Delta\Gamma^{\rm CGC}\) can then be computed from
the same \(\Gamma(r,Y)\).  In the baseline large-\(\Nc\)/Gaussian reduction one obtains the
dipole products in Eq.~\eqref{eq:channel-correlators}.  A future fully finite-\(\Nc\)
Gaussian implementation would replace these products by the singlet-basis matrix
elements described above.  In a pure target-dipole implementation, one uses a common
dipole input \(S_F^h(r,Y)\) for each target ensemble \(h=p,A\).  This input controls
both the inclusive CGC production kernel and the Wilson-line resolution factor in the
multiplicity kernel.
This shared input is what a fully matched target-ensemble calculation must enforce.
The numerical baseline in Sec.~\ref{sec:numerical-status} is more limited.  Its
production rows use an analytic-tail momentum-space table with a phenomenological
\(Q_s(x_A)\), whereas the internal multiplicity evolution uses the coordinate-space
MV/rcBK-inspired ansatz of Eq.~\eqref{eq:mv-dipole} with
Eq.~\eqref{eq:qs-y}.  Nuclear-PDF ratios are then applied as an effective
channel-dependent production reweighting.  The baseline is therefore a diagnostic
implementation of the matching structure, not a common-ensemble rcBK calculation or a
complete separate-\(S_p\)/\(S_A\) target-dipole comparison.

The limitation is important.  The Gaussian truncation does not give the exact
finite-\(\Nc\) JIMWLK correlators for all multi-parton channels.  It is a controlled
ensemble model.  Therefore any claim about a first-principles \(\Qs\)-dependent
multiplicity anomalous dimension beyond the small-dipole limit must be stated with this
ensemble dependence.

\section{Mapping to NLO hard coefficients}
\label{sec:hard-coefficients}

The NLO hard coefficients contain diagonal and off-diagonal channels.  In the notation
of Ref.~\cite{Wang:2022zla}, each \(\sigma_{ab}^n\) denotes one finite or subtraction
sector in the NLO forward-jet cross section, where the first index labels the observed
jet parton after the splitting and the second labels the incoming projectile parton.
Our import map does not redefine these coefficients.  It identifies the finite,
real-resolved pieces that can be multiplied by the local multiplicity weight
\({\cal M}^{\rm raw}-{\cal M}^{\rm coh}\).  It also identifies the pieces absorbed
into PDF, BK/JIMWLK, Sudakov, or SiJF evolution.  With this convention, the off-diagonal
finite pieces used here are
\begin{align}
  q\to g:&\qquad
  \sigma_{gq}^{\rm finite}
  =
  \sigma_{gq}^{2}
  +
  \sigma_{gq}^{3}
  +
  \sigma_{gq}^{5},
  \\
  g\to q:&\qquad
  \sigma_{qg}^{\rm finite}
  =
  \sigma_{qg}^{2}
  +
  \sigma_{qg}^{3}
  +
  \sigma_{qg}^{5}.
  \label{eq:offdiag-finite}
\end{align}
The logarithmic \(\sigma^1\) and \(\sigma^4\) pieces belong to PDF/SiJF evolution and
are not imported as finite multiplicity kernels.

For the dominant diagonal gluon channel we use the notation of
Ref.~\cite{Wang:2022zla}.  The inclusive denominator is built from the LO term and the
finite, non-evolution part of the NLO gluon-channel coefficients.  The full channel is
listed in Appendix~\ref{app:hard-coefficients} as
\(\dd\sigma_{gg}^{\rm LO}+\sum_{i=1}^{11}\dd\sigma_{gg}^i\).  Not every numbered term
is a daughter-resolved hard kernel.  In the scheme of Ref.~\cite{Wang:2022zla},
\(\dd\sigma_{gg}^1\) contains the projectile-collinear \(P_{gg}\) counterterm.
The term \(\dd\sigma_{gg}^{11}\) contains the jet-radius logarithm assigned to the
SiJF sector.  LO-like running-coupling and scale pieces, such as
\(\dd\sigma_{gg}^2\), enter the inclusive denominator and the scale-variation check.
They are not assigned a resolved \(g\to gg\) multiplicity weight.  After these sectors
are separated, the real-resolved
finite part that would enter a fully daughter-differential gluon numerator is
\begin{equation}
  \dd\sigma_{gg}^{\rm fin}
  =
  \dd\sigma_{gg}^{3}
  +\dd\sigma_{gg}^{4}
  +\dd\sigma_{gg}^{5}
  +\dd\sigma_{gg}^{6}
  +\frac{1}{2}\dd\sigma_{gg}^{7a}
  +\dd\sigma_{gg}^{7b}
  +\dd\sigma_{gg}^{8}
  +\dd\sigma_{gg}^{9}
  +\dd\sigma_{gg}^{10}.
  \label{eq:gg-finite}
\end{equation}
Equation~\eqref{eq:gg-finite} is therefore not a claim that the omitted terms are absent
from the inclusive NLO cross section.  It is an import map for the part of the NLO
coefficient that can be paired with the local cone measurement.  The omitted logarithmic
and endpoint sectors remain in the inclusive denominator through their assigned
evolution factors.  These are the PDF, running-coupling, Sudakov/threshold, and SiJF
sectors.  They are checked in the \(s=0\) denominator consistency test.
The measured numerator is not a new hard coefficient.  At the fully differential level
it is obtained by inserting the channel-dependent local multiplicity difference before
the daughter phase space is integrated, before singular limits are subtracted, and
before the final Fourier transform.  The parent-gluon coefficient contains both
\(g\to gg\) and \(g\to q\bar q\) cuts, so a single \(g\to gg\) weight cannot multiply
the complete finite kernel.  We therefore define
\begin{align}
  D_{gg}^{\rm incl}
  &=
  \int \dd\Pi_{gg}^{\rm LO}\,
  {\cal K}_{gg}^{\rm LO}
  +
  \int \dd\Pi_{gg}^{\rm fin}\,
  {\cal K}_{gg}^{\rm fin},
  \nonumber\\
  N_{g}^{\rm mult}(s)
  &=
  \sum_{ab\in\{gg,q\bar q\}}
  \int \dd\Pi_{g\to ab}^{\rm fin}\,
  {\cal K}_{g\to ab}^{\rm fin,diff}
  \left[\mathcal M_{g\to ab}^{\rm raw}(s)-Z_g(s,\omega)\right],
  \nonumber\\
  \delta_{g}^{\rm match}(s)
  &=
  \frac{N_{g}^{\rm mult}(s)}{D_{gg}^{\rm incl}}.
  \label{eq:delta-gg-match}
\end{align}
Here \({\cal K}_{g\to ab}^{\rm fin,diff}\) denotes the corresponding cut, daughter-
differential component of the finite combination in Eq.~\eqref{eq:gg-finite}; it is not
an additional hard coefficient beyond Ref.~\cite{Wang:2022zla}.  Equation
\eqref{eq:delta-gg-match} is the formal matching definition.  The reduced
\(K_{82}\)--\(K_{85}\) representation of Ref.~\cite{Wang:2022zla} has already
undergone part of the inclusive daughter integration and therefore does not define
\({\cal K}_{g\to ab}^{\rm fin,diff}\) uniquely.  It is excluded from the displayed
baseline.

In practice, the measured Wilson-line operator is most naturally inserted in coordinate
space, whereas Appendix~\ref{app:hard-coefficients} lists the finite hard coefficients in
the momentum-space notation of Ref.~\cite{Wang:2022zla}.  Their formal consistency
condition is
\begin{equation}
  \left.
  \int\dd\Pi_{\rm coord}\,
  {\cal K}_{i}^{\rm coord}\,
  {\cal M}_{i}^{\rm coh}(s)
  \right|_{s=0}
  =
  \int\dd\Pi_{\rm mom}\,
  {\cal K}_{i}^{\rm mom}
  \label{eq:coord-mom-validation}
\end{equation}
Equation~\eqref{eq:coord-mom-validation} is a consistency requirement, not a definition
of the coordinate kernels.  A reduced coordinate-space implementation of the
parent-gluon terms was tested, but an independent quadrature refinement changed the
integrated result by factors ranging from \(0.73\) to \(15.69\).  It therefore fails
the integral-convergence requirement, and neither its denominator correction nor its
candidate measured numerator is used in the displayed baseline.  The momentum-space expressions in
Appendix~\ref{app:hard-coefficients} remain the formal hard-coefficient bookkeeping
target; no numerical coordinate-to-momentum closure is claimed.

The ratio \(\delta_{g}^{\rm match}(s)\) is a measurement-dependent finite NLO
cross-section correction; it is not an anomalous dimension and is therefore not
multiplied into the branching kernel.  Its proper use is the additive matched
combination
\begin{equation}
  \dd\sigma_{g}^{\rm mult}(s)
  =D_{gg}^{\rm incl}{\cal M}_{g}^{\rm coh}(s)+N_{g}^{\rm mult}(s)
  =D_{gg}^{\rm incl}\left[{\cal M}_{g}^{\rm coh}(s)+\delta_{g}^{\rm match}(s)\right].
  \label{eq:gg-cross-section-match}
\end{equation}
Because \({\cal M}_{g}^{\rm coh}(0)=1\) and
\(N_{g}^{\rm mult}(0)=0\), Eq.~\eqref{eq:gg-cross-section-match} reduces to
\(D_{gg}^{\rm incl}\) in the inclusive limit.  The target-dependent correction to the
branching kernel is instead the local, measurement-independent quantity
\(\Delta\Gamma^h\) defined in Sec.~\ref{sec:delta-gamma}.  The same separation between
finite cross-section matching and anomalous-dimension evolution applies to the quark
and flavor-transfer channels.

\section{Numerical implementation and baseline validation}
\label{sec:numerical-status}

The numerical implementation presented here tests the internal consistency of the
factorized construction with a common inclusive denominator and illustrates the size and direction of the
multiplicity-dependent effects.  In particular, the diagonal quark finite remainder
exposes a fixed-order stability issue at the upper end of the \(P_T\) scan, and we keep
this limitation explicit throughout the section.
The implementation has two tasks.  First, it must record the projectile PDF, effective
nuclear ratios, analytic-tail production input, \(z\)-convolution, and channel
bookkeeping that define each inclusive production row.  Second, once this common
denominator has been fixed, it must convert the
matched kernels into baseline observables that can be compared between pp and pA.  We
therefore organize the numerical illustration around the per-nucleon inclusive baseline
ratio
\begin{equation}
  {\cal R}_{\rm base}^{\rm incl}(P_T,\eta)
  =
  \frac{D^{pA}(P_T,\eta)}{D^{pp}(P_T,\eta)},
  \qquad
  D^{pA}=\sum_i D_i^{pA},
  \label{eq:incl-rpa-phen}
\end{equation}
and its high-multiplicity counterpart
\begin{equation}
  {\cal R}_{\rm base}(n_{\rm cut};P_T,\eta)
  =
  \frac{
  \sum_i D_i^{pA}(P_T,\eta)\sum_{n\ge n_{\rm cut}}P_i^{pA}(n;P_T,\eta)}
  {\sum_i D_i^{pp}(P_T,\eta)\sum_{n\ge n_{\rm cut}}P_i^{pp}(n;P_T,\eta)} .
  \label{eq:tail-ratio-final}
\end{equation}
The numerator and denominator in Eq.~\eqref{eq:tail-ratio-final} use the same
production rows as Eq.~\eqref{eq:incl-rpa-phen}; the only new ingredient is the
multiplicity evolution.  This common-denominator construction is the main numerical
consistency condition in this section.
The symbols in Eqs.~\eqref{eq:incl-rpa-phen} and \eqref{eq:tail-ratio-final} are
deliberately distinguished from the physical nuclear modification factor
\(R_{pA}\) of Eq.~\eqref{eq:rpa-n}.  EPPS21 ratios are defined per bound nucleon, so
\(D^{pA}\) in the displayed phenomenological baseline is constructed directly as a
per-nucleon reweighted production weight.  No absolute transverse-area normalization
is claimed.

The production rows used by the displayed numerical baseline are more restricted than
the formal matched construction.  They are the LO-like hybrid weights defined
explicitly in Eqs.~\eqref{eq:implemented-production-weights} and
\eqref{eq:analytic-tail-production}; the reduced
\(K_{82}\)--\(K_{85}\) combination and the off-diagonal finite grids are not added to
them.  The only finite hard-coefficient modification promoted to these rows is the
diagonal quark factor, treated cautiously:
after correcting the \(\sigma_{qq}^{5}\) prefactor and the local \(\sigma_{qq}^{4}\)
bookkeeping, it is included only where the \(z\)-convoluted matched denominator remains
positive and perturbatively stable.  When the diagonal-quark finite remainder fails this
integrated acceptance test, that unstable channel contribution is excluded at the
corresponding kinematic point; the remaining restricted ingredients are shown only as a
reduced-channel baseline.
More explicitly, Figs.~\ref{fig:incl-rpa-final}--\ref{fig:npdf-set-comparison} use
\begin{equation}
  \dd\sigma_{\rm base}(n)
  =\sum_{i=q,g}D_i^{\rm base}\,P_i^{\Delta\Gamma}(n),
  \label{eq:implemented-baseline-cross-section}
\end{equation}
where \(D_i^{\rm base}\) denotes precisely the restricted production weights described
above and \(P_i^{\Delta\Gamma}\) is obtained by evolving the generating function with the
local Wilson-line correction of Sec.~\ref{sec:delta-gamma}.  No inverse transform of
\(N_i^{\rm mult}(s)\) is added to Eq.~\eqref{eq:implemented-baseline-cross-section}.
This distinction is summarized in Table~\ref{tab:measured-implementation-map}.

\begin{table}[t]
\centering
\scriptsize
\begin{tabularx}{\textwidth}{L{0.28\textwidth}L{0.18\textwidth}L{0.22\textwidth}Y}
\toprule
Contribution & Production weight & Multiplicity insertion & Status \\
\midrule
LO-like hybrid weights & yes & \(P_i^{\Delta\Gamma}\) & used \\
accepted diagonal-\(q\) factor & yes, accepted points & no & stability-audited \\
local \(\Delta\Gamma_{i\to ab}^{\rm CGC}\) & -- & yes & used \\
reduced \(K_{82}\)--\(K_{85}\) terms & no & no & excluded after failed convergence test \\
finite measured numerator \(N_i^{\rm mult}\) & no & no & not implemented in the plotted baseline \\
off-diagonal finite pieces & no & no & not implemented in the plotted baseline \\
\bottomrule
\end{tabularx}
\caption{Binary implementation map for the displayed numerical baseline.  The finite
measured numerators are defined by the matching construction but are not added to the
published curves; the only multiplicity-dependent nuclear insertion in those curves is
the local \(\Delta\Gamma\) evolution.}
\label{tab:measured-implementation-map}
\end{table}
The figures in this section are therefore not independent diagnostics.  They form a
single validation chain: first the transverse-momentum support of the
Wilson-line-resolution correction is checked, then the correction is propagated to
multiplicity distributions, and only after these two steps are the inclusive and
multiplicity-conditioned baseline ratios constructed.  The later figures
and the stability table then identify which part of the effect is driven by flavor
composition, which part is controlled by nuclear-input uncertainties, and where the
fixed-order hard-coefficient bookkeeping still limits the numerical claims.

\subsection{Detailed implementation workflow}
\label{sec:detailed-implementation}

The numerical calculation is organized in five modules that mirror the factorized
structure derived above.  The first module supplies the production PDF and saturation
inputs.  The scan uses \(P_T=5,8,12,16,20\) GeV,
\(\eta=3.0,3.2,3.4\), and \(\sqrt{s}=5.02\) TeV.  For
\(i=q,g\), the rows actually used in the figures are
\begin{align}
 D_i^{pp}
 &=\int_\tau^1\frac{\dd z}{z^2}\,
   x_p f_i^{p}(x_p,\mu)\,{\cal F}_i^{p}(P_T/z,x_A),
 \nonumber\\
 D_i^{pA}
 &=A^{-1/3}\int_\tau^1\frac{\dd z}{z^2}\,
   x_p f_i^{p}(x_p,\mu)\,R_i^A(x_A,\mu)\,
   {\cal F}_i^{A}(P_T/z,x_A),
 \label{eq:implemented-production-weights}
\end{align}
where \(f_q^p\) and \(f_g^p\) are interpolated from a CT18NNLO projectile grid,
\({\cal F}_{q}^{h}={\cal F}_{F}^{h}\), and
\({\cal F}_{g}^{h}={\cal F}_{A}^{h}\).  The effective target ratios are constructed
with \path{EPPS21nlo_CT18Anlo_Pb208} divided by \texttt{CT18ANLO}; all 107 EPPS21
members are retained.  The momentum-space saturation input is the tabulated analytic
tail
\begin{align}
 {\cal F}_{R}^{h}(k,x_A)
 &=\frac{Q_{s,R,h}^2(x_A)}{\pi[k^2+Q_{s,R,h}^2(x_A)]^2},
 \nonumber\\
 Q_{s,F,A}^2(x_A)
 &=(1~{\rm GeV})^2\left(\frac{0.01}{x_A}\right)^{0.20},
 \qquad
 Q_{s,A,A}^2(x_A)=\frac{\CA}{\CF}Q_{s,F,A}^2(x_A),
 \label{eq:analytic-tail-production}
\end{align}
Here \(A=208\) for the lead target, and
\(Q_{s,R,p}^2=A^{-1/3}Q_{s,R,A}^2\).  Thus the production input is an
MV/rcBK-inspired analytic-tail parametrization, not a numerical solution of
Eq.~\eqref{eq:rcbk}.  The \(z\) integral is segmented at the nonempty points in
\(\{\tau,0.08,0.15,0.30,0.60,1\}\), with 12 Gauss--Legendre nodes per segment.
The scale choices \(\mu=P_T/2,P_T,2P_T\) are stored row by row.  The correlated
replacements \(Q_s\to0.95Q_s,Q_s,1.05Q_s\) are stored in the same way.  Equation
\eqref{eq:implemented-production-weights} is an effective per-nucleon reweighting.  It
is not a strict hybrid-CGC factorization formula, because it combines an effective nPDF
ratio with fundamental or adjoint saturation tails according to the projectile channel.
A pure CGC production
calculation would instead use separately normalized proton and nuclear dipoles with a
common impact-parameter prescription and no additional target-side nPDF multiplier.
In the baseline setup, the present choice fixes the flavor fractions
\(f_i^h=D_i^h/(D_q^h+D_g^h)\) for \(h=pp,pA\).  These fractions enter the
production-weighted multiplicity distribution and
Eq.~\eqref{eq:implemented-baseline-cross-section}.

\begin{figure}[t]
\centering
\includegraphics[width=0.78\textwidth]{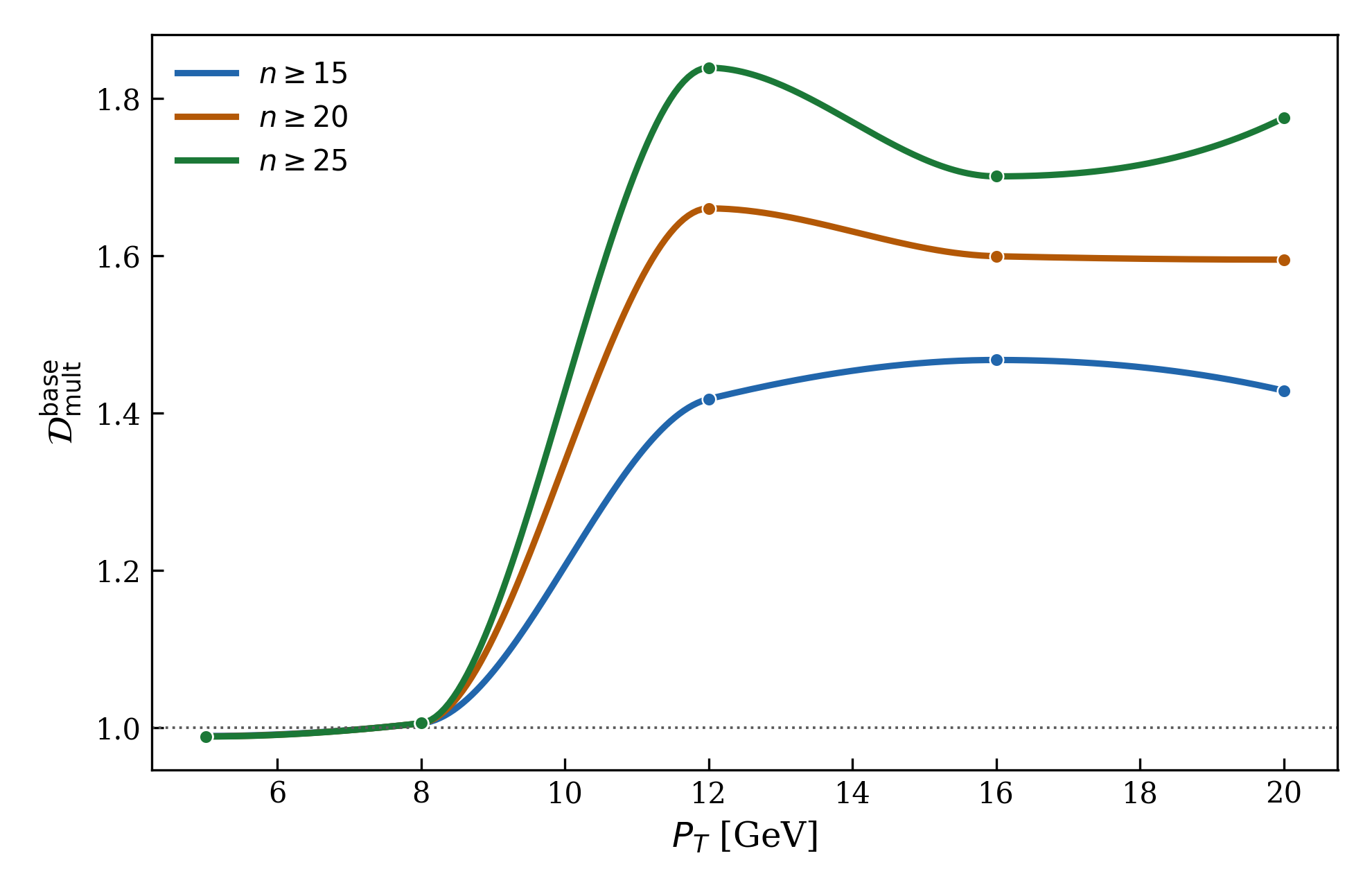}
\caption{Double ratio \({\cal D}_{\rm mult}^{\rm base}\) at \(\eta=3.2\) for three
multiplicity cuts.  Values above unity indicate that the multiplicity selection
enhances the high-\(n\) tail relative to the inclusive baseline suppression.  The
finite-dipole implementation gives a small effect at \(P_T=5,8\) GeV and a
larger enhancement for \(P_T\geq12\) GeV, where the perturbative in-cone phase space
opens while the Wilson-line resolution correction is still sizable.  At \(P_T=16,20\)
GeV the plotted values use the reduced-channel baseline specified in
Sec.~\ref{sec:uncertainties}.  The markers are the computed scan points and the curves
are interpolations.}
\label{fig:double-ratio-final}
\end{figure}

The second module evolves the multiplicity generating functions
\(Z_i(s,\omega_T,\rho)\).  It uses anti-\(k_T\) jets with \(R=0.4\),
\(Q_0=1.0\) GeV, \(n_f=3\), and the fixed coupling \(\alpha_s=0.25\) in all
multiplicity splitting kernels.  No running or infrared freezing prescription is
therefore applied in this module.  The coupling and jet radius are held fixed and are
not included in the displayed uncertainty band.  The implementation keeps an explicit
transverse-energy grid and evaluates the daughter functions at \(x\omega_T\) and
\((1-x)\omega_T\).  We use the hadron-collider angular variable
\(\rho=(\Delta R)^2/2\), with 60 logarithmic steps from
\(\rho_0=\tfrac12(Q_0/P_T)^2\) to \(\rho_R=R^2/2\), together with 24
Gauss--Legendre nodes in \(x\) and 16 logarithmically spaced nodes in \(\omega_T\).
The nonperturbative boundary condition at \(Q_0\) is an energy-dependent
negative-binomial distribution (NBD).  With \(u=e^{-s}\), the code initializes
\begin{equation}
  Z_i^{\rm IC}(u,\omega_T,Q_0)
  =\left[1+\frac{\bar n_i(\omega_T)}{k_i}(1-u)\right]^{-k_i},
  \label{eq:numerical-nbd-boundary}
\end{equation}
We use \(k_q=5\) and \(k_g=6\).  The corresponding means are
\(\bar n_q=2.5(\omega_T/10~{\rm GeV})^{0.12}\) and
\(\bar n_g=3.8(\omega_T/10~{\rm GeV})^{0.12}\).  These central NBD parameters are
illustrative model inputs, not a fit to forward-jet multiplicity data or a parton-shower
tune.  They are chosen to provide a broader, higher-mean gluon boundary than the quark
boundary; their role is to test the stability of ratios under a controlled quark/gluon
separation.  The same boundary condition is used in pp and pA, so
the displayed nuclear modification is not generated by a target-dependent
hadronization ansatz.  For each \(P_T\), the logarithmic transverse-energy grid spans
\begin{equation}
  \omega_T\in
  \left[\max(1.05Q_0,P_TR),\ \max(P_T,1.1Q_0)\right],
  \qquad N_\omega=16.
  \label{eq:numerical-omega-grid}
\end{equation}
Daughter functions at \(x\omega_T\) and \((1-x)\omega_T\) are interpolated linearly
in \(\ln\omega_T\), separately for their real and imaginary parts.  Arguments outside
the tabulated interval are frozen to the nearest endpoint.  Equation
\eqref{eq:numerical-nbd-boundary} is a model input rather than a perturbative
prediction; its parameter sensitivity is quantified separately in
Sec.~\ref{sec:uncertainties} and is not folded into the displayed band.
This is the transverse form of Eq.~\eqref{eq:hadron-collider-angular-map}; it does not
identify the physical opening angle with the jet radius at forward rapidity.  The local
constraint \(k_\perp>Q_0\) is imposed at every \((x,\rho,\omega_T)\) node.  At fixed
\(\omega_T\), the largest in-cone splitting momentum is
\(k_{\perp,\max}=R\omega_T/4\); hence an active splitting requires
\(\omega_T>4Q_0/R\).  We use \(R=0.4\) and \(Q_0=1\) GeV.  Over the full scan
\(P_T\leq20\) GeV, the lower boundary in Eq.~\eqref{eq:numerical-omega-grid} obeys
\(\max(1.05Q_0,P_TR)\leq8~{\rm GeV}<4Q_0/R=10~{\rm GeV}\).  Therefore no daughter
energy capable of satisfying \(k_\perp>Q_0\) lies below the tabulated interval; the
endpoint freezing acts only in the inactive region.  The earlier
\(\omega_T\)-independent scan is used only as a speed check; the production plots use
the explicit \(\omega_T\)-grid implementation.  Residual
\(\omega_T\)-grid dependence is treated as a numerical stability check rather than as a
separately propagated uncertainty band.  Increasing \(N_\omega\) from 16 to 24 changes
the \(n\ge20\) double ratio by at most \(1.7\times10^{-3}\) over the central
\(\eta=3.2\) scan.  Since the evolution is written in \(\ln\mu^2\) and
\(k_\perp^2\propto\rho\) at fixed \(x,\omega_T\), one has
\(\dd\ln k_\perp^2=\dd\ln\rho\); no extra factor \(1/2\) enters the Runge--Kutta
step.  At each angular step the code computes
\[
  k_\perp=\sqrt{2\rho}\,x(1-x)\omega_T
\]
and removes the contribution when \(k_\perp<Q_0\).  The numerical implementation uses
the coordinate-space MV/rcBK-inspired dipole of Eqs.~\eqref{eq:mv-dipole} and
\eqref{eq:qs-y}.  Its parameters are \(Q_{s0}=0.50\) GeV,
\(\Lambda_{\rm IR}=0.20\) GeV, and \(\lambda_s=0.28\).  We also impose
\(S_A=S_F^{\CA/\CF}\).  This internal-evolution input is
distinct from the analytic-tail production table in
Eq.~\eqref{eq:analytic-tail-production}.  It enters
the full dipole ratio
\[
  {\cal R}_{i\to ab}^{\rm dip}
  =
  C_{\rm daughters}/C_{\rm parent},
\]
rather than the small-dipole expansion Eq.~\eqref{eq:small-dipole-resolution}; the
latter is presented for analytic transparency only.  Explicitly,
\begin{align}
  {\cal R}_{q\to qg}^{\rm dip}
  &=
  \frac{S_F((1-x)/k_\perp,Y)S_A(x/k_\perp,Y)}
       {S_F(1/k_\perp,Y)},
  \nonumber\\
  {\cal R}_{g\to gg}^{\rm dip}
  &=
  \frac{S_A(x/k_\perp,Y)S_A((1-x)/k_\perp,Y)}
       {S_A(1/k_\perp,Y)},
  \nonumber\\
  {\cal R}_{g\to q\bar q}^{\rm dip}
  &=
  \frac{S_F(1/k_\perp,Y)}
       {S_A(1/k_\perp,Y)}.
  \label{eq:numerical-dipole-ratios}
\end{align}
The effective branching factor used for the nuclear-side finite-dipole diagnostic is
\begin{equation}
  \frac{\Gamma_{i\to ab}^{pA}}{\Gamma_{i\to ab}^{\rm vac}}
  =
  1+
  \left({\cal R}_{i\to ab}^{\rm dip}-1\right)
  \left(1-e^{-L_{{\rm eff},T}/t_{f,T}}\right),
  \qquad
  t_{f,T}=\frac{2x(1-x)\omega_T}{k_\perp^2},
  \label{eq:numerical-effective-ratio}
\end{equation}
with \(L_{{\rm eff},T}=5.0~{\rm GeV}^{-1}\).  The \(k_\perp<Q_0\) cutoff and the
formation-time gate are applied locally in the \((x,\rho,\omega_T)\) integration.  We
use this value of \(L_{{\rm eff},T}\) as a representative transverse overlap parameter for the baseline
calculation.  It is not included in the displayed uncertainty band.  Instead, we vary
\(L_{{\rm eff},T}\), \(Q_0\), and the rapidity prescription as multiplicity-kernel
model systematics in Table~\ref{tab:uncertainty-budget}.
Consequently the size of the finite-dipole enhancement should not be interpreted as a
precision prediction of the target-length dependence.  The pp comparison used in the
displayed baseline keeps the internal multiplicity evolution vacuum-like.  Thus the
finite-dipole part of Figs.~\ref{fig:tail-rpa-final}--\ref{fig:double-ratio-final}
should be read as a nuclear-side finite-dipole versus vacuum diagnostic.  A fully
physical target-dipole comparison would evolve both the proton and nuclear sides with
Eq.~\eqref{eq:cgc-gamma} and would insert the difference
\(\Delta\Gamma^A-\Delta\Gamma^p\).  That extension requires separately normalized
proton and nuclear dipoles and a common impact-parameter or transverse-area
prescription, and is left outside the displayed numerical baseline.  We implement two
rapidity prescriptions.  In the parent-\(Y\) prescription, \(x_A\) is held fixed during
the branching.  In the dynamic-\(Y\) prescription, the daughter energy fraction mildly
changes the \(x_A\) used in \(Q_s(Y)\).
The central curves use the dynamic-\(Y\) prescription; the parent-\(Y\) result is
reported as a model sensitivity in Sec.~\ref{sec:uncertainties}.
The operator-level correction is therefore tested through the physical observables
below rather than through a separate diagnostic plot.  In the accepted implementation,
the correction is active only above the nonperturbative cutoff and is concentrated in
the semihard region where \(k_{\perp,{\rm split}}\) is comparable to the target
saturation scale.  This support property is what allows the multiplicity-dependent
effect to be interpreted as Wilson-line resolution of an early in-cone splitting,
rather than as an arbitrary deformation of the shower model.

The third module audits the inclusive NLO hard-coefficient denominator and promotes
only the accepted diagonal-quark factor to the restricted production rows.  The
corresponding resolved finite numerator is defined formally by the local replacement
\({\cal K}^{\rm real}({\cal M}^{\rm raw}-{\cal M}^{\rm coh})\).  Before this
replacement, we remove the subtraction terms assigned to PDFs, BK/JIMWLK, Sudakov
evolution, SiJFs, and \(\Delta\Gamma\).  None of the finite measured numerators is added
to the displayed curves; they use
Eq.~\eqref{eq:implemented-baseline-cross-section}.  The fourth module performs the
resummed evolution and inverse transform
of the generating functions on the complex \(u=e^{-s}\) circle.  This evolution resums
the leading angular-ordered multiplicity logarithms generated by repeated in-cone
branchings between the jet scale \(P_TR\) and the infrared cutoff \(Q_0\).  The formal
denominator retains the Sudakov/threshold sector of Ref.~\cite{Wang:2022zla}; in the
restricted rows, its explicit finite bookkeeping enters only through the accepted
diagonal-quark factor rather than through a complete all-channel resummed hard factor.
No claim is made here of a complete NNLL or
all-channel precision-resummed phenomenology; the accuracy of the multiplicity
evolution is the LL nonlinear generating-function accuracy of
Ref.~\cite{Duan:2025hmjets}, combined with the restricted production weights in
Table~\ref{tab:measured-implementation-map}.
The inverse transform uses a contour
\(|u|=r_0=0.96\) with \(N_s=192\) Fourier nodes and reconstructing \(P_i(n)\) up to
\(N_{\rm max}=90\).  Negative roundoff-level bins are clipped and the distributions are
renormalized.  The total clipped probability mass is monitored a posteriori; its
largest value in the contour scan is \(1.8\times10^{-12}\), well below the smallest
reported tail probability.  The
central spectra shown below are based on the
full contour inversion.  Negative-binomial fits from the first two derivatives of
\(\ln Z_i(s)\) at \(s=0\) are used only for exploratory scans and as a speed check
against the full-contour result.  The fifth module forms
the observables
\[
  P_h(n)=f_q^hP_q^h(n)+f_g^hP_g^h(n),
  \qquad
  {\cal R}_{\rm base}(n)={\cal R}_{\rm base}^{\rm incl}
  \frac{P_{pA}(n)}{P_{pp}(n)},
\]
and the tail ratios in Eq.~\eqref{eq:tail-ratio-final}.  Since the corrected
\(\lambda\)-coefficients are implemented through the full finite-dipole ratios,
Wilson-line resolution can enhance resolved \(k_\perp\sim Q_s\) branchings in the
perturbative in-cone phase space.  The
following figures therefore focus on three directly interpretable quantities: the
inclusive baseline, the multiplicity-selected tail, and the double ratio that divides
out the inclusive normalization.

\begin{figure}[t]
\centering
\includegraphics[width=0.98\textwidth]{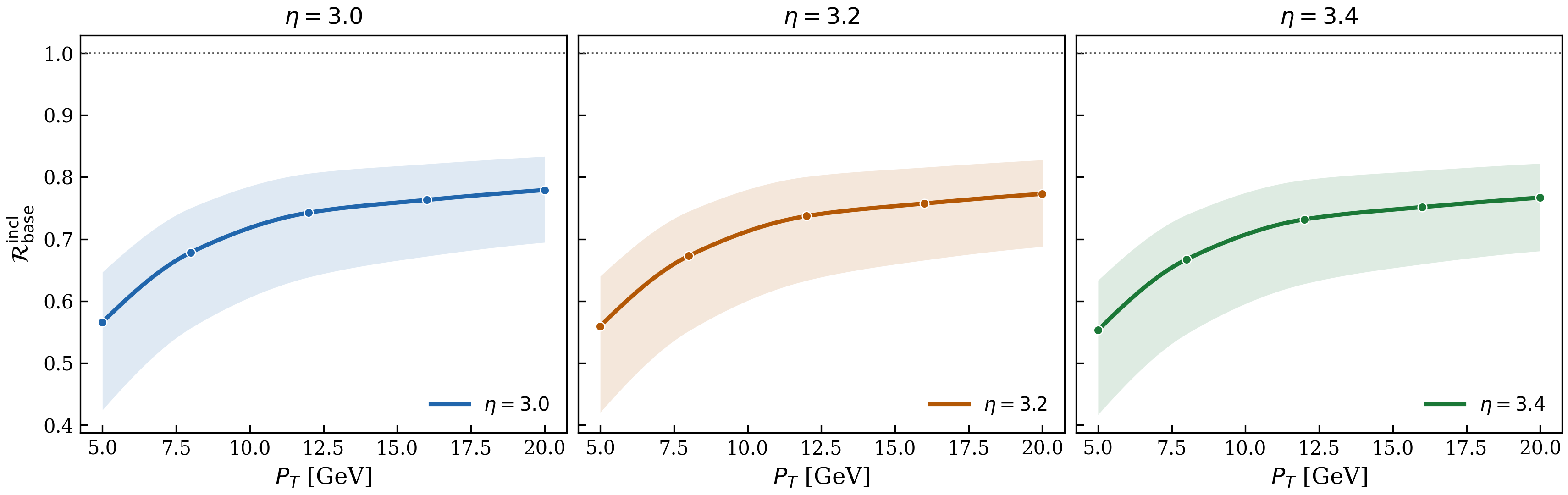}
\caption{Phenomenological per-nucleon inclusive forward-jet baseline ratio with the
restricted production denominator used in the numerical implementation.  The diagonal
\(q\to q\) finite correction is included only at \(P_T=5,8,12\) GeV.  At these points,
the integrated matched denominator passes the stability test.  The correction is
omitted at larger \(P_T\).  The three panels separate the rapidity slices so that the
EPPS21, scale, and saturation-scale-normalization envelope can be read without overlapping the
neighboring \(\eta\) curves.}
\label{fig:incl-rpa-final}
\end{figure}

\begin{figure}[t]
\centering
\includegraphics[width=0.98\textwidth]{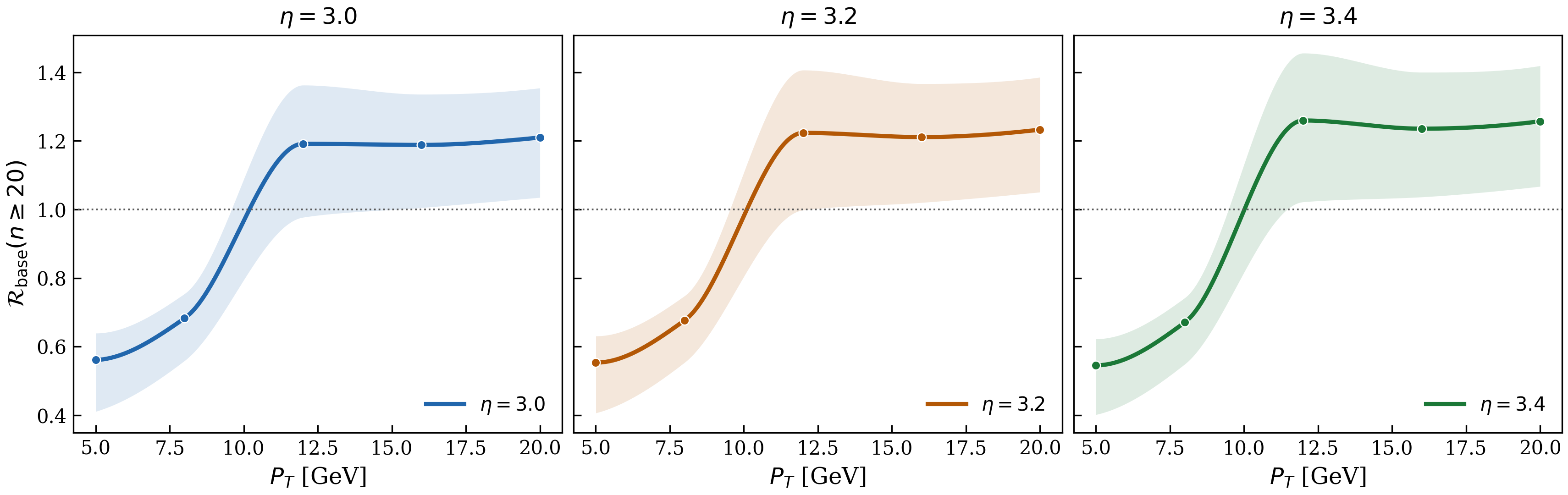}
\caption{Multiplicity-tail baseline ratio
\({\cal R}_{\rm base}(n\geq20)\) computed with the same restricted denominator as
Fig.~\ref{fig:incl-rpa-final} and the full finite-dipole baseline multiplicity
evolution.
The separate rapidity panels make visible that the high-multiplicity tail follows a
different \(P_T\) dependence from the inclusive baseline.  At \(P_T=16,20\) GeV the
curves are the reduced-channel baseline defined in Sec.~\ref{sec:uncertainties},
because the diagonal \(q\to q\) finite remainder does not pass the stability criterion
in Eq.~\eqref{eq:dqq-match-stability}.  The markers are the computed scan points; the
smooth curves are interpolations to guide the eye.}
\label{fig:tail-rpa-final}
\end{figure}

In the phenomenological per-nucleon baseline, the pA denominator is evaluated by
multiplying the quark- and gluon-labeled production weights by nuclear PDF ratios
\begin{equation}
  R_q^A(x_A,\mu)
  =
  \frac{\sum_f x_A[f_f+\bar f_f]_{p/A}(x_A,\mu)}
       {\sum_f x_A[f_f+\bar f_f]_{p}(x_A,\mu)},
  \qquad
  R_g^A(x_A,\mu)
  =
  \frac{x_A g_{p/A}(x_A,\mu)}{x_A g_p(x_A,\mu)} .
  \label{eq:npdf-ratios-final}
\end{equation}
The labels \(q\) and \(g\) on the production weights identify the projectile channel,
whereas the ratios in Eq.~\eqref{eq:npdf-ratios-final} are target-side collinear inputs.
Pairing \(R_q^A\) with the projectile-quark weight and \(R_g^A\) with the
projectile-gluon weight is therefore an effective channel-dependent reweighting, not a
consequence of strict hybrid CGC factorization.  In the latter, the target enters the
two projectile channels through fundamental and adjoint Wilson-line correlators.  We
retain the reweighting only as the stated phenomenological baseline and test it by
applying a common \(R_g^A\) to both channels.  The result is reported in
Table~\ref{tab:boundary-inversion-sensitivity}; the high-\(P_T\) double ratio changes
by only a few percent, while the low-\(P_T\) flavor-composition effect is visibly model
dependent.
The baseline curves below use EPPS21 at its native 90\% confidence level.  We propagate
all EPPS21 Hessian members.  We vary the scale over
\(\mu=P_T/2,P_T,2P_T\).  The saturation-scale choices are
\(Q_s\to 0.95Q_s,Q_s,1.05Q_s\), correlated between the analytic production tail and the
coordinate-space multiplicity ansatz.  We then compare the baseline band to
nCTEQ15 as a robustness check.  The scan points used for the uncertainty bands are
\(P_T=5,8,12,16,20\) GeV and \(\eta=3.0,3.2,3.4\).
The uncertainty bands in Figs.~\ref{fig:incl-rpa-final},
\ref{fig:tail-rpa-final}, and \ref{fig:npdf-set-comparison} are constructed as
follows.  For each central scale and \(Q_s\) choice we evaluate the EPPS21 Hessian
members and form the native 90\% confidence-level Hessian uncertainty.  We then take
the envelope over the three common perturbative scales and the three correlated
\(Q_s\)-normalization choices.  The common scale \(\mu\) is used in the projectile PDF
and the nuclear PDF ratio.  The same scale also enters the accepted diagonal-quark
factor and the SiJF/Sudakov factors.  The jet scale is tied to \(P_TR\) in the
multiplicity evolution.  This correlated variation is not a full independent variation
of \(\mu_H\), \(\mu_J\), and \(\mu_{\rm mult}\).  We use it for the baseline because the present calculation is
designed to test the matched construction rather than to deliver a precision global
uncertainty analysis.  The EPPS21 Hessian uncertainty should therefore be interpreted
as the uncertainty of the phenomenological nPDF-reweighted baseline, not as the full
nuclear uncertainty of a pure CGC target calculation.  Numerically, the nPDF component
dominates the inclusive and absolute tail bands at large rapidity.  The \(Q_s\)
variation dominates at the lowest \(P_T\).  Much of the nPDF normalization cancels in
the double ratio.  Its remaining band is controlled mainly by the correlated
\(Q_s\)-normalization and common-scale variations in the semihard region.

\begin{table}[t]
\centering
\small
\renewcommand{\arraystretch}{0.90}
\begin{tabularx}{\textwidth}{L{0.23\textwidth}L{0.20\textwidth}Y}
\toprule
Source & Treatment in the displayed band & Comment \\
\midrule
nPDF Hessian & included & EPPS21 native 90\% confidence-level Hessian envelope; nCTEQ15 shown separately \\
common scale \(\mu\) & included & correlated \(\mu=P_T/2,P_T,2P_T\) variation in PDF, nPDF ratio, the accepted diagonal-quark factor, and Sudakov/SiJF factors \\
saturation-scale normalization & included & correlated \(Q_s\to0.95Q_s,Q_s,1.05Q_s\) envelope in the analytic production tail and coordinate-space multiplicity ansatz \\
jet radius \(R\) & fixed, not included & central anti-\(k_T\) value \(R=0.4\); no radius envelope is claimed \\
multiplicity coupling & fixed, not included & \(\alpha_s=0.25\) and \(n_f=3\); no running-coupling or freezing variation is included \\
\(\mu_H,\mu_J,\mu_{\rm mult}\) independent variation & not included & deferred to a precision phenomenology study; would enlarge the scale uncertainty \\
\(Q_0\), \(L_{{\rm eff},T}\), parent-\(Y\) versus dynamic-\(Y\) & quantified separately & model sensitivity reported in Table~\ref{tab:multiplicity-model-sensitivity}, not folded into the band \\
finite-dipole versus small-dipole kernel & quantified separately & comparison reported in Table~\ref{tab:multiplicity-model-sensitivity} \\
nonfactorizing finite multiplicity-SiJF matching terms & not included in the band & treated as an unquantified finite matching systematic beyond the product ansatz of Eq.~\eqref{eq:sijf-z} \\
nonperturbative multiplicity boundary & quantified separately & NBD-parameter scan and absolute tail sensitivity reported in Table~\ref{tab:boundary-inversion-sensitivity}; not folded into the band \\
reduced \(K_{82}\)--\(K_{85}\) quadrature & fails; excluded & independent refinement changes the integral by factors \(0.73\)--\(15.69\) \\
contour inversion & acceptance criterion & varying \(r_0,N_s,N_{\max}\) changes the double ratio by less than \(3.9\times10^{-9}\) and the absolute tail by less than \(9.1\times10^{-8}\) \\
effective nPDF channel mapping & quantified separately & common target \(R_g^A\) test reported in Table~\ref{tab:boundary-inversion-sensitivity} \\
diagonal-\(qq\) stability thresholds & monitored separately & varied in the stability discussion below rather than folded into the plotted band \\
\bottomrule
\end{tabularx}
\caption{Uncertainty budget for the baseline numerical implementation.  The displayed
bands include the nPDF, common-scale, and saturation-scale-normalization sources.  The remaining entries are matching or model
systematics that are stated explicitly but not claimed to be part of the plotted
precision uncertainty.}
\label{tab:uncertainty-budget}
\end{table}

The inclusive observable is shown first because it is the denominator of every tail
ratio.  If this baseline were unstable, any subsequent statement about multiplicity
dependence would be ambiguous.

Figure~\ref{fig:incl-rpa-final} shows the inclusive nPDF-reweighted baseline.  The suppression is
strongest at \(P_T=5\) GeV and weakens as the hard scale increases.  At \(\eta=3.2\), the
central values are \({\cal R}_{\rm base}^{\rm incl}=0.559,0.673,0.737,0.757,0.773\) for
\(P_T=5,8,12,16,20\) GeV.  These numbers set the normalization used in the
multiplicity-dependent observables below.  The important point is that the inclusive
denominator gives a monotonic and physically reasonable suppression pattern over the
forward kinematic range.  We therefore use it as the reference against which
multiplicity selection is measured.  This ordering is essential: the tail and double
ratios below should be interpreted only after the common inclusive normalization has
been fixed.

Figure~\ref{fig:tail-rpa-final} displays \({\cal R}_{\rm base}(n\ge20)\).  The high-tail observable
does not simply inherit the inclusive suppression.  At \(\eta=3.2\), the central values are
\[
\begin{array}{c|ccccc}
P_T~[{\rm GeV}] & 5 & 8 & 12 & 16 & 20 \\
\hline
{\cal R}_{\rm base}(n\ge20) & 0.553 & 0.677 & 1.214 & 1.210 & 1.229
\end{array}
\]
The low-\(P_T\) points remain close to the inclusive baseline, while the moderate-\(P_T\)
points show a sizable high-multiplicity enhancement within the displayed model.  This
pattern arises from the combination of flavor reweighting and the finite-dipole
Wilson-line correction to the in-cone multiplicity evolution.  Since the pp side is
kept vacuum-like in the displayed baseline, the finite-dipole component should be
understood as a qualitative nuclear-side diagnostic rather than as a completed
\(\Delta\Gamma^A-\Delta\Gamma^p\) calculation.  The cleanest way to isolate the
multiplicity-conditioned effect is the double ratio
\begin{equation}
  {\cal D}_{\rm mult}^{\rm base}
  =
  \frac{{\cal R}_{\rm base}(n\ge n_{\rm cut})}
       {{\cal R}_{\rm base}^{\rm incl}} .
  \label{eq:double-ratio-final}
\end{equation}

\begin{figure}[t]
\centering
\includegraphics[width=0.78\textwidth]{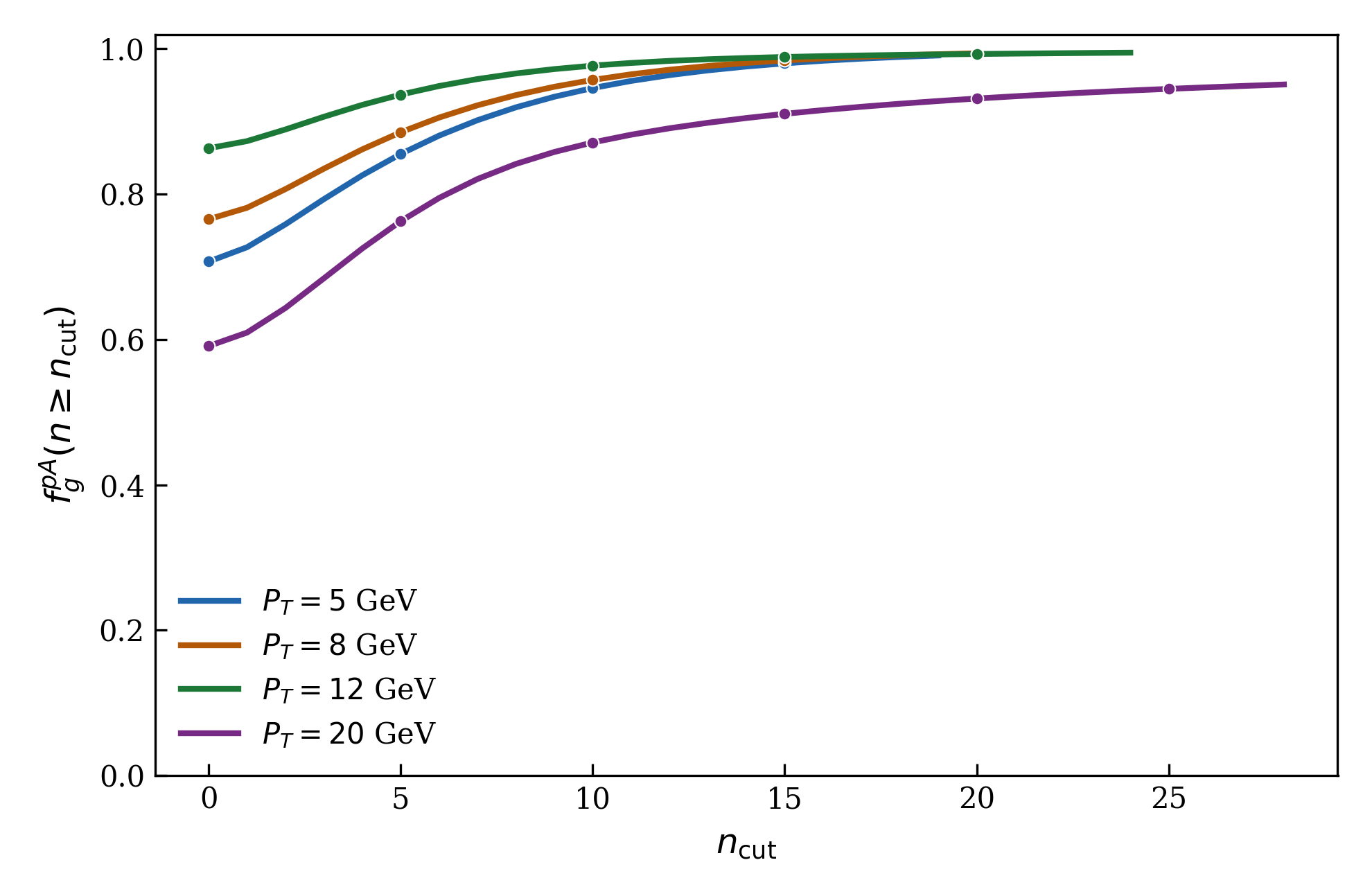}
\caption{Tail-selected gluon fraction \(f_g^{pA}(n\ge n_{\rm cut})\) at \(\eta=3.2\)
in the baseline calculation.  The high-multiplicity tail is increasingly gluon
dominated, which explains why
\({\cal R}_{\rm base}(n\ge n_{\rm cut})\) is not a trivial rescaling of
\({\cal R}_{\rm base}^{\rm incl}\).  The markers are the computed scan points and the curves are
interpolations.}
\label{fig:gluon-fraction-final}
\end{figure}

Figure~\ref{fig:double-ratio-final} shows the multiplicity-conditioned effect after the
inclusive suppression has been divided out.  To separate a pure production-weight
effect from a modification of the in-cone cascade, we also define the diagnostic
\begin{equation}
  {\cal D}_{\rm prod}^{\rm base}
  =
  \left.
  {\cal D}_{\rm mult}^{\rm base}
  \right|_{\Delta\Gamma^{A}=0},
  \label{eq:production-only-double-ratio}
\end{equation}
where the \(pA\) and \(pp\) production weights and quark/gluon fractions are kept, but
the multiplicity generating functions are evolved with the vacuum kernels.  This
production-only diagnostic remains much closer to unity than the finite-dipole
diagnostic result.  The
enhancement in Fig.~\ref{fig:double-ratio-final} should therefore not be attributed to
denominator normalization alone, nor to the finite-dipole kernel alone; it is generated
by the combined effect of multiplicity-induced quark/gluon reweighting and the
finite-dipole Wilson-line modification of the in-cone cascade.  This motivates the
flavor-composition plot in Fig.~\ref{fig:gluon-fraction-final}.

\begin{table}[t]
\centering
\begin{tabular}{cccc}
\toprule
\(P_T\) [GeV] & \(n\ge15\) & \(n\ge20\) & \(n\ge25\) \\
\midrule
5  & 0.990 & 0.989 & 0.989 \\
8  & 1.006 & 1.006 & 1.006 \\
12 & 1.418 & 1.660 & 1.839 \\
16 & 1.467 & 1.599 & 1.701 \\
20 & 1.429 & 1.595 & 1.775 \\
\bottomrule
\end{tabular}
\caption{Double ratio
\({\cal D}_{\rm mult}^{\rm base}={\cal R}_{\rm base}(n\ge n_{\rm cut})/
{\cal R}_{\rm base}^{\rm incl}\) at
\(\eta=3.2\), obtained from the full finite-dipole baseline multiplicity evolution.  The
moderate-\(P_T\) enhancement is substantially larger than in the small-dipole
estimate, indicating that the high-multiplicity tail is sensitive to the
finite-dipole Wilson-line ratio in the region where \(k_{\perp,{\rm split}}\) remains
comparable to the target saturation scale.  The central entries in this table do not
include variations of \(Q_0\) or \(L_{{\rm eff},T}\).  They also omit a proton
finite-dipole subtraction.  The first two effects are quantified separately in
Table~\ref{tab:multiplicity-model-sensitivity}.  The entries should therefore be read
as a model-dependent baseline illustration.}
\label{tab:double-ratio-corrected}
\end{table}

The central values collected in Table~\ref{tab:double-ratio-corrected} quantify the
cut dependence visible in Fig.~\ref{fig:double-ratio-final}.  In particular, the
increase with \(n_{\rm cut}\) at moderate \(P_T\) is the expected consequence of
selecting progressively farther into the gluon-dominated multiplicity tail within this
finite-dipole baseline.

The same evolution shifts the gluon-jet mean only after appreciable perturbative
in-cone phase space opens above \(Q_0\).  At \(\eta=3.2\), the ratios
\(\langle n\rangle_g^{pA}/\langle n\rangle_g^{pp}\) are (1.000) at
\(P_T=5,8\) GeV and \(1.027, 1.063, 1.073\) at
\(P_T=12,16,20\) GeV, respectively.  These values are finite-dipole diagnostics;
the pp entries use vacuum internal evolution.

The partonic origin of the multiplicity effect is shown in
Fig.~\ref{fig:gluon-fraction-final}.  We plot the tail fraction rather than the
pointwise \(f_g(n)\), because the latter is numerically unstable once the probability
distribution becomes exponentially small at very large \(n\).  Since \(P_g(n)\) has a
harder tail than \(P_q(n)\), selecting large \(n_{\rm cut}\) enhances the
gluon-initiated component.  This composition effect is present already in pp, and the
nuclear target further changes it through the relative quark and gluon production
weights and through \(\Delta\Gamma^{\rm CGC}\).  The figure therefore separates two
effects that would otherwise be conflated: a high-multiplicity cut can change
the baseline ratio by selecting a more gluon-rich sample, while the Wilson-line correction
modifies the in-cone evolution of that sample.

\section{Uncertainty checks and final caveats}
\label{sec:uncertainties}

The displayed bands do not include the model choices internal to the finite-dipole
evolution.  We therefore vary them separately at \(\eta=3.2\) and \(n_{\rm cut}=20\).
Table~\ref{tab:multiplicity-model-sensitivity} compares the central dynamic-\(Y\),
full-dipole result with variations of the overlap length, infrared cutoff, rapidity
prescription, dipole expansion, and energy-grid density.

\begin{table}[t]
\centering
\small
\setlength{\tabcolsep}{4pt}
\begin{tabular}{ccccccc}
\toprule
\(P_T\) [GeV] & central & \(L_{{\rm eff},T}\in[2.5,10]\) & \(Q_0\in[0.8,1.2]\) GeV & parent \(Y\) & small dipole & \(N_\omega=24\) \\
\midrule
12 & 1.660 & 1.416--1.897 & 1.007--2.096 & 1.849 & 1.376 & 1.660 \\
16 & 1.599 & 1.383--1.812 & 1.459--2.034 & 1.722 & 1.399 & 1.600 \\
20 & 1.595 & 1.372--1.826 & 1.414--1.921 & 1.703 & 1.415 & 1.594 \\
\bottomrule
\end{tabular}
\caption{Sensitivity of \({\cal D}_{\rm mult}^{\rm base}(n\ge20)\) at \(\eta=3.2\)
to model and numerical choices.  The central result uses
\(L_{{\rm eff},T}=5~{\rm GeV}^{-1}\), \(Q_0=1~{\rm GeV}\), dynamic \(Y\), the full
finite-dipole ratio, and \(N_\omega=16\).  The \(L_{{\rm eff},T}\) and \(Q_0\) columns
show envelopes over the indicated endpoint choices.}
\label{tab:multiplicity-model-sensitivity}
\end{table}

The nonperturbative NBD boundary in Eq.~\eqref{eq:numerical-nbd-boundary} is tested by
varying both mean normalizations by \(\pm10\%\), both shape parameters by \(\pm20\%\),
and the energy exponent over \(0.08\)--\(0.16\).  Table
\ref{tab:boundary-inversion-sensitivity} shows that the absolute tail probability is
strongly boundary dependent, as expected, while the pp/pA double ratio is more stable
because the same boundary is used on both sides.  This cancellation is incomplete and
the boundary-condition envelope is therefore a genuine model systematic.

\begin{table}[t]
\centering
\small
\setlength{\tabcolsep}{3.5pt}
\begin{tabular}{cccccc}
\toprule
\(P_T\) & central \(P_{pp}(n\ge20)\) & NBD range & central \({\cal D}_{\rm mult}^{\rm base}\) & NBD range & common-\(R_g^A\) \\
\midrule
5  & \(9.33\times10^{-6}\) & \((2.85\!-!25.84)\times10^{-6}\) & 0.989 & 0.989--0.989 & 1.000 \\
8  & \(1.82\times10^{-5}\) & \((0.58\!-!4.89)\times10^{-5}\) & 1.006 & 1.006--1.006 & 1.000 \\
12 & \(8.02\times10^{-5}\) & \((2.87\!-!19.25)\times10^{-5}\) & 1.660 & 1.530--1.764 & 1.649 \\
16 & \(2.46\times10^{-4}\) & \((0.94\!-!5.52)\times10^{-4}\) & 1.599 & 1.542--1.640 & 1.568 \\
20 & \(5.89\times10^{-4}\) & \((2.40\!-!12.51)\times10^{-4}\) & 1.595 & 1.533--1.645 & 1.560 \\
\bottomrule
\end{tabular}
\caption{Sensitivity to the nonperturbative multiplicity boundary at \(\eta=3.2\)
and \(n_{\rm cut}=20\).  The final column replaces the channel-dependent target nPDF
ratios by the common gluon ratio \(R_g^A\) while retaining the same finite-dipole
evolution.  It tests whether the enhancement is manufactured by the phenomenological
\(R_q^A/R_g^A\) flavor assignment.}
\label{tab:boundary-inversion-sensitivity}
\end{table}

The contour inversion is much better controlled than the boundary model.  Varying
\(r_0=0.93\)--0.98, \(N_s=128\)--256, and \(N_{\max}=70\)--110 changes the smallest
reported pp tail probability by less than \(9.1\times10^{-8}\) relatively and changes
the double ratio by less than \(3.9\times10^{-9}\).  The largest clipped mass in this
scan is \(1.8\times10^{-12}\), well below the minimum tail probability.  These are
tail-level convergence tests, not merely normalization checks.

The \(N_\omega\) comparison confirms numerical convergence, whereas the substantial
spread under \(Q_0\), \(L_{{\rm eff},T}\), and the finite-/small-dipole choice is a genuine
model dependence.  The enhancement above unity persists for most variations, but its
magnitude is not a precision prediction; in particular, the \(P_T=12\) GeV result can
approach unity for the larger cutoff.  This explicit sensitivity is why the finite-dipole
curves are described as a diagnostic baseline rather than a completed nuclear
phenomenology result.

\begin{figure}[t]
\centering
\includegraphics[width=0.98\textwidth]{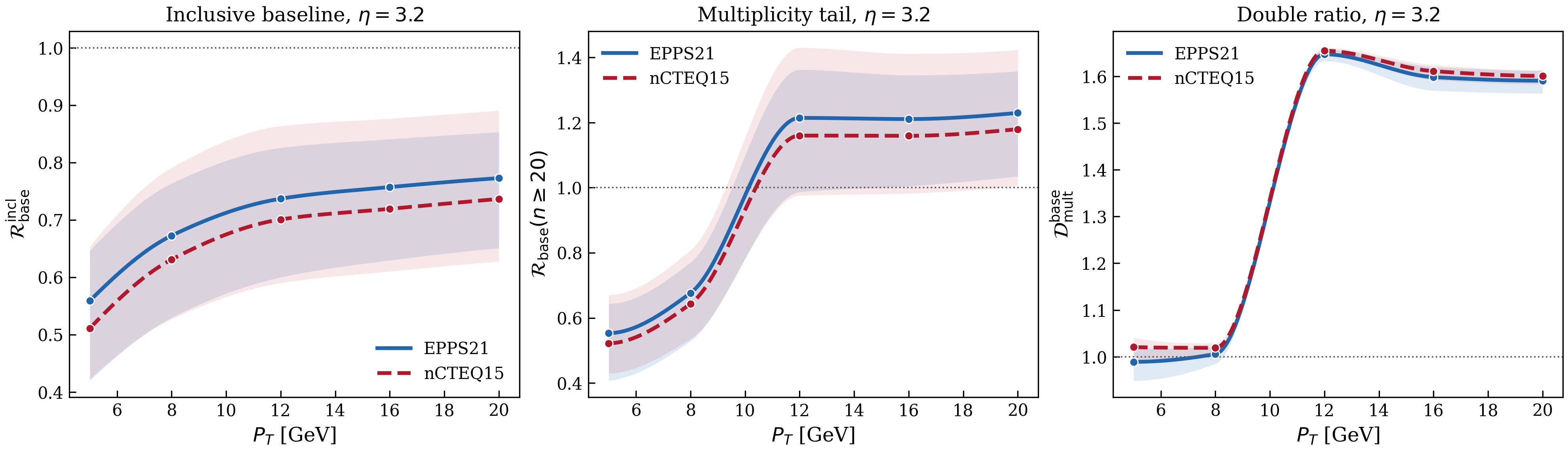}
\caption{Comparison between EPPS21 and nCTEQ15 at \(\eta=3.2\), using the same
full finite-dipole baseline evolution as in Figs.~\ref{fig:tail-rpa-final} and
\ref{fig:double-ratio-final}.
The panels show the inclusive baseline, the \(n\ge20\) tail, and the double ratio.
The bands are the native 90\% confidence-level Hessian uncertainties of each set.}
\label{fig:npdf-set-comparison}
\end{figure}

Among the sources included in the displayed band, the largest uncertainty is generally
the nuclear PDF uncertainty.  To
verify that the qualitative behavior is not an artifact of a single nPDF fit, we repeat
the denominator calculation with nCTEQ15 and compare to EPPS21.  Both sets give
suppression in the inclusive observable and a similar pattern for the multiplicity
tail, although nCTEQ15 gives a somewhat lower central value and a larger upper
uncertainty at large \(P_T\).  This comparison is a robustness check, not a replacement
for a full global uncertainty analysis.  After the mechanism and the flavor-composition
effect have been isolated in the previous figures, the purpose of
Fig.~\ref{fig:npdf-set-comparison} is to test whether the qualitative picture survives
a change of nuclear PDF input.

Figure~\ref{fig:npdf-set-comparison} addresses the nuclear-PDF ambiguity.  EPPS21 and
nCTEQ15 give different inclusive normalizations and different absolute tail
normalizations, as expected for forward small-\(x_A\) production.  However, the double
ratio is much more stable between the two sets.  This is a useful internal check: the
multiplicity-conditioned enhancement is not generated primarily by the nPDF
normalization, but by the multiplicity evolution and flavor selection after the
inclusive suppression has divided out.

We also perform a numerical stability check of the low-\(P_T\) points by doubling the
number of \(z\)-integration nodes from 12 to 24.  The relative changes in
\({\cal R}_{\rm base}^{\rm incl}\) and in \({\cal R}_{\rm base}(n\ge20)\) are below \(5\times10^{-5}\) for
\(P_T=5\) and below \(10^{-5}\) for \(P_T=8\) over the three rapidities.  Thus the large
uncertainty of the \(P_T=5\) double ratio is not a numerical quadrature instability of
the production denominator; it is a genuine sensitivity to the low hard scale, the nPDF
band, and the multiplicity-kernel correction.

We also tested a reduced coordinate-space implementation of the parent-gluon kernels.
Independent quadrature refinement changed the integrated result by factors
ranging from \(0.73\) to \(15.69\).  The corresponding terms were therefore excluded from
Eq.~\eqref{eq:implemented-baseline-cross-section}; no numerical result in this work
depends on these reduced kernels.

The finite hard-coefficient contribution retained in the production rows is the
diagonal \(q\to q\) factor at accepted points.  A direct production-level,
prefactor-corrected stability check of the matched denominator
\begin{equation}
  D_{qq}^{\rm match}
  =
  D_{\rm LO}+D_{\rm PDF}+D_{\rm Sudakov}+D_{\rm SiJF}
  +\sigma_{qq}^{3}+\sigma_{qq}^{4}
  +\frac{1}{2}\sigma_{qq}^{5a}+\sigma_{qq}^{5b}
  +\sigma_{qq}^{6}+\sigma_{qq}^{7}+\sigma_{qq}^{8}
  +\sigma_{qq}^{10}+\sigma_{qq}^{11}
\label{eq:dqq-match-stability}
\end{equation}
uses the original \(\alpha_s/(2\pi^2)\) normalization of the
\(\sigma_{qq}^{3}/T_{qq}^{(1)}\) term.  It also uses the original
\(-2\alpha_s/\pi^2\) normalization of the double-convolution part of
\(\sigma_{qq}^{5}\), together with the local \(\int_0^1\dd\xi'\) plus prescription of
\(\sigma_{qq}^{4}\).  In Eq.~\eqref{eq:dqq-match-stability}, \(D_{\rm LO}\) is the
Born quark production weight.  The projectile-collinear counterterm is
\(D_{\rm PDF}\), while \(D_{\rm Sudakov}\) collects the soft-recoil/threshold
subtraction assigned to the Sudakov factor.  The term \(D_{\rm SiJF}\) is the
jet-collinear subtraction assigned to the semi-inclusive jet function.  The numbered
\(\sigma_{qq}^i\) are the finite quark-channel pieces of
Ref.~\cite{Wang:2022zla}, evaluated with the same normalization conventions.
Local subtraction bins can be negative.  We therefore impose the acceptance test only
after the \(z\)-convolution.  We require \(D_{qq}^{\rm match}>0\),
\(D_{qq}^{\rm match}/D_{qq}^{\rm LO}>0.2\), and
\(|N_{qq}^{\rm meas}/D_{qq}^{\rm match}|<0.25\).  These numerical thresholds are
operational stability cuts, not new physical constants.  The first condition enforces a
positive physical denominator.  The second excludes points where the finite remainder
nearly cancels the Born contribution and the fixed-order expansion is no longer
perturbatively controlled.  The third prevents a measured finite numerator from
becoming larger than the denominator to which it is matched.  Although
\(N_{qq}^{\rm meas}\) is not inserted into
Eq.~\eqref{eq:implemented-baseline-cross-section}, this ratio tests whether neglecting
the finite measured correction is parametrically consistent with the denominator-only
restricted baseline.  It is therefore an acceptance diagnostic for that approximation,
not an additional contribution to the plotted observable.  When any criterion fails,
the point is recomputed with the denser integration grid used for the denominator
closure test.  If it still fails, the result at that kinematic point is reported only
as a reduced-channel baseline.  It is not presented as a complete matched prediction.
The central rapidity
slice is shown
in Table~\ref{tab:qq-prefactor-audit}.  The table serves as a stability check of the
hard-coefficient bookkeeping.  Its purpose is to make transparent
which diagonal-quark points enter the baseline and which are excluded by the stability
criterion.  We include it in the main text rather than hiding it in an appendix because
it directly controls the scope of the numerical claims made above.

\begin{table}[t]
\centering
\begin{tabular}{ccccc}
\toprule
\(P_T\) [GeV] &
\(D_{qq}^{\rm match}/D_{qq}^{\rm LO}\) &
\(\sigma_{qq}^{3}/D_{qq}^{\rm LO}\) &
\((\sigma_{qq}^{5}-\sigma_{qq}^{5a}/2)/D_{qq}^{\rm LO}\) &
status \\
\midrule
5  & 1.353  & -0.448 & 0.624  & accepted \\
8  & 0.862  & -0.753 & 0.262  & accepted \\
12 & 0.352  & -1.045 & -0.089 & accepted \\
16 & 0.045  & -1.251 & -0.281 & excluded \\
20 & -0.085 & -1.390 & -0.349 & excluded \\
\bottomrule
\end{tabular}
\caption{Prefactor-corrected diagonal \(q\to q\) matched-denominator stability check at
\(\eta=3.2\).  The accepted points are inserted into the restricted baseline defined
in Eq.~\eqref{eq:implemented-baseline-cross-section} and
Table~\ref{tab:measured-implementation-map}; the
high-\(P_T\) points are not inserted because the matched denominator is too small or
negative after the finite remainder is included.}
\label{tab:qq-prefactor-audit}
\end{table}

The baseline result therefore uses a mixed but explicit rule.  The checked
diagonal-\(qq\) finite remainder is included for \(P_T=5,8,12\) GeV.  It is omitted at
\(P_T=16,20\) GeV because the fixed-order quark-channel denominator is numerically
unstable there.  This classification is unchanged for
\((c_D,c_N)=(0.10,0.30),(0.20,0.25),(0.30,0.20)\); only the deliberately stricter
choice \((0.40,0.20)\) removes the \(P_T=12\) GeV point.  At \(P_T=5,8,12\) GeV, the
accepted baseline points have
\(|N_{qq}^{\rm meas}/D_{qq}^{\rm match}|=7.2\times10^{-5},\,1.34\times10^{-2},\,
6.14\times10^{-2}\), respectively.  They remain below even the more restrictive
numerator cut \(0.20\).  The high-\(P_T\) points fail because the
matched denominator becomes too small or negative, not because of an oscillatory
measured numerator.

The pattern in Table~\ref{tab:qq-prefactor-audit} and the cut variation above identifies the specific obstruction to a more
ambitious numerical prediction: the \(\sigma_{qq}^{3}\) contribution grows into a large
negative finite remainder at high \(P_T\), so the diagonal-quark channel requires a
subtraction-local or resummed treatment before it can be used over the full scan.  It
also shows why the present numerical section is reported as a restricted baseline
phenomenology rather than as a precision all-channel NLO prediction.  A global analysis
with a
complete scan over rcBK initial
conditions, nuclear PDFs, jet-radius systematics, and nonperturbative multiplicity
boundary conditions is left for a dedicated phenomenology study.

\section{Conclusions}
\label{sec:conclusions}

We have developed a multiplicity-dependent extension of forward jet production in the
hybrid CGC framework.  The construction combines the inclusive NLO/resummed CGC
forward-jet calculation with multiplicity generating functions and flavor-transfer
SiJFs.  The central formal result is a generic add--subtract matching structure.  Its
inclusive-limit constraints preserve the \(s=0\) result through the stated
\(O(\alpha_s)\) matching order.  At nonzero \(s\), the high-multiplicity measurement
can reweight the quark and gluon production channels.  When the target resolves an early
in-cone splitting, it can also modify the multiplicity anomalous dimension through
Wilson-line correlators.  A fully differential channel-by-channel implementation of all finite
measured hard terms remains beyond the restricted numerical baseline.

This separation leads to a clear physical interpretation.  In the full CGC target-dipole
framework, a leading nuclear effect on high-multiplicity forward jets can arise from
the production kernel changing the quark/gluon mixture.  A more direct modification of
the internal multiplicity evolution is possible but requires the formation time or
transverse size of the splitting to make the two daughters resolvable by the target
field.  In a Gaussian CGC ensemble this effect can be expressed in terms of the same
target dipole that enters the inclusive production kernel for the same ensemble.  The
baseline numerical implementation used here is more restricted: its production
normalization is an nPDF-reweighted per-nucleon baseline, while its finite-dipole
component is a nuclear-side diagnostic relative to vacuum internal evolution.  With
this restriction understood, the framework separates three effects that are often
entangled phenomenologically.  These are nuclear modification of the production rate,
multiplicity-induced flavor reweighting, and genuine target resolution of the in-cone
cascade.
Within the central finite-dipole/Gaussian parameter choice, the Wilson-line resolution
mechanism produces an enhancement of the high-multiplicity forward-jet ratio in the
resolved kinematic region.  Its magnitude depends strongly on \(Q_0\),
\(L_{{\rm eff},T}\), and the finite-dipole prescription.  This sensitivity is
particularly pronounced near \(P_T=12\) GeV, as shown in
Table~\ref{tab:multiplicity-model-sensitivity}.  This statement is therefore a
model-dependent baseline result, not a universal sign or precision-magnitude claim.
The double ratio approaches unity only once the typical
in-cone splitting scale becomes parametrically larger than the saturation scale and the
Wilson-line resolution correction decouples.

The numerical implementation presented here propagates nPDF uncertainties, tests
the sensitivity of the low-\(P_T\) region, and audits the hard-kernel channels.  Its
role is to demonstrate the restricted baseline assembly and to expose the
hard-coefficient stability issues that must be addressed in future phenomenology.  In
particular, the parent-gluon reduced-kernel quadrature and the high-\(P_T\)
diagonal-quark remainder do not satisfy the production-level criteria and are not
promoted into the displayed curves.  The numerical section should therefore be viewed
as a controlled illustration of the operator-level mechanism rather than as the
endpoint of the uncertainty program.
This conservative interpretation is important.  The present calculation identifies the
operator structure and matching constraints needed for multiplicity-dependent forward
jets.  A precision phenomenological analysis will additionally require a
subtraction-local treatment of all finite hard coefficients.  It will also require a
systematic scan of rcBK initial conditions and a more complete propagation of
nuclear-PDF and jet-radius uncertainties.

Several extensions are natural.  The first is to combine this CGC framework with
energy-flow observables inside the same forward jet.  Multiplicity-conditioned EECs
have recently been shown to acquire a
\(\nu\)-dependent anomalous dimension in the perturbative collinear region
\cite{Duan:2026eecmultiplicity}.  Embedding such observables in the present pA
framework would make it possible to ask whether saturation modifies only the
quark/gluon mixture and the multiplicity distribution, or whether it also changes the
angular energy correlations among particles inside the jet.  This would connect
forward saturation physics to the rapidly developing EEC program in pp, DIS, and
nuclear collisions.

A second direction is to include final-state medium effects beyond the cold nuclear
target.  In hot or dense nuclear environments, jet quenching can reshape both the
multiplicity distribution and the EEC at fixed multiplicity.  Medium response and
energy loss can produce additional modifications
\cite{Chien:2015hda,Kang:2017frl,Casalderrey-Solana:2016jvj,
Mehtar-Tani:2016aco,Tachibana:2017syd,Yang:2023dwc,Barata:2023bhh,Bossi:2024qho,
Ke:2025ibt}.  The operator viewpoint developed here is useful for this problem because
it keeps the measured jet evolution separate from the hard production kernel and makes
the inclusive-limit constraint explicit.  It should therefore provide a controlled
starting point for future studies of these effects.  Combined measurements of jet
multiplicity, energy correlations, and nuclear modification factors may help disentangle
saturation, multiplicity bias, and medium modification.

Finally, the present calculation should not be interpreted as a complete NLO
factorization of the multiplicity-dependent cross section.  Establishing that
factorization is a central objective of future work.  Its implementation must include
subtraction-local finite measured hard coefficients in every partonic channel and must
match them consistently to the resummed evolution.

\acknowledgments
This work is supported in part by the National Natural Science Foundation of China
(NSFC) under Grant No.~1234710148, and in part by the China
Postdoctoral Science Foundation under Grant No.~2023M742098.

\appendix

\section{Scheme conventions and inclusive-limit checks}
\label{app:scheme}

This appendix records the scheme conventions used throughout the construction.  The
purpose is to make the matching testable term by term.  We follow the organization of
the inclusive NLO/resummed forward-jet calculation and do not introduce a new
factorization scheme.  The semi-hard scale \(\Lamb\), the jet scale
\(\mu_J\simeq P_T R\), and the target rapidity \(Y=\ln(1/x_A)\) are treated as
independent bookkeeping variables.  The multiplicity generating functions evolve in the
angular evolution variable, not in \(\Lamb\).

The full multiplicity-dependent matched cross section may be written as
\begin{equation}
  \dd\sigma^{\rm match}(s)
  =
  \dd\sigma^{\rm resum}(s)
  +
  \left[
  \dd\sigma^{\rm NLO}(s)
  -
  \left.\dd\sigma^{\rm resum}(s)\right|_{\rm NLO}
  \right].
  \label{eq:app-add-subtract}
\end{equation}
The inclusive limit requires
\begin{equation}
  Z_i(0)=1,\qquad
  Z_a(0)Z_b(0)-Z_i(0)=0,
  \label{eq:app-z0}
\end{equation}
and therefore
\begin{equation}
  \left.\dd\sigma^{\rm match}(s)\right|_{s=0}
  =
  \dd\sigma_{\rm incl}^{\rm resum+NLO}.
  \label{eq:app-inclusive-limit}
\end{equation}
This condition is the main non-negotiable check of the framework.  It fixes the
normalization of all measured numerators and forbids any residual multiplicity weight in
PDF, BK/JIMWLK, or inclusive virtual subtraction terms.
Equation~\eqref{eq:app-inclusive-limit} refers to the formal matched construction.  The
restricted numerical baseline described in Sec.~\ref{sec:uncertainties} implements this
identity only for the channels and kinematic points passing the stated stability
criteria.

The logarithmic sectors are assigned as follows:
\begin{align}
  \text{PDF counterterms:}\quad&
  \dd\sigma_{\rm PDF}^{(1)}(s)
  =
  \dd\sigma_{\rm PDF}^{(1)}(0)\,Z_i(s),
  \label{eq:app-pdf-sector}
  \\
  \text{BK/JIMWLK rapidity subtraction:}\quad&
  \dd\sigma_{\rm BK}^{(1)}(s)
  =
  \dd\sigma_{\rm BK}^{(1)}(0)\,Z_i(s),
  \label{eq:app-bk-sector}
  \\
  \text{SiJF cone logarithms:}\quad&
  \begin{aligned}[t]
  J_i(z,\Lamb)
  &\to
  \sum_j \int_z^1\frac{\dd z_J}{z_J}\,
  \SiJF_{ji}\!\left(\frac{z}{z_J},\Lamb,\mu_J\right)
  \\
  &\quad\times
  Z_j(s,z_J\omega,\mu_J),
  \end{aligned}
  \label{eq:app-sijf-sector}
  \\
  \text{in-cone multiplicity evolution:}\quad&
  \begin{aligned}[t]
  \Gamma_{i\to ab}
  \big[
  &Z_a(s,x\omega,\mu_J)Z_b(s,(1-x)\omega,\mu_J)
  \\
  &-
  Z_i(s,\omega,\mu_J)
  \big].
  \end{aligned}
  \label{eq:app-z-sector}
\end{align}
Equations~\eqref{eq:app-pdf-sector} and \eqref{eq:app-bk-sector} state that unresolved
initial-state collinear radiation and target rapidity subtraction are not counted as
charged particles inside the reconstructed final-state jet.  Equation~\eqref{eq:app-sijf-sector}
assigns one-out-of-cone final-state splittings to the flavor-transfer SiJF matrix.
Equation~\eqref{eq:app-z-sector} assigns both-in-cone unresolved branchings to the
nonlinear generating-function evolution.  This separation is what prevents the cone
logarithms in the hard coefficients from being counted a second time in \(Z_i\).

For the finite NLO remainders imported from the inclusive hard coefficients, the
inclusive forward-jet organization of Ref.~\cite{Wang:2022zla} gives
\begin{align}
  \dd\sigma_{\rm fin,qq}^{(1)}(s)
  &=
  \left[
  \dd\sigma_{qq}^{3}
  +\dd\sigma_{qq}^{4}
  +\frac{1}{2}\dd\sigma_{qq}^{5a}
  +\dd\sigma_{qq}^{5b}
  +\dd\sigma_{qq}^{6}
  +\dd\sigma_{qq}^{7}
  +\dd\sigma_{qq}^{8}
  +\dd\sigma_{qq}^{10}
  +\dd\sigma_{qq}^{11}
  \right] Z_q(s),
  \label{eq:app-fin-qq}
  \\
  \dd\sigma_{\rm fin,gg}^{(1)}(s)
  &=
  \left[
  \dd\sigma_{gg}^{3}
  +\dd\sigma_{gg}^{4}
  +\dd\sigma_{gg}^{5}
  +\dd\sigma_{gg}^{6}
  +\frac{1}{2}\dd\sigma_{gg}^{7a}
  +\dd\sigma_{gg}^{7b}
  +\dd\sigma_{gg}^{8}
  +\dd\sigma_{gg}^{9}
  +\dd\sigma_{gg}^{10}
  \right] Z_g(s),
  \label{eq:app-fin-gg}
  \\
  \dd\sigma_{\rm fin,gq}^{(1)}(s)
  &=
  \left[
  \dd\sigma_{gq}^{2}
  +\dd\sigma_{gq}^{3}
  +\dd\sigma_{gq}^{5}
  \right] Z_g(s),
  \label{eq:app-fin-gq}
  \\
  \dd\sigma_{\rm fin,qg}^{(1)}(s)
  &=
  \left[
  \dd\sigma_{qg}^{2}
  +\dd\sigma_{qg}^{3}
  +\dd\sigma_{qg}^{5}
  \right] Z_q(s).
  \label{eq:app-fin-qg}
\end{align}
For the isolated matched numerator of a resolved real-emission kernel, one inserts the
local difference
\begin{equation}
  \Mmeas_{i\to ab}
  =
  Z_a(s,\omega_a)Z_b(s,\omega_b)-Z_i(s,\omega_i),
  \label{eq:app-local-weight}
\end{equation}
with the same subtraction term evaluated in the same variables.  This difference does
not replace the coherent contribution in the full cross section: the complete measured
term is the coherent baseline \(D_i^{\rm incl}Z_i\) plus the isolated matched numerator,
as in Eq.~\eqref{eq:gg-cross-section-match}.  The insertion is made only for
real-emission pieces that contain a resolved daughter configuration.
Endpoint, virtual, PDF, BK/JIMWLK, Sudakov, running-coupling, and SiJF subtraction
pieces are evaluated with the parent-coherent measurement, or equivalently with
\(Z=1\) in the isolated multiplicity numerator.  This is the operational meaning of
the measured numerator \(N_i^{\rm mult}\) and of the \(s=0\) consistency check.
In particular, \(\dd\sigma_{qq}^{9}\) is not included in
Eq.~\eqref{eq:app-fin-qq}: its jet-radius logarithm is assigned to the SiJF sector, in
agreement with the import map of Appendix~\ref{app:hard-coefficients}.  The
\(\Lambda_{\rm IR}\) factors displayed in Appendix~\ref{app:hard-coefficients} are
infrared regulators of the hard-coefficient formulae and should not be confused with
the semi-hard factorization scale \(\Lamb\) used in the main text.

\section{Complete NLO hard coefficients and import map}
\label{app:hard-coefficients}

\input{appendix_hard_coefficients_full}

\nocite{Altarelli:1977zs,Catani:1996vz,Cacciari:2008gp,Cacciari:2011ma,
Koba:1972ng,Konishi:1978yx,Marchesini:1987cf,Catani:1990rr,
Kharzeev:2003wz,Albacete:2010bs,Albacete:2012xq,Lappi:2013zma,
Ducloue:2016shw,Ducloue:2017dit,Becher:2006mr,Sterman:1986aj,Catani:1989ne}

\bibliographystyle{JHEP}
\bibliography{references}

\end{document}

%% file: appendix_hard_coefficients_full.tex
\subsection{Purpose and notation}
\newcommand{\dsig}[2]{\mathrm d\sigma_{#1}^{#2}}

This appendix collects the momentum-space NLO hard coefficients of
Ref.~\cite{Wang:2022zla} in a form useful for the multiplicity-dependent CGC--SiJF
framework. The formulae below are reformatted from Sec.~III of that work. They are not
new results; their role here is to define precisely which terms are imported into the
finite NLO matching piece and which terms are subtracted into PDFs, BK evolution,
Sudakov factors, SiJFs, or multiplicity evolution.

The jet transverse momentum satisfies $\bm P_{J\perp}=z\bm q$. The lower limit of
the $z$ integral is $\tau=P_{J\perp}e^\eta/\sqrt{s}$. To keep the displayed hard
coefficients readable at the same font size as the main text, this appendix abbreviates
\(q\equiv q_\perp\) and \(q_i\equiv q_{i\perp}\); all scalar products and squares are
understood in the transverse plane.  In the long Sudakov-sensitive terms we further use
\(K_i=q-q_i\) and \(Q_i=q-q_i+q_3\).  The transverse dipole amplitude is
\(F(\bm q,x_A)\), with the \(x_A\) argument suppressed below.  We also write
\(\dsig{ab}{i}\equiv \dd\sigma_{ab}^{i}/(\dd\eta\,\dd^2P_J)\) for compactness.

\subsection{\texorpdfstring{$q\to q$}{q to q} channel}

The quark channel is written as
\begin{equation}
  \dsig{qq}{\rm NLO}
  =
  \dsig{qq}{\rm LO}
  +
  \sum_{i=1}^{11}
  \dsig{qq}{i}.
\end{equation}
The LO term is
\begin{equation}
  \dsig{qq}{\rm LO}
  =
  \Sperp
  \int_\tau^1\frac{\dd z}{z^2}
  J_q^{(0)}(z)\,xq(x,\mu^2)F(q).
\end{equation}
The NLO terms are
\begin{align*}
  \dsig{qq}{1}
  &=
  \frac{\as}{2\pi}C_F\Sperp
  \int_\tau^1\frac{\dd z}{z^2}J_q^{(0)}(z)
  \int_x^1\dd\xi\,
  \frac{x}{\xi}q\!\left(\frac{x}{\xi},\mu^2\right)
  P_{qq}(\xi)\ln\frac{\Lambda_{\rm IR}^2}{\mu^2}\,
  F(q),
  \\
  \dsig{qq}{2}
  &=
  \frac{3}{2}\frac{\as}{2\pi}C_F\Sperp
  \int_\tau^1\frac{\dd z}{z^2}J_q^{(0)}(z)
  xq(x,\mu^2)\ln\frac{q^2}{\Lambda_{\rm IR}^2}F(q),
  \\
  \dsig{qq}{3}
  &=
  \frac{\as}{2\pi^2}C_F\Sperp
  \int_\tau^1\frac{\dd z}{z^2}J_q^{(0)}(z)
  \int_x^1\dd\xi
  \int\dd^2q_1\dd^2q_2\,
  \frac{x}{\xi}q\!\left(\frac{x}{\xi},\mu^2\right)
  \frac{1+\xi^2}{(1-\xi)_+}
  \Tqq(\xi,q_1,q_2,q),
  \\
  \dsig{qq}{4}
  &=
  -\frac{\as}{\pi}C_F\Sperp
  \int_\tau^1\frac{\dd z}{z^2}J_q^{(0)}(z)
  \int_0^1\dd\xi'\int\dd^2q_1\,
  xq(x,\mu^2)
  \frac{1+\xi'^2}{(1-\xi')_+}
  \ln\frac{(q_1-\xi' q)^2}{q^2}
  F(q_1)F(q),
  \\
  \dsig{qq}{5}
  &=
  \frac{2\as}{\pi^2}C_F\Sperp
  \int_\tau^1\frac{\dd z}{z^2}J_q^{(0)}(z)
  \int\dd^2q_1\,
  xq(x,\mu^2)
  \frac{1}{q_1^2}\ln\frac{q^2}{q_1^2}
  \left[
  F(K_1)
  -
  \theta(q^2-q_1^2)F(q)
  \right]
  \nonumber\\
  &\hspace{-3em}\quad+
  \frac{\as}{\pi}C_F\Sperp
  \int_\tau^1\frac{\dd z}{z^2}J_q^{(0)}(z)
  \int\dd^2q_1\,
  xq(x,\mu^2)
  F(q_1)F(q)
  \ln\frac{q^2}{K_1^2}
  \nonumber\\
  &\hspace{-3em}\quad-
  \frac{2\as}{\pi}C_F\Sperp
  \int_\tau^1\frac{\dd z}{z^2}J_q^{(0)}(z)
  \int\dd^2q_1\dd^2q_2\,
  xq(x,\mu^2)
  F(q_1)F(q_2)
  \ln\frac{q^2}{K_1^2}
  \nonumber\\
  &\hspace{-3em}\quad\times
  \frac{K_1\cdot K_2}{K_1^2}\,\frac{1}{K_2^2},
  \\
  \dsig{qq}{6}
  &=
  \frac{\as}{2\pi}C_F\Sperp
  \left(6-\frac{4\pi^2}{3}\right)
  \int_\tau^1\frac{\dd z}{z^2}
  J_q(z)xq(x,\mu^2)F(q),
  \\
  \dsig{qq}{7}
  &=
  -\frac{\as}{2\pi}C_F\Sperp
  \int_\tau^1\frac{\dd z}{z^2}J_q^{(0)}(z)
  \int_x^1\dd\xi\,
  \frac{x}{\xi}q\!\left(\frac{x}{\xi},\mu^2\right)
  \left[
  \frac{\ln[(1-\xi)^2/\xi^2]}{1-\xi}
  \right]_+
  F(q/\xi),
  \\
  \dsig{qq}{8}
  &=
  \frac{3}{2}\frac{\as}{2\pi}C_F\Sperp
  \int_\tau^1\frac{\dd z}{z^2}J_q^{(0)}(z)
  xq(x,\mu^2)\ln\frac{q^2}{\Lambda_{\rm IR}^2}F(q),
  \\
  \dsig{qq}{9}
  &=
  \frac{\as}{2\pi}C_F\Sperp
  \int_\tau^1\frac{\dd z}{z^2}J_q^{(0)}(z)
  \int_x^1\dd\xi\,
  \frac{x}{\xi}q\!\left(\frac{x}{\xi},\mu^2\right)
  \frac{1}{\xi^2}P_{qq}(\xi)
  \ln\frac{\Lambda_{\rm IR}^2}{q^2R^2}
  F(q/\xi),
  \\
  \dsig{qq}{10}
  &=
  \frac{\as}{2\pi}C_F\Sperp
  \int_\tau^1\frac{\dd z}{z^2}J_q^{(0)}(z)
  \int_x^1\dd\xi\,
  \frac{x}{\xi}q\!\left(\frac{x}{\xi},\mu^2\right)
  (1-\xi)F(q),
  \\
  \dsig{qq}{11}
  &=
  -\frac{1}{2}\frac{\as}{2\pi}C_F\Sperp
  \int_\tau^1\frac{\dd z}{z^2}J_q^{(0)}(z)
  xq(x,\mu^2)F(q).
\end{align*}
The auxiliary kernel is
\begin{align*}
  \Tqq(\xi,q_1,q_2,q)
  &=
  \frac{(q_2-q_1/\xi)^2}
  {(q+q_1)^2(q/\xi+q_2)^2}
  F(q_1)F(q_2)
  \nonumber\\
  &\hspace{-3em}\quad-
  \frac{1}{(q+q_1)^2}
  \frac{\Lambda_{\rm IR}^2}{\Lambda_{\rm IR}^2+(q+q_1)^2}
  F(q_2)F(q)
  \nonumber\\
  &\hspace{-3em}\quad-
  \frac{1}{(q+\xi q_2)^2}
  \frac{\Lambda_{\rm IR}^2}{\Lambda_{\rm IR}^2+(q/\xi+q_2)^2}
  F(q/\xi)F(q_1).
\end{align*}
The Sudakov split is
\begin{equation}
  \dsig{qq}{5}
  =
  \dsig{qq}{5a}
  +
  \dsig{qq}{5b},
\end{equation}
with
\begin{equation}
  \dsig{qq}{5a}
  =
  -\frac{\as}{2\pi}C_F\Sperp
  \int_\tau^1\frac{\dd z}{z^2}J_q^{(0)}(z)
  xq(x,\mu^2)F(q)
  \ln\frac{2q^2}{\Lambda_{\rm IR}^2},
  \qquad
  \sigma_{qq}^{5b}\equiv\sigma_{qq}^{5}-\sigma_{qq}^{5a}.
\end{equation}

\subsection{\texorpdfstring{$g\to g$}{g to g} channel}

The gluon channel is
\begin{equation}
  \dsig{gg}{\rm NLO}
  =
  \dsig{gg}{\rm LO}
  +
  \sum_{i=1}^{11}
  \dsig{gg}{i},
\end{equation}
with
\begin{align*}
  \dsig{gg}{\rm LO}
  &=
  \Sperp
  \int_\tau^1\frac{\dd z}{z^2}J_g^{(0)}(z)xg(x,\mu^2)
  \int\dd^2q_1F(q_1)F(q-q_1),
  \\
  \dsig{gg}{1}
  &=
  \frac{\as}{2\pi}N_c\Sperp
  \int_\tau^1\frac{\dd z}{z^2}J_g^{(0)}(z)
  \int_x^1\dd\xi\int\dd^2q_1\,
  g\!\left(\frac{x}{\xi},\mu^2\right)
  P_{gg}(\xi)\ln\frac{\Lambda_{\rm IR}^2}{\mu^2}
  F(q-q_1)F(q_1),
  \\
  \dsig{gg}{2}
  &=
  2\beta_0\frac{\as}{2\pi}N_c\Sperp
  \int_\tau^1\frac{\dd z}{z^2}J_g^{(0)}(z)xg(x,\mu^2)
  \int\dd^2q_1F(q-q_1)F(q_1)
  \ln\frac{q^2}{\Lambda_{\rm IR}^2},
  \\
  \dsig{gg}{3}
  &=
  -\frac{1}{3}\frac{\as}{2\pi}N_fT_R\Sperp
  \int_\tau^1\frac{\dd z}{z^2}J_g^{(0)}(z)xg(x,\mu^2)
  \int\dd^2q_1F(q-q_1)F(q_1),
  \\
  \dsig{gg}{4}
  &=
  \frac{\as}{\pi^2}N_c\Sperp
  \int_\tau^1\frac{\dd z}{z^2}J_g^{(0)}(z)
  \int_x^1\dd\xi\,
  g\!\left(\frac{x}{\xi},\mu^2\right)
  \frac{[1-\xi(1-\xi)]^2}{\xi(1-\xi)_+}
  \nonumber\\
  &\hspace{-3em}\quad\times
  \int\dd^2q_1\dd^2q_2\dd^2q_3\,
  \Tgg(\xi,q_1,q_2,q_3,q),
  \\
  \dsig{gg}{5}
  &=
  -2N_fT_R\Sperp\frac{\as}{2\pi}
  \int_\tau^1\frac{\dd z}{z^2}J_g^{(0)}(z)xg(x,\mu^2)
  \int_0^1\dd\xi'\int\dd^2q_1
  [\xi'^2+(1-\xi')^2]
  \nonumber\\
  &\hspace{-3em}\quad\times
  F(q_1)F(q-q_1)
  \ln\frac{(q_1-\xi' q)^2}{q^2},
  \\
  \dsig{gg}{6}
  &=
  -4N_c\Sperp\frac{\as}{2\pi}
  \int_\tau^1\frac{\dd z}{z^2}J_g^{(0)}(z)xg(x,\mu^2)
  \int_0^1\dd\xi'\int\dd^2q_1\dd^2q_2
  \left[
  \frac{\xi'}{(1-\xi')_+}+\frac{\xi'(1-\xi')}{2}
  \right]
  \nonumber\\
  &\hspace{-3em}\quad\times
  F(q_1)F(q_2)F(q-q_1)
  \ln\frac{(q_1+q_2-\xi' q)^2}{q^2},
  \\
  \dsig{gg}{8}
  &=
  \frac{\as}{2\pi}\Sperp N_c
  \left(\frac{67}{9}-\frac{4\pi^2}{3}\right)
  \int_\tau^1\frac{\dd z}{z^2}J_g^{(0)}(z)xg(x)
  \int\dd^2q_1F(q_1)F(q-q_1)
  \nonumber\\
  &\hspace{-3em}\quad-
  \frac{\as}{2\pi}\Sperp N_fT_R\frac{26}{9}
  \int_\tau^1\frac{\dd z}{z^2}J_g^{(0)}(z)xg(x)
  \int\dd^2q_1F(q_1)F(q-q_1),
  \\
  \dsig{gg}{9}
  &=
  -\frac{\as}{2\pi}\Sperp N_c
  \int_\tau^1\frac{\dd z}{z^2}J_g^{(0)}(z)
  \int_x^1\dd\xi\,
  g\!\left(\frac{x}{\xi}\right)
  \left[
  \frac{\ln[(1-\xi)^2/\xi^2]}{1-\xi}
  \right]_+
  \nonumber\\
  &\hspace{-3em}\quad\times
  \frac{2[1-\xi(1-\xi)]^2}{\xi^3}
  \int\dd^2q_1F(q_1)F(q/\xi-q_1),
  \\
  \dsig{gg}{10}
  &=
  2\beta_0\frac{\as}{2\pi}N_c\Sperp
  \int_\tau^1\frac{\dd z}{z^2}J_g^{(0)}(z)
  xg(x)
  \int\dd^2q_1F(q-q_1)F(q_1)
  \ln\frac{q^2}{\Lambda_{\rm IR}^2},
  \\
  \dsig{gg}{11}
  &=
  \frac{\as}{2\pi}N_c\Sperp
  \int_\tau^1\frac{\dd z}{z^2}J_g^{(0)}(z)
  \int_x^1\dd\xi\,
  g\!\left(\frac{x}{\xi}\right)
  \frac{1}{\xi^2}P_{gg}(\xi)
  \int\dd^2q_1
  F(q/\xi-q_1)F(q_1)
  \ln\frac{\Lambda_{\rm IR}^2}{q^2R^2}.
\end{align*}
The auxiliary three-dipole kernel is
\begin{align*}
  \Tgg(\xi,q_1,q_2,q_3,q)
  &=
  \frac{1}{\xi^2}
  \frac{\big[(1-\xi)q_1+q_3-\xi q_2\big]^2}
  {(q_1+q_3-q)^2
  (q_1+q_2-q/\xi)^2}
  F(q_1)F(q_2)F(q_3)
  \nonumber\\
  &\hspace{-3em}\quad
  -
  \frac{1}{(q_1+q_3-q)^2}
  \frac{\Lambda_{\rm IR}^2}{\Lambda_{\rm IR}^2+(q_1+q_3-q)^2}
  F(q-q_1)F(q_2)F(q_3)
  \nonumber\\
  &\hspace{-3em}\quad
  -
  \frac{1}{\xi^2}
  \frac{1}{(q_1+q_2-q/\xi)^2}
  \frac{\Lambda_{\rm IR}^2}{\Lambda_{\rm IR}^2+(q_1+q_2-q/\xi)^2}
  \nonumber\\
  &\hspace{-3em}\quad\times
  F(q/\xi-q_2)F(q_2)F(q_3).
\end{align*}
The long Sudakov-sensitive term is
\begin{align*}
  \dsig{gg}{7}
  &=
  \frac{2\as}{\pi^2}N_c\Sperp
  \int_\tau^1\frac{\dd z}{z^2}J_g^{(0)}(z)xg(x,\mu^2)
  \int\dd^2q_1\dd^2q_2
  \frac{1}{q_2^2}\ln\frac{q^2}{q_2^2}
  F(q-q_1)
  \nonumber\\
  &\hspace{-3em}\quad\times
  \left[
  F(q_1+q_2)
  -
  \theta(q^2-q_2^2)F(q_1)
  \right]
  \nonumber\\
  &\hspace{-3em}\quad+
  \frac{\as}{\pi}N_c\Sperp
  \int_\tau^1\frac{\dd z}{z^2}J_g^{(0)}(z)xg(x,\mu^2)
  \int\dd^2q_1\dd^2q_2
  F(q_1)F(q_2)F(q-q_2)
  \ln\frac{q^2}{(q_1+q_2-q)^2}
  \nonumber\\
  &\hspace{-3em}\quad-
  \frac{2\as}{\pi}N_c\Sperp
  \int_\tau^1\frac{\dd z}{z^2}J_g^{(0)}(z)xg(x,\mu^2)
  \int\dd^2q_1\dd^2q_2\dd^2q_3
  F(q_1)F(q_2)F(q_3)
  \nonumber\\
  &\hspace{-3em}\quad\times
  \frac{Q_1\cdot Q_2}{Q_1^2}\,
  \frac{1}{Q_2^2}
  \ln\frac{q^2}{Q_1^2}.
\end{align*}
Its Sudakov split is
\begin{align*}
  \sigma_{gg}^{7}&=\sigma_{gg}^{7a}+\sigma_{gg}^{7b},
  \nonumber\\
  \dsig{gg}{7a}
  &=
  -\frac{\as}{2\pi}N_c\Sperp
  \int_\tau^1\frac{\dd z}{z^2}J_g^{(0)}(z)xg(x,\mu^2)
  \nonumber\\
  &\hspace{-3em}\quad\times
  \ln\frac{2q^2}{\Lambda_{\rm IR}^2}
  \int\dd^2q_1F(q-q_1)F(q_1),
\end{align*}
and $\sigma_{gg}^{7b}\equiv\sigma_{gg}^{7}-\sigma_{gg}^{7a}$. Equivalently,
\begin{align*}
  \dsig{gg}{7b}
  &=
  \frac{2\as}{\pi^2}N_c\Sperp
  \int_\tau^1\frac{\dd z}{z^2}J_g^{(0)}(z)xg(x,\mu^2)
  \int\dd^2q_1\dd^2q_2
  \frac{1}{q_2^2}\ln\frac{q^2}{q_2^2}
  F(q-q_1)
  \nonumber\\
  &\hspace{-3em}\quad\times
  \left[
  F(q_1+q_2)
  -
  \theta(\Lambda_{\rm IR}^2-q_2^2)F(q_1)
  \right]
  \nonumber\\
  &\hspace{-3em}\quad-
  \frac{\as}{2\pi}N_c\Sperp
  \int_\tau^1\frac{\dd z}{z^2}J_g^{(0)}(z)xg(x,\mu^2)
  \ln\frac{q^2}{\Lambda_{\rm IR}^2}
  \int\dd^2q_1F(q-q_1)F(q_1)
  \nonumber\\
  &\hspace{-3em}\quad+
  \frac{\as}{\pi}N_c\Sperp
  \int_\tau^1\frac{\dd z}{z^2}J_g^{(0)}(z)xg(x,\mu^2)
  \int\dd^2q_1\dd^2q_2
  F(q_1)F(q_2)F(q-q_2)
  \ln\frac{q^2}{(q_1+q_2-q)^2}
  \nonumber\\
  &\hspace{-3em}\quad-
  \frac{2\as}{\pi}N_c\Sperp
  \int_\tau^1\frac{\dd z}{z^2}J_g^{(0)}(z)xg(x,\mu^2)
  \int\dd^2q_1\dd^2q_2\dd^2q_3
  F(q_1)F(q_2)F(q_3)
  \nonumber\\
  &\hspace{-3em}\quad\times
  \frac{Q_1\cdot Q_2}{Q_1^2}\,
  \frac{1}{Q_2^2}
  \ln\frac{q^2}{Q_1^2}.
\end{align*}

\subsection{Off-diagonal channels}

For $q\to g$,
\begin{equation}
  \dsig{gq}{\rm NLO}
  =
  \sum_{i=1}^{5}
  \dsig{gq}{i},
\end{equation}
with
\begin{align*}
  \dsig{gq}{1}
  &=
  \frac{\as}{2\pi}C_F\Sperp
  \int_\tau^1\frac{\dd z}{z^2}J_g^{(0)}(z)
  \int_x^1\dd\xi
  \int\dd^2q_1\,
  q\!\left(\frac{x}{\xi},\mu^2\right)
  P_{gq}(\xi)\ln\frac{\Lambda_{\rm IR}^2}{\mu^2}
  F(q_1)F(q-q_1),
  \\
  \dsig{gq}{2}
  &=
  \frac{\as}{2\pi}C_F\Sperp
  \int_\tau^1\frac{\dd z}{z^2}J_g^{(0)}(z)
  \int_x^1\dd\xi\int\dd^2q_1\,
  q\!\left(\frac{x}{\xi},\mu^2\right)
  \xi F(q_1)F(q-q_1),
  \\
  \dsig{gq}{3}
  &=
  \frac{\as}{2\pi^2}C_F\Sperp
  \int_\tau^1\frac{\dd z}{z^2}J_g^{(0)}(z)
  \int_x^1\dd\xi
  \int\dd^2q_1\dd^2q_2\,
  xq(x,\mu^2)P_{gq}(\xi)
  \Tgq(\xi,q_1,q_2,q),
  \\
  \dsig{gq}{4}
  &=
  \frac{\as}{2\pi}C_F\Sperp
  \int_\tau^1\frac{\dd z}{z^2}J_g^{(0)}(z)
  \int_x^1\dd\xi\,
  q\!\left(\frac{x}{\xi}\right)
  \frac{1}{\xi^2}P_{gq}(\xi)
  \ln\frac{\Lambda_{\rm IR}^2}{q^2R^2}F(q/\xi),
  \\
  \dsig{gq}{5}
  &=
  -\frac{\as}{2\pi}C_F\Sperp
  \int_\tau^1\frac{\dd z}{z^2}J_g^{(0)}(z)
  \int_x^1\dd\xi\,
  q\!\left(\frac{x}{\xi}\right)
  P_{gq}(\xi)
  \frac{1}{\xi^2}F(q/\xi)
  \ln\frac{(1-\xi)^2}{\xi^2}.
\end{align*}
For $g\to q$,
\begin{equation}
  \dsig{qg}{\rm NLO}
  =
  \sum_{i=1}^{5}
  \dsig{qg}{i},
\end{equation}
with
\begin{align*}
  \dsig{qg}{1}
  &=
  \frac{\as}{2\pi}\Sperp T_R
  \int_\tau^1\frac{\dd z}{z^2}J_q^{(0)}(z)
  \int_x^1\dd\xi\,
  g\!\left(\frac{x}{\xi},\mu^2\right)
  P_{qg}(\xi)\ln\frac{\Lambda_{\rm IR}^2}{\mu^2}F(q),
  \\
  \dsig{qg}{2}
  &=
  2\frac{\as}{2\pi}\Sperp T_R
  \int_\tau^1\frac{\dd z}{z^2}J_q^{(0)}(z)
  \int_x^1\dd\xi\,
  g\!\left(\frac{x}{\xi},\mu^2\right)
  \xi(1-\xi)F(q),
  \\
  \dsig{qg}{3}
  &=
  \frac{\as}{2\pi^2}\Sperp T_R
  \int_\tau^1\frac{\dd z}{z^2}J_q^{(0)}(z)
  \int_x^1\dd\xi
  \int\dd^2q_1\dd^2q_2\,
  g\!\left(\frac{x}{\xi},\mu^2\right)
  P_{qg}(\xi)
  \Tqg(\xi,q_1,q_2,q),
  \\
  \dsig{qg}{4}
  &=
  \frac{\as}{2\pi}\Sperp T_R
  \int_\tau^1\frac{\dd z}{z^2}J_q^{(0)}(z)
  \int_x^1\dd\xi\int\dd^2q_1\,
  g\!\left(\frac{x}{\xi},\mu^2\right)
  \frac{1}{\xi^2}P_{qg}(\xi)
  \ln\frac{\Lambda_{\rm IR}^2}{q^2R^2}
  F(q_1)F(q/\xi-q_1),
  \\
  \dsig{qg}{5}
  &=
  -\frac{\as}{2\pi}\Sperp T_R
  \int_\tau^1\frac{\dd z}{z^2}J_q^{(0)}(z)
  \int_x^1\dd\xi\int\dd^2q_1\,
  g\!\left(\frac{x}{\xi},\mu^2\right)
  \frac{1}{\xi^2}P_{qg}(\xi)
  F(q_1)F(q/\xi-q_1)
  \nonumber\\
  &\hspace{-3em}\quad\times
  \ln\frac{(1-\xi)^2}{\xi^2}.
\end{align*}
The corresponding finite kernels are
\begin{align*}
  \Tgq(\xi,q_1,q_2,q)
  &=
  \left[
  \frac{q-q_1-q_2}
  {(q-q_1-q_2)^2}
  -
  \frac{q-\xi q_2}
  {(q-\xi q_2)^2}
  \right]^2
  F(q_1)F(q_2)
  \nonumber\\
  &\hspace{-3em}\quad
  -
  \frac{\Lambda_{\rm IR}^2}{\Lambda_{\rm IR}^2+(q-q_1-q_2)^2}
  \frac{1}{(q-q_1-q_2)^2}
  F(q_2)F(q-q_2)
  \nonumber\\
  &\hspace{-3em}\quad
  -
  \frac{\Lambda_{\rm IR}^2}{\Lambda_{\rm IR}^2+(q/\xi-q_2)^2}
  \frac{1}{(q-\xi q_2)^2}
  F(q_1)F(q/\xi),
  \\
  \Tqg(\xi,q_1,q_2,q)
  &=
  \left[
  \frac{q-\xi q_1-\xi q_2}
  {(q-\xi q_1-\xi q_2)^2}
  -
  \frac{q-q_2}
  {(q-q_2)^2}
  \right]^2
  F(q_1)F(q_2)
  \nonumber\\
  &\hspace{-3em}\quad
  -
  \frac{1}{(q-\xi q_1-\xi q_2)^2}
  \frac{\Lambda_{\rm IR}^2}{\Lambda_{\rm IR}^2+(q/\xi-q_1-q_2)^2}
  F(q_2)F(q/\xi-q_2)
  \nonumber\\
  &\hspace{-3em}\quad
  -
  \frac{1}{(q-q_2)^2}
  \frac{\Lambda_{\rm IR}^2}{\Lambda_{\rm IR}^2+(q-q_2)^2}
  F(q_1)F(q).
\end{align*}

\subsection{Import map}

\begin{center}
\begin{tabular}{lll}
\toprule
Channel & Imported finite pieces & Subtracted/evolved pieces \\
\midrule
$q\to q$ &
$\sigma_{qq}^{3,4,5b,6,7,8,10,11}+\frac12\sigma_{qq}^{5a}$ &
$\sigma_{qq}^{1,2,9}$ and $Z_q^{(1)}$ \\
$g\to g$ &
$\sigma_{gg}^{3,4,5,6,7b,8,9,10}+\frac12\sigma_{gg}^{7a}$ &
$\sigma_{gg}^{1,2,11}$ and $Z_g^{(1)}$ \\
$q\to g$ &
$\sigma_{gq}^{2,3,5}$ &
$\sigma_{gq}^{1,4}$ \\
$g\to q$ &
$\sigma_{qg}^{2,3,5}$ &
$\sigma_{qg}^{1,4}$ \\
\bottomrule
\end{tabular}
\end{center}

%% file: references.bib
@article{McLerran:1993ni,
  author = {McLerran, Larry D. and Venugopalan, Raju},
  title = {Computing quark and gluon distribution functions for very large nuclei},
  journal = {Phys. Rev. D},
  volume = {49},
  pages = {2233},
  year = {1994},
  doi = {10.1103/PhysRevD.49.2233}
}

@article{McLerran:1993ka,
  author = {McLerran, Larry D. and Venugopalan, Raju},
  title = {Gluon distribution functions for very large nuclei at small transverse momentum},
  journal = {Phys. Rev. D},
  volume = {49},
  pages = {3352},
  year = {1994},
  doi = {10.1103/PhysRevD.49.3352}
}

@article{Balitsky:1995ub,
  author = {Balitsky, I.},
  title = {Operator expansion for high-energy scattering},
  journal = {Nucl. Phys. B},
  volume = {463},
  pages = {99},
  year = {1996},
  eprint = {hep-ph/9509348},
  archivePrefix = {arXiv},
  doi = {10.1016/0550-3213(95)00638-9}
}

@article{Kovchegov:1999yj,
  author = {Kovchegov, Yuri V.},
  title = {Small-x F2 structure function of a nucleus including multiple pomeron exchanges},
  journal = {Phys. Rev. D},
  volume = {60},
  pages = {034008},
  year = {1999},
  eprint = {hep-ph/9901281},
  archivePrefix = {arXiv},
  doi = {10.1103/PhysRevD.60.034008}
}

@article{Gelis:2010nm,
  author = {Gelis, Francois and Iancu, Edmond and Jalilian-Marian, Jamal and Venugopalan, Raju},
  title = {The Color Glass Condensate},
  journal = {Ann. Rev. Nucl. Part. Sci.},
  volume = {60},
  pages = {463},
  year = {2010},
  eprint = {1002.0333},
  archivePrefix = {arXiv},
  primaryClass = {hep-ph},
  doi = {10.1146/annurev.nucl.010909.083629}
}

@article{Chirilli:2011km,
  author = {Chirilli, Giovanni A. and Xiao, Bo-Wen and Yuan, Feng},
  title = {One-loop factorization for inclusive hadron production in pA collisions in the saturation formalism},
  journal = {Phys. Rev. Lett.},
  volume = {108},
  pages = {122301},
  year = {2012},
  eprint = {1112.1061},
  archivePrefix = {arXiv},
  primaryClass = {hep-ph},
  doi = {10.1103/PhysRevLett.108.122301}
}

@article{Mueller:2013wwa,
  author = {Mueller, Alfred H. and Xiao, Bo-Wen and Yuan, Feng},
  title = {Sudakov double logarithms resummation in hard processes in the small-x saturation formalism},
  journal = {Phys. Rev. D},
  volume = {88},
  pages = {114010},
  year = {2013},
  eprint = {1308.2993},
  archivePrefix = {arXiv},
  primaryClass = {hep-ph},
  doi = {10.1103/PhysRevD.88.114010}
}

@article{Kang:2016mcy,
  author = {Kang, Zhong-Bo and Ringer, Felix and Vitev, Ivan},
  title = {The semi-inclusive jet function in SCET and small radius resummation for inclusive jet production},
  journal = {JHEP},
  volume = {10},
  number = {10},
  pages = {125},
  year = {2016},
  eprint = {1606.06732},
  archivePrefix = {arXiv},
  primaryClass = {hep-ph},
  doi = {10.1007/JHEP10(2016)125}
}

@article{Duan:2025hmjets,
  author = {Duan, Pi and Ke, Weiyao and Qin, Guang-You and Wang, Lei},
  title = {Internal multiplicity distributions of jets from nonlinear evolution within the jet function framework},
  journal = {JHEP},
  volume = {03},
  number = {03},
  pages = {243},
  year = {2026},
  eprint = {2510.04895},
  archivePrefix = {arXiv},
  primaryClass = {hep-ph},
  doi = {10.1007/JHEP03(2026)243}
}

@article{Wang:2022zla,
  author = {Wang, Lei and Chen, Lin and Gao, Zhan and Shi, Yu and Wei, Shu-Yi and Xiao, Bo-Wen},
  title = {Forward Inclusive Jet Productions in pA Collisions},
  journal = {Phys. Rev. D},
  volume = {107},
  pages = {016016},
  year = {2023},
  eprint = {2211.08322},
  archivePrefix = {arXiv},
  primaryClass = {hep-ph},
  doi = {10.1103/PhysRevD.107.016016}
}

@article{BFKL:1977,
  author = {Kuraev, E. A. and Lipatov, L. N. and Fadin, V. S.},
  title = {Multi-Reggeon Processes in the Yang-Mills Theory},
  journal = {Sov. Phys. JETP},
  volume = {44},
  pages = {443},
  year = {1976},
  url = {http://www.jetp.ras.ru/cgi-bin/e/index/e/44/2/p443?a=list}
}

@article{GLR:1983,
  author = {Gribov, L. V. and Levin, E. M. and Ryskin, M. G.},
  title = {Semihard Processes in QCD},
  journal = {Phys. Rept.},
  volume = {100},
  pages = {1},
  year = {1983},
  doi = {10.1016/0370-1573(83)90022-4}
}

@article{MuellerQiu:1986,
  author = {Mueller, Alfred H. and Qiu, Jian-wei},
  title = {Gluon Recombination and Shadowing at Small Values of x},
  journal = {Nucl. Phys. B},
  volume = {268},
  pages = {427},
  year = {1986},
  doi = {10.1016/0550-3213(86)90164-1}
}

@article{JIMWLK:1997,
  author = {Jalilian-Marian, Jamal and Kovner, Alex and Leonidov, Andrei and Weigert, Heribert},
  title = {The BFKL equation from the Wilson renormalization group},
  journal = {Nucl. Phys. B},
  volume = {504},
  pages = {415},
  year = {1997},
  eprint = {hep-ph/9701284},
  archivePrefix = {arXiv},
  doi = {10.1016/S0550-3213(97)00440-9}
}

@article{JIMWLK:2001,
  author = {Weigert, Heribert},
  title = {Unitarity at small Bjorken x},
  journal = {Nucl. Phys. A},
  volume = {703},
  pages = {823},
  year = {2002},
  eprint = {hep-ph/0004044},
  archivePrefix = {arXiv},
  doi = {10.1016/S0375-9474(01)01668-2}
}

@article{Iancu:2001ad,
  author = {Iancu, Edmond and Leonidov, Andrei and McLerran, Larry D.},
  title = {The renormalization group equation for the color glass condensate},
  journal = {Phys. Lett. B},
  volume = {510},
  pages = {133},
  year = {2001},
  eprint = {hep-ph/0102009},
  archivePrefix = {arXiv},
  doi = {10.1016/S0370-2693(01)00524-X}
}

@book{KovchegovLevin:2012,
  author = {Kovchegov, Yuri V. and Levin, Eugene},
  title = {Quantum Chromodynamics at High Energy},
  publisher = {Cambridge University Press},
  year = {2012},
  doi = {10.1017/CBO9781139022187}
}

@article{Dumitru:2002qt,
  author = {Dumitru, Adrian and Jalilian-Marian, Jamal},
  title = {Forward quark jets from protons shattering the colored glass},
  journal = {Phys. Rev. Lett.},
  volume = {89},
  pages = {022301},
  year = {2002},
  eprint = {hep-ph/0204028},
  archivePrefix = {arXiv},
  doi = {10.1103/PhysRevLett.89.022301}
}

@article{Dominguez:2011wm,
  author = {Dominguez, Fabio and Xiao, Bo-Wen and Yuan, Feng},
  title = {$k_t$-factorization for hard processes in nuclei},
  journal = {Phys. Rev. Lett.},
  volume = {106},
  pages = {022301},
  year = {2011},
  eprint = {1009.2141},
  archivePrefix = {arXiv},
  primaryClass = {hep-ph},
  doi = {10.1103/PhysRevLett.106.022301}
}

@article{Dominguez:2011br,
  author = {Dominguez, Fabio and Marquet, Cyrille and Xiao, Bo-Wen and Yuan, Feng},
  title = {Universality of Unintegrated Gluon Distributions at small x},
  journal = {Phys. Rev. D},
  volume = {83},
  pages = {105005},
  year = {2011},
  eprint = {1101.0715},
  archivePrefix = {arXiv},
  primaryClass = {hep-ph},
  doi = {10.1103/PhysRevD.83.105005}
}

@article{Chirilli:2012jd,
  author = {Chirilli, Giovanni A. and Xiao, Bo-Wen and Yuan, Feng},
  title = {Inclusive Hadron Productions in pA Collisions},
  journal = {Phys. Rev. D},
  volume = {86},
  pages = {054005},
  year = {2012},
  eprint = {1203.6139},
  archivePrefix = {arXiv},
  primaryClass = {hep-ph},
  doi = {10.1103/PhysRevD.86.054005}
}

@article{Stasto:2013cha,
  author = {Stasto, Anna M. and Xiao, Bo-Wen and Zaslavsky, Dima},
  title = {Towards the Test of Saturation Physics Beyond Leading Logarithm},
  journal = {Phys. Rev. Lett.},
  volume = {112},
  pages = {012302},
  year = {2014},
  eprint = {1307.4057},
  archivePrefix = {arXiv},
  primaryClass = {hep-ph},
  doi = {10.1103/PhysRevLett.112.012302}
}

@article{Watanabe:2015tja,
  author = {Watanabe, Kazuhiro and Xiao, Bo-Wen and Yuan, Feng and Zaslavsky, Dima},
  title = {Implementing the exact kinematical constraint in the saturation formalism},
  journal = {Phys. Rev. D},
  volume = {92},
  pages = {034026},
  year = {2015},
  eprint = {1505.05183},
  archivePrefix = {arXiv},
  primaryClass = {hep-ph},
  doi = {10.1103/PhysRevD.92.034026}
}

@article{Shi:2021hwx,
  author = {Shi, Yu and Wang, Lei and Wei, Shu-Yi and Xiao, Bo-Wen},
  title = {Threshold resummation in forward hadron productions},
  journal = {Phys. Rev. Lett.},
  volume = {128},
  pages = {202302},
  year = {2022},
  eprint = {2112.06975},
  archivePrefix = {arXiv},
  primaryClass = {hep-ph},
  doi = {10.1103/PhysRevLett.128.202302}
}

@article{Bauer:2000ew,
  author = {Bauer, Christian W. and Fleming, Sean and Luke, Michael E.},
  title = {Summing Sudakov logarithms in B to Xs gamma in effective field theory},
  journal = {Phys. Rev. D},
  volume = {63},
  pages = {014006},
  year = {2000},
  eprint = {hep-ph/0005275},
  archivePrefix = {arXiv},
  doi = {10.1103/PhysRevD.63.014006}
}

@article{Bauer:2001yt,
  author = {Bauer, Christian W. and Fleming, Sean and Pirjol, Dan and Stewart, Iain W.},
  title = {An effective field theory for collinear and soft gluons},
  journal = {Phys. Rev. D},
  volume = {63},
  pages = {114020},
  year = {2001},
  eprint = {hep-ph/0011336},
  archivePrefix = {arXiv},
  doi = {10.1103/PhysRevD.63.114020}
}

@article{Bauer:2002nz,
  author = {Bauer, Christian W. and Pirjol, Dan and Stewart, Iain W.},
  title = {Soft-Collinear Factorization in Effective Field Theory},
  journal = {Phys. Rev. D},
  volume = {65},
  pages = {054022},
  year = {2002},
  eprint = {hep-ph/0109045},
  archivePrefix = {arXiv},
  doi = {10.1103/PhysRevD.65.054022}
}

@article{Beneke:2002ph,
  author = {Beneke, M. and Chapovsky, A. P. and Diehl, M. and Feldmann, T.},
  title = {Soft collinear effective theory and heavy to light currents beyond leading power},
  journal = {Nucl. Phys. B},
  volume = {643},
  pages = {431},
  year = {2002},
  eprint = {hep-ph/0206152},
  archivePrefix = {arXiv},
  doi = {10.1016/S0550-3213(02)00687-9}
}

@article{Kang:2016ehg,
  author = {Kang, Zhong-Bo and Ringer, Felix and Vitev, Ivan},
  title = {Jet substructure using semi-inclusive jet functions in SCET},
  journal = {JHEP},
  volume = {11},
  number = {11},
  pages = {155},
  year = {2016},
  eprint = {1606.07063},
  archivePrefix = {arXiv},
  primaryClass = {hep-ph},
  doi = {10.1007/JHEP11(2016)155}
}

@article{Dasgupta:2014yra,
  author = {Dasgupta, Mrinal and Dreyer, Fr{\'e}d{\'e}ric A. and Salam, Gavin P. and Soyez, Gregory},
  title = {Small-radius jets to all orders in QCD},
  journal = {JHEP},
  volume = {04},
  number = {04},
  pages = {039},
  year = {2015},
  eprint = {1411.5182},
  archivePrefix = {arXiv},
  primaryClass = {hep-ph},
  doi = {10.1007/JHEP04(2015)039}
}

@article{Dasgupta:2016bnd,
  author = {Dasgupta, Mrinal and Dreyer, Fr{\'e}d{\'e}ric A. and Salam, Gavin P. and Soyez, Gregory},
  title = {Inclusive jet spectrum for small-radius jets},
  journal = {JHEP},
  volume = {06},
  number = {06},
  pages = {057},
  year = {2016},
  eprint = {1602.01110},
  archivePrefix = {arXiv},
  primaryClass = {hep-ph},
  doi = {10.1007/JHEP06(2016)057}
}

@article{CMS:HighMultiplicityJets,
  author = {Tumasyan, Armen and others},
  collaboration = {CMS},
  title = {Observation of enhanced long-range elliptic anisotropies inside high-multiplicity jets in pp collisions at \(\sqrt{s}=13\) TeV},
  journal = {Phys. Rev. Lett.},
  volume = {133},
  pages = {142301},
  year = {2024},
  eprint = {2312.17103},
  archivePrefix = {arXiv},
  primaryClass = {hep-ex},
  doi = {10.1103/PhysRevLett.133.142301}
}

@article{CMS:Ridge:2010,
  author = {Khachatryan, Vardan and others},
  collaboration = {CMS},
  title = {Observation of Long-Range Near-Side Angular Correlations in Proton-Proton Collisions at the LHC},
  journal = {JHEP},
  volume = {09},
  number = {09},
  pages = {091},
  year = {2010},
  eprint = {1009.4122},
  archivePrefix = {arXiv},
  primaryClass = {hep-ex},
  doi = {10.1007/JHEP09(2010)091}
}

@article{CMS:Collectivity:2017,
  author = {Khachatryan, Vardan and others},
  collaboration = {CMS},
  title = {Evidence for collectivity in pp collisions at the LHC},
  journal = {Phys. Lett. B},
  volume = {765},
  pages = {193},
  year = {2017},
  eprint = {1606.06198},
  archivePrefix = {arXiv},
  primaryClass = {nucl-ex},
  doi = {10.1016/j.physletb.2016.12.009}
}

@article{ALICE:IntraJetMultiplicity,
  author = {Acharya, Shreyasi and others},
  collaboration = {ALICE},
  title = {Multiplicity dependence of charged-particle intra-jet properties in pp collisions at \(\sqrt{s}=13\) TeV},
  journal = {Eur. Phys. J. C},
  volume = {84},
  pages = {1079},
  year = {2024},
  doi = {10.1140/epjc/s10052-024-13228-0}
}

@article{DokshitzerWebber:1991,
  author = {Dokshitzer, Yuri L. and Webber, B. R.},
  title = {Hadron multiplicity moments in QCD},
  journal = {Phys. Lett. B},
  volume = {352},
  pages = {451},
  year = {1995},
  doi = {10.1016/0370-2693(95)00548-Y}
}

@book{Dokshitzer:1991wu,
  author = {Dokshitzer, Yuri L. and Khoze, Valery A. and Mueller, Alfred H. and Troian, S. I.},
  title = {Basics of Perturbative QCD},
  publisher = {Editions Frontieres},
  year = {1991},
  url = {https://inspirehep.net/literature/325492}
}

@article{Bassetto:1982ma,
  author = {Bassetto, Antonio and Ciafaloni, Marcello and Marchesini, Giuseppe},
  title = {Jet Structure and Infrared Sensitive Quantities in Perturbative QCD},
  journal = {Phys. Rept.},
  volume = {100},
  pages = {201},
  year = {1983},
  doi = {10.1016/0370-1573(83)90083-2}
}

@article{Altarelli:1977zs,
  author = {Altarelli, Guido and Parisi, G.},
  title = {Asymptotic Freedom in Parton Language},
  journal = {Nucl. Phys. B},
  volume = {126},
  pages = {298},
  year = {1977},
  doi = {10.1016/0550-3213(77)90384-4}
}

@article{Catani:1996vz,
  author = {Catani, S. and Seymour, M. H.},
  title = {A general algorithm for calculating jet cross sections in NLO QCD},
  journal = {Nucl. Phys. B},
  volume = {485},
  pages = {291},
  year = {1997},
  eprint = {hep-ph/9605323},
  archivePrefix = {arXiv},
  doi = {10.1016/S0550-3213(96)00589-5}
}

@article{Cacciari:2008gp,
  author = {Cacciari, Matteo and Salam, Gavin P. and Soyez, Gregory},
  title = {The anti-\(k_t\) jet clustering algorithm},
  journal = {JHEP},
  volume = {04},
  number = {04},
  pages = {063},
  year = {2008},
  eprint = {0802.1189},
  archivePrefix = {arXiv},
  primaryClass = {hep-ph},
  doi = {10.1088/1126-6708/2008/04/063}
}

@article{Cacciari:2011ma,
  author = {Cacciari, Matteo and Salam, Gavin P. and Soyez, Gregory},
  title = {FastJet User Manual},
  journal = {Eur. Phys. J. C},
  volume = {72},
  pages = {1896},
  year = {2012},
  eprint = {1111.6097},
  archivePrefix = {arXiv},
  primaryClass = {hep-ph},
  doi = {10.1140/epjc/s10052-012-1896-2}
}

@article{Koba:1972ng,
  author = {Koba, Z. and Nielsen, Holger Bech and Olesen, P.},
  title = {Scaling of multiplicity distributions in high-energy hadron collisions},
  journal = {Nucl. Phys. B},
  volume = {40},
  pages = {317},
  year = {1972},
  doi = {10.1016/0550-3213(72)90551-2}
}

@article{Konishi:1978yx,
  author = {Konishi, K. and Ukawa, A. and Veneziano, G.},
  title = {Jet Calculus: A Simple Algorithm for Resolving QCD Jets},
  journal = {Nucl. Phys. B},
  volume = {157},
  pages = {45},
  year = {1979},
  doi = {10.1016/0550-3213(79)90053-1}
}

@article{Marchesini:1987cf,
  author = {Marchesini, G. and Webber, B. R.},
  title = {Simulation of QCD Jets Including Soft Gluon Interference},
  journal = {Nucl. Phys. B},
  volume = {238},
  pages = {1},
  year = {1984},
  doi = {10.1016/0550-3213(84)90463-2}
}

@article{Catani:1990rr,
  author = {Catani, S. and Webber, B. R. and Marchesini, G.},
  title = {QCD coherent branching and semiinclusive processes at large x},
  journal = {Nucl. Phys. B},
  volume = {349},
  pages = {635},
  year = {1991},
  doi = {10.1016/0550-3213(91)90275-B}
}

@article{Kharzeev:2003wz,
  author = {Kharzeev, Dmitri and Kovchegov, Yuri V. and Tuchin, Kirill},
  title = {Cronin effect and high-\(p_T\) suppression in pA collisions},
  journal = {Phys. Rev. D},
  volume = {68},
  pages = {094013},
  year = {2003},
  eprint = {hep-ph/0307037},
  archivePrefix = {arXiv},
  doi = {10.1103/PhysRevD.68.094013}
}

@article{Albacete:2010bs,
  author = {Albacete, Javier L. and Marquet, Cyrille},
  title = {Single inclusive hadron production at RHIC and the LHC from the color glass condensate},
  journal = {Phys. Lett. B},
  volume = {687},
  pages = {174},
  year = {2010},
  eprint = {1001.1378},
  archivePrefix = {arXiv},
  primaryClass = {hep-ph},
  doi = {10.1016/j.physletb.2010.02.073}
}

@article{Albacete:2012xq,
  author = {Albacete, Javier L. and Dumitru, Adrian and Fujii, Hirotsugu and Nara, Yasushi},
  title = {CGC predictions for p+Pb collisions at the LHC},
  journal = {Nucl. Phys. A},
  volume = {897},
  pages = {1},
  year = {2013},
  eprint = {1209.2001},
  archivePrefix = {arXiv},
  primaryClass = {hep-ph},
  doi = {10.1016/j.nuclphysa.2012.09.012}
}

@article{Lappi:2013zma,
  author = {Lappi, T. and Mantysaari, H.},
  title = {Single inclusive particle production at high energy from HERA data to proton-nucleus collisions},
  journal = {Phys. Rev. D},
  volume = {88},
  pages = {114020},
  year = {2013},
  eprint = {1309.6963},
  archivePrefix = {arXiv},
  primaryClass = {hep-ph},
  doi = {10.1103/PhysRevD.88.114020}
}

@article{Ducloue:2016shw,
  author = {Duclou{\'e}, B. and Lappi, T. and Zhu, Y.},
  title = {Single inclusive forward hadron production at next-to-leading order},
  journal = {Phys. Rev. D},
  volume = {93},
  pages = {114016},
  year = {2016},
  eprint = {1604.00225},
  archivePrefix = {arXiv},
  primaryClass = {hep-ph},
  doi = {10.1103/PhysRevD.93.114016}
}

@article{Ducloue:2017dit,
  author = {Duclou{\'e}, B. and Iancu, E. and Lappi, T. and Mueller, A. H. and Soyez, G. and Triantafyllopoulos, D. N.},
  title = {Use of a running coupling in the NLO calculation of forward hadron production},
  journal = {Phys. Rev. D},
  volume = {97},
  pages = {054020},
  year = {2018},
  eprint = {1712.07480},
  archivePrefix = {arXiv},
  primaryClass = {hep-ph},
  doi = {10.1103/PhysRevD.97.054020}
}

@article{Becher:2006mr,
  author = {Becher, Thomas and Neubert, Matthias},
  title = {Toward a NNLO calculation of the anti-B to X(s) gamma decay rate with a cut on photon energy. II},
  journal = {Phys. Lett. B},
  volume = {637},
  pages = {251},
  year = {2006},
  eprint = {hep-ph/0603140},
  archivePrefix = {arXiv},
  doi = {10.1016/j.physletb.2006.04.046}
}

@article{Sterman:1986aj,
  author = {Sterman, George F.},
  title = {Summation of Large Corrections to Short Distance Hadronic Cross-Sections},
  journal = {Nucl. Phys. B},
  volume = {281},
  pages = {310},
  year = {1987},
  doi = {10.1016/0550-3213(87)90258-6}
}

@article{Catani:1989ne,
  author = {Catani, S. and Trentadue, L.},
  title = {Resummation of the QCD Perturbative Series for Hard Processes},
  journal = {Nucl. Phys. B},
  volume = {327},
  pages = {323},
  year = {1989},
  doi = {10.1016/0550-3213(89)90273-3}
}

@article{Procura:2009vm,
  author = {Procura, Massimiliano and Stewart, Iain W.},
  title = {Quark Fragmentation within an Identified Jet},
  journal = {Phys. Rev. D},
  volume = {81},
  pages = {074009},
  year = {2010},
  eprint = {0911.4980},
  archivePrefix = {arXiv},
  primaryClass = {hep-ph},
  doi = {10.1103/PhysRevD.81.074009}
}

@article{Jain:2011xz,
  author = {Jain, Ambar and Procura, Massimiliano and Waalewijn, Wouter J.},
  title = {Parton Fragmentation within an Identified Jet at NNLL},
  journal = {JHEP},
  volume = {05},
  number = {05},
  pages = {035},
  year = {2011},
  eprint = {1101.4953},
  archivePrefix = {arXiv},
  primaryClass = {hep-ph},
  doi = {10.1007/JHEP05(2011)035}
}

@article{Benic:2016uku,
  author = {Benic, Sanjin and Fukushima, Kenji and Garcia-Montero, Oscar and Venugopalan, Raju},
  title = {Probing gluon saturation with next-to-leading order photon production at central rapidities in proton-nucleus collisions},
  journal = {JHEP},
  volume = {01},
  pages = {115},
  year = {2017},
  eprint = {1609.09424},
  archivePrefix = {arXiv},
  primaryClass = {hep-ph},
  doi = {10.1007/JHEP01(2017)115}
}

@article{Iancu:2020jch,
  author = {Iancu, Edmond and Mueller, Alfred H. and Triantafyllopoulos, Dimitri N. and Wei, Shu-Yi},
  title = {Saturation effects in SIDIS at very forward rapidities},
  journal = {JHEP},
  volume = {07},
  pages = {196},
  year = {2021},
  eprint = {2012.08562},
  archivePrefix = {arXiv},
  primaryClass = {hep-ph},
  doi = {10.1007/JHEP07(2021)196}
}

@article{Hatta:2022lzj,
  author = {Hatta, Yoshitaka and Xiao, Bo-Wen and Yuan, Feng},
  title = {Semi-inclusive diffractive deep inelastic scattering at small x},
  journal = {Phys. Rev. D},
  volume = {106},
  pages = {094015},
  year = {2022},
  eprint = {2205.08060},
  archivePrefix = {arXiv},
  primaryClass = {hep-ph},
  doi = {10.1103/PhysRevD.106.094015}
}

@article{Caucal:2021ent,
  author = {Caucal, Paul and Salazar, Farid and Venugopalan, Raju},
  title = {Dijet impact factor in DIS at next-to-leading order in the Color Glass Condensate},
  journal = {JHEP},
  volume = {11},
  pages = {222},
  year = {2021},
  eprint = {2108.06347},
  archivePrefix = {arXiv},
  primaryClass = {hep-ph},
  doi = {10.1007/JHEP11(2021)222}
}

@article{Caucal:2022ulg,
  author = {Caucal, Paul and Salazar, Farid and Schenke, Bjoern and Venugopalan, Raju},
  title = {Back-to-back inclusive dijets in DIS at small x: Sudakov suppression and gluon saturation at NLO},
  journal = {JHEP},
  volume = {11},
  pages = {169},
  year = {2022},
  eprint = {2208.13872},
  archivePrefix = {arXiv},
  primaryClass = {hep-ph},
  doi = {10.1007/JHEP11(2022)169}
}

@article{Taels:2022tza,
  author = {Taels, Pieter and Altinoluk, Tolga and Beuf, Guillaume and Marquet, Cyrille},
  title = {Dijet photoproduction at low x at next-to-leading order and its back-to-back limit},
  journal = {JHEP},
  volume = {10},
  pages = {184},
  year = {2022},
  eprint = {2204.11650},
  archivePrefix = {arXiv},
  primaryClass = {hep-ph},
  doi = {10.1007/JHEP10(2022)184}
}

@article{Bergabo:2022tcu,
  author = {Bergabo, Francisco and Jalilian-Marian, Jamal},
  title = {One-loop corrections to dihadron production in DIS at small x},
  journal = {Phys. Rev. D},
  volume = {106},
  pages = {054035},
  year = {2022},
  eprint = {2207.03606},
  archivePrefix = {arXiv},
  primaryClass = {hep-ph},
  doi = {10.1103/PhysRevD.106.054035}
}

@article{Bergabo:2022zhe,
  author = {Bergabo, Francisco and Jalilian-Marian, Jamal},
  title = {Single Inclusive Hadron Production in DIS at Small x: Next to Leading Order Corrections},
  journal = {Phys. Rev. D},
  volume = {107},
  pages = {054036},
  year = {2023},
  eprint = {2210.03208},
  archivePrefix = {arXiv},
  primaryClass = {hep-ph},
  doi = {10.1103/PhysRevD.107.054036}
}

@article{Tong:2022zwp,
  author = {Tong, Xuan-Bo and Xiao, Bo-Wen and Zhang, Yong-Yao},
  title = {Harmonics of parton saturation in lepton-jet correlations at the EIC},
  journal = {Phys. Rev. Lett.},
  volume = {130},
  pages = {151902},
  year = {2023},
  eprint = {2211.01647},
  archivePrefix = {arXiv},
  primaryClass = {hep-ph},
  doi = {10.1103/PhysRevLett.130.151902}
}

@article{vanHameren:2014lna,
  author = {van Hameren, A. and Kotko, P. and Kutak, K. and Marquet, C. and Sapeta, S.},
  title = {Saturation effects in forward-forward dijet production in p+Pb collisions},
  journal = {Phys. Rev. D},
  volume = {89},
  pages = {094014},
  year = {2014},
  eprint = {1402.5065},
  archivePrefix = {arXiv},
  primaryClass = {hep-ph},
  doi = {10.1103/PhysRevD.89.094014}
}

@article{Stasto:2014sea,
  author = {Stasto, Anna M. and Xiao, Bo-Wen and Yuan, Feng and Zaslavsky, Dima},
  title = {Matching collinear and small x factorization calculations for inclusive hadron production in pA collisions},
  journal = {Phys. Rev. D},
  volume = {90},
  pages = {014047},
  year = {2014},
  eprint = {1405.6311},
  archivePrefix = {arXiv},
  primaryClass = {hep-ph},
  doi = {10.1103/PhysRevD.90.014047}
}

@article{Altinoluk:2014eka,
  author = {Altinoluk, Tolga and Armesto, Nestor and Beuf, Guillaume and Kovner, Alex and Lublinsky, Michael},
  title = {Single-inclusive particle production in proton-nucleus collisions at next-to-leading order in the hybrid formalism},
  journal = {Phys. Rev. D},
  volume = {91},
  pages = {094016},
  year = {2015},
  eprint = {1411.2869},
  archivePrefix = {arXiv},
  primaryClass = {hep-ph},
  doi = {10.1103/PhysRevD.91.094016}
}

@article{Stasto:2016wrf,
  author = {Stasto, Anna M. and Zaslavsky, Dima},
  title = {Saturation in inclusive production beyond leading logarithm accuracy},
  journal = {Int. J. Mod. Phys. A},
  volume = {31},
  pages = {1630039},
  year = {2016},
  eprint = {1608.02285},
  archivePrefix = {arXiv},
  primaryClass = {hep-ph},
  doi = {10.1142/S0217751X16300398}
}

@article{Iancu:2016vyg,
  author = {Iancu, Edmond and Mueller, Alfred H. and Triantafyllopoulos, Dimitri N.},
  title = {CGC factorization for forward particle production in proton-nucleus collisions at next-to-leading order},
  journal = {JHEP},
  volume = {12},
  pages = {041},
  year = {2016},
  eprint = {1608.05293},
  archivePrefix = {arXiv},
  primaryClass = {hep-ph},
  doi = {10.1007/JHEP12(2016)041}
}

@article{Liu:2020mpy,
  author = {Liu, Hao-Yu and Kang, Zhong-Bo and Liu, Xiaohui},
  title = {Threshold resummation for hadron production in the small-x region},
  journal = {Phys. Rev. D},
  volume = {102},
  pages = {051502},
  year = {2020},
  eprint = {2004.11990},
  archivePrefix = {arXiv},
  primaryClass = {hep-ph},
  doi = {10.1103/PhysRevD.102.051502}
}

@article{Liu:2022ijp,
  author = {Liu, Hao-Yu and Xie, Kexin and Kang, Zhong-Bo and Liu, Xiaohui},
  title = {Single inclusive jet production in pA collisions at NLO in the small-x regime},
  journal = {JHEP},
  volume = {07},
  pages = {041},
  year = {2022},
  eprint = {2204.03026},
  archivePrefix = {arXiv},
  primaryClass = {hep-ph},
  doi = {10.1007/JHEP07(2022)041}
}

@article{Becher:2008cf,
  author = {Becher, Thomas and Schwartz, Matthew D.},
  title = {A precise determination of {$\alpha_s$} from LEP thrust data using effective field theory},
  journal = {JHEP},
  volume = {07},
  pages = {034},
  year = {2008},
  eprint = {0803.0342},
  archivePrefix = {arXiv},
  primaryClass = {hep-ph},
  doi = {10.1088/1126-6708/2008/07/034}
}

@article{Ellis:2010rwa,
  author = {Ellis, Stephen D. and Vermilion, Christopher K. and Walsh, Jonathan R. and Hornig, Andrew and Lee, Christopher},
  title = {Jet Shapes and Jet Algorithms in SCET},
  journal = {JHEP},
  volume = {11},
  pages = {101},
  year = {2010},
  eprint = {1001.0014},
  archivePrefix = {arXiv},
  primaryClass = {hep-ph},
  doi = {10.1007/JHEP11(2010)101}
}

@article{Becher:2009cu,
  author = {Becher, Thomas and Neubert, Matthias},
  title = {Infrared singularities of scattering amplitudes in perturbative QCD},
  journal = {Phys. Rev. Lett.},
  volume = {102},
  pages = {162001},
  year = {2009},
  eprint = {0901.0722},
  archivePrefix = {arXiv},
  primaryClass = {hep-ph},
  doi = {10.1103/PhysRevLett.102.162001}
}

@article{Becher:2009th,
  author = {Becher, Thomas and Schwartz, Matthew D.},
  title = {Direct photon production with effective field theory},
  journal = {JHEP},
  volume = {02},
  pages = {040},
  year = {2010},
  eprint = {0911.0681},
  archivePrefix = {arXiv},
  primaryClass = {hep-ph},
  doi = {10.1007/JHEP02(2010)040}
}

@article{Baty:2021ugw,
  author = {Baty, Austin and Gardner, Peter and Li, Wei},
  title = {Novel observables for exploring QCD collective evolution and quantum entanglement within individual jets},
  journal = {Phys. Rev. C},
  volume = {107},
  pages = {064908},
  year = {2023},
  eprint = {2104.11735},
  archivePrefix = {arXiv},
  primaryClass = {hep-ph},
  doi = {10.1103/PhysRevC.107.064908}
}

@techreport{CMS-PAS-HIN-24-024,
  author = {{CMS Collaboration}},
  title = {Unveiling the dynamics of long-range correlations in high-multiplicity jets through substructure engineering in pp collisions at {$\sqrt{s}=13$ TeV}},
  type = {Tech. Rep.},
  number = {CMS-PAS-HIN-24-024},
  institution = {CERN},
  address = {Geneva},
  year = {2025},
  url = {https://cds.cern.ch/record/2930797}
}

@article{ALICE:2023lyr,
  author = {Acharya, Shreyasi and others},
  collaboration = {ALICE},
  title = {Multiplicity and event-scale dependent flow and jet fragmentation in pp collisions at sqrt(s)=13 TeV and in p-Pb collisions at sqrt(sNN)=5.02 TeV},
  journal = {JHEP},
  volume = {03},
  pages = {092},
  year = {2024},
  eprint = {2308.16591},
  archivePrefix = {arXiv},
  primaryClass = {nucl-ex},
  doi = {10.1007/JHEP03(2024)092}
}

@article{Zhao:2024wqs,
  author = {Zhao, Wenbin and Lin, Zi-Wei and Wang, Xin-Nian},
  title = {Collectivity inside high-multiplicity jets in high-energy proton-proton collisions},
  year = {2024},
  eprint = {2401.13137},
  archivePrefix = {arXiv},
  primaryClass = {hep-ph}
}

@article{Dokshitzer:2025owq,
  author = {Dokshitzer, Yuri L. and Webber, Bryan R.},
  title = {Hadron multiplicity fluctuations in perturbative QCD},
  journal = {JHEP},
  volume = {08},
  pages = {168},
  year = {2025},
  eprint = {2505.00652},
  archivePrefix = {arXiv},
  primaryClass = {hep-ph},
  doi = {10.1007/JHEP08(2025)168}
}

@article{Lam:1983vw,
  author = {Lam, C. S. and Walton, M. A.},
  title = {A Proposal for the Origin of KNO Scaling},
  journal = {Phys. Lett. B},
  volume = {140},
  pages = {246},
  year = {1984},
  doi = {10.1016/0370-2693(84)90928-6}
}

@article{Hinz:1985wq,
  author = {Hinz, D. C. and Lam, C. S.},
  title = {A Dynamical Basis for KNO Scaling and Its Violation},
  journal = {Phys. Rev. D},
  volume = {33},
  pages = {3256},
  year = {1986},
  doi = {10.1103/PhysRevD.33.3256}
}

@article{Vertesi:2020utz,
  author = {Vertesi, Robert and Gemes, Adam and Barnafoldi, Gergely Gabor},
  title = {Koba-Nielsen-Olesen-like scaling within a jet in proton-proton collisions at LHC energies},
  journal = {Phys. Rev. D},
  volume = {103},
  pages = {L051503},
  year = {2021},
  eprint = {2012.01132},
  archivePrefix = {arXiv},
  primaryClass = {hep-ph},
  doi = {10.1103/PhysRevD.103.L051503}
}

@article{Liu:2022bru,
  author = {Liu, Yizhuang and Nowak, Maciej A. and Zahed, Ismail},
  title = {Mueller's dipole wave function in QCD: Emergent Koba-Nielsen-Olesen scaling in the double logarithm limit},
  journal = {Phys. Rev. D},
  volume = {108},
  pages = {034017},
  year = {2023},
  eprint = {2211.05169},
  archivePrefix = {arXiv},
  primaryClass = {hep-ph},
  doi = {10.1103/PhysRevD.108.034017}
}

@article{Germano:2024ier,
  author = {Germano, Guilherme R. and Navarra, Fernando S. and Wilk, Grzegorz and Wlodarczyk, Zbigniew},
  title = {Emergence of Koba-Nielsen-Olsen scaling in multiplicity distributions in jets produced at the LHC},
  journal = {Phys. Rev. D},
  volume = {110},
  pages = {034026},
  year = {2024},
  eprint = {2406.04856},
  archivePrefix = {arXiv},
  primaryClass = {hep-ph},
  doi = {10.1103/PhysRevD.110.034026}
}

@article{Duan:2026eecmultiplicity,
  author = {Duan, Pi and Ke, Weiyao and Qin, Guang-You and Wang, Lei},
  title = {Uncover the correlation between jet energy correlators and multiplicity fluctuations},
  year = {2026},
  eprint = {2604.01102},
  archivePrefix = {arXiv},
  primaryClass = {hep-ph}
}

@article{Basham:1977iq,
  author = {Basham, C. Louis and Brown, Lowell S. and Ellis, Stephen D. and Love, Sherwin T.},
  title = {Electron-positron annihilation energy pattern in quantum chromodynamics: asymptotically free perturbation theory},
  journal = {Phys. Rev. D},
  volume = {17},
  pages = {2298},
  year = {1978},
  doi = {10.1103/PhysRevD.17.2298}
}

@article{Basham:1978bw,
  author = {Basham, C. Louis and Brown, Lowell S. and Ellis, Stephen D. and Love, Sherwin T.},
  title = {Energy correlations in electron-positron annihilation: testing QCD},
  journal = {Phys. Rev. Lett.},
  volume = {41},
  pages = {1585},
  year = {1978},
  doi = {10.1103/PhysRevLett.41.1585}
}

@article{Basham:1978zq,
  author = {Basham, C. Louis and Brown, Lowell S. and Ellis, Stephen D. and Love, Sherwin T.},
  title = {Energy correlations in electron-positron annihilation in quantum chromodynamics: asymptotically free perturbation theory},
  journal = {Phys. Rev. D},
  volume = {19},
  pages = {2018},
  year = {1979},
  doi = {10.1103/PhysRevD.19.2018}
}

@article{PhysRevD.24.2383,
  author = {Brown, Lowell S. and Ellis, Stephen D.},
  title = {Energy-energy correlations in electron-positron annihilation},
  journal = {Phys. Rev. D},
  volume = {24},
  pages = {2383},
  year = {1981},
  doi = {10.1103/PhysRevD.24.2383}
}

@article{Moult:2018jzp,
  author = {Moult, Ian and Zhu, Hua Xing},
  title = {Simplicity from recoil: the three-loop soft function and factorization for the energy-energy correlation},
  journal = {JHEP},
  volume = {08},
  pages = {160},
  year = {2018},
  eprint = {1801.02627},
  archivePrefix = {arXiv},
  primaryClass = {hep-ph},
  doi = {10.1007/JHEP08(2018)160}
}

@article{Dixon:2019uzg,
  author = {Dixon, Lance J. and Moult, Ian and Zhu, Hua Xing},
  title = {Collinear limit of the energy-energy correlator},
  journal = {Phys. Rev. D},
  volume = {100},
  pages = {014009},
  year = {2019},
  eprint = {1905.01310},
  archivePrefix = {arXiv},
  primaryClass = {hep-ph},
  doi = {10.1103/PhysRevD.100.014009}
}

@article{Chen:2020vvp,
  author = {Chen, Hao and Moult, Ian and Zhang, XiaoYuan and Zhu, Hua Xing},
  title = {Rethinking jets with energy correlators: tracks, resummation, and analytic continuation},
  journal = {Phys. Rev. D},
  volume = {102},
  pages = {054012},
  year = {2020},
  eprint = {2004.11381},
  archivePrefix = {arXiv},
  primaryClass = {hep-ph},
  doi = {10.1103/PhysRevD.102.054012}
}

@article{Duhr:2022yyp,
  author = {Duhr, Claude and Mistlberger, Bernhard and Vita, Gherardo},
  title = {Four-loop rapidity anomalous dimension and event shapes to fourth logarithmic order},
  journal = {Phys. Rev. Lett.},
  volume = {129},
  pages = {162001},
  year = {2022},
  eprint = {2205.02242},
  archivePrefix = {arXiv},
  primaryClass = {hep-ph},
  doi = {10.1103/PhysRevLett.129.162001}
}

@article{Li:2020bub,
  author = {Li, Hai Tao and Vitev, Ivan and Zhu, Yu Jiao},
  title = {Transverse-energy-energy correlations in deep inelastic scattering},
  journal = {JHEP},
  volume = {11},
  pages = {051},
  year = {2020},
  eprint = {2006.02437},
  archivePrefix = {arXiv},
  primaryClass = {hep-ph},
  doi = {10.1007/JHEP11(2020)051}
}

@article{Li:2021txc,
  author = {Li, Hai Tao and Makris, Yiannis and Vitev, Ivan},
  title = {Energy-energy correlators in deep inelastic scattering},
  journal = {Phys. Rev. D},
  volume = {103},
  pages = {094005},
  year = {2021},
  eprint = {2102.05669},
  archivePrefix = {arXiv},
  primaryClass = {hep-ph},
  doi = {10.1103/PhysRevD.103.094005}
}

@article{Liu:2022wop,
  author = {Liu, Xiaohui and Zhu, Hua Xing},
  title = {Nucleon energy correlators},
  journal = {Phys. Rev. Lett.},
  volume = {130},
  pages = {091901},
  year = {2023},
  eprint = {2209.02080},
  archivePrefix = {arXiv},
  primaryClass = {hep-ph},
  doi = {10.1103/PhysRevLett.130.091901}
}

@article{Liu:2023aqb,
  author = {Liu, Hao-Yu and Liu, Xiaohui and Pan, Ji-Chen and Yuan, Feng and Zhu, Hua Xing},
  title = {Nucleon energy correlators for the Color Glass Condensate},
  journal = {Phys. Rev. Lett.},
  volume = {130},
  pages = {181901},
  year = {2023},
  eprint = {2301.01788},
  archivePrefix = {arXiv},
  primaryClass = {hep-ph},
  doi = {10.1103/PhysRevLett.130.181901}
}

@article{Yang:2023dwc,
  author = {Yang, Zhong and He, Yayun and Moult, Ian and Wang, Xin-Nian},
  title = {Probing the short-distance structure of the quark-gluon plasma with energy correlators},
  journal = {Phys. Rev. Lett.},
  volume = {132},
  pages = {011901},
  year = {2024},
  eprint = {2310.01500},
  archivePrefix = {arXiv},
  primaryClass = {hep-ph},
  doi = {10.1103/PhysRevLett.132.011901}
}

@article{Barata:2023bhh,
  author = {Barata, Jo{\~a}o and Caucal, Paul and Soto-Ontoso, Alba and Szafron, Robert},
  title = {Advancing the understanding of energy-energy correlators in heavy-ion collisions},
  journal = {JHEP},
  volume = {11},
  pages = {060},
  year = {2024},
  eprint = {2312.12527},
  archivePrefix = {arXiv},
  primaryClass = {hep-ph},
  doi = {10.1007/JHEP11(2024)060}
}

@article{Bossi:2024qho,
  author = {Bossi, Hannah and Kudinoor, Arjun Srinivasan and Moult, Ian and Pablos, Daniel and Rai, Ananya and Rajagopal, Krishna},
  title = {Imaging the wakes of jets with energy-energy-energy correlators},
  journal = {JHEP},
  volume = {12},
  pages = {073},
  year = {2024},
  eprint = {2407.13818},
  archivePrefix = {arXiv},
  primaryClass = {hep-ph},
  doi = {10.1007/JHEP12(2024)073}
}

@article{Ke:2025ibt,
  author = {Ke, Weiyao and Mecaj, Bianka and Vitev, Ivan},
  title = {Renormalization group evolution for in-medium energy correlators},
  year = {2025},
  eprint = {2512.11952},
  archivePrefix = {arXiv},
  primaryClass = {hep-ph}
}

@article{CMS:2024mlf,
  author = {Hayrapetyan, Aram and others},
  collaboration = {CMS},
  title = {Measurement of energy correlators inside jets and determination of the strong coupling \(\alpha_s(m_Z)\)},
  journal = {Phys. Rev. Lett.},
  volume = {133},
  pages = {071903},
  year = {2024},
  eprint = {2402.13864},
  archivePrefix = {arXiv},
  primaryClass = {hep-ex},
  doi = {10.1103/PhysRevLett.133.071903}
}

@article{ATLAS:2023tgo,
  author = {Aad, Georges and others},
  collaboration = {ATLAS},
  title = {Determination of the strong coupling constant from transverse energy-energy correlations in multijet events at \(\sqrt{s}=13\) TeV with the ATLAS detector},
  journal = {JHEP},
  volume = {07},
  pages = {085},
  year = {2023},
  eprint = {2301.09351},
  archivePrefix = {arXiv},
  primaryClass = {hep-ex},
  doi = {10.1007/JHEP07(2023)085}
}

@article{ALICE:2024dfl,
  author = {Acharya, Shreyasi and others},
  collaboration = {ALICE},
  title = {Exposing the parton-hadron transition within jets with energy-energy correlators in pp collisions at \(\sqrt{s}=5.02\) TeV},
  year = {2024},
  eprint = {2409.12687},
  archivePrefix = {arXiv},
  primaryClass = {hep-ex}
}

@article{STAR:2025jut,
  author = {Aboona, B. E. and others},
  collaboration = {STAR},
  title = {Measurement of two-point energy correlators within jets in \(p+p\) collisions at \(\sqrt{s}=200\) GeV},
  journal = {Phys. Rev. Lett.},
  volume = {135},
  pages = {111901},
  year = {2025},
  eprint = {2502.15925},
  archivePrefix = {arXiv},
  primaryClass = {hep-ex},
  doi = {10.1103/wv2t-dkgn}
}

@article{Chien:2015hda,
  author = {Chien, Yang-Ting and Vitev, Ivan},
  title = {Towards the understanding of jet shapes and cross sections in heavy ion collisions using soft-collinear effective theory},
  journal = {JHEP},
  volume = {05},
  pages = {023},
  year = {2016},
  eprint = {1509.07257},
  archivePrefix = {arXiv},
  primaryClass = {hep-ph},
  doi = {10.1007/JHEP05(2016)023}
}

@article{Kang:2017frl,
  author = {Kang, Zhong-Bo and Ringer, Felix and Vitev, Ivan},
  title = {Inclusive production of small radius jets in heavy-ion collisions},
  journal = {Phys. Lett. B},
  volume = {769},
  pages = {242},
  year = {2017},
  eprint = {1701.05839},
  archivePrefix = {arXiv},
  primaryClass = {hep-ph},
  doi = {10.1016/j.physletb.2017.03.067}
}

@article{Casalderrey-Solana:2016jvj,
  author = {Casalderrey-Solana, Jorge and Gulhan, Doga and Milhano, Guilherme and Pablos, Daniel and Rajagopal, Krishna},
  title = {Angular structure of jet quenching within a hybrid strong/weak coupling model},
  journal = {JHEP},
  volume = {03},
  pages = {135},
  year = {2017},
  eprint = {1609.05842},
  archivePrefix = {arXiv},
  primaryClass = {hep-ph},
  doi = {10.1007/JHEP03(2017)135}
}

@article{Mehtar-Tani:2016aco,
  author = {Mehtar-Tani, Yacine and Tywoniuk, Konrad},
  title = {Groomed jets in heavy-ion collisions: sensitivity to medium-induced bremsstrahlung},
  journal = {JHEP},
  volume = {04},
  pages = {125},
  year = {2017},
  eprint = {1610.08930},
  archivePrefix = {arXiv},
  primaryClass = {hep-ph},
  doi = {10.1007/JHEP04(2017)125}
}

@article{Tachibana:2017syd,
  author = {Tachibana, Yasuki and Chang, Ning-Bo and Qin, Guang-You},
  title = {Full jet in quark-gluon plasma with hydrodynamic medium response},
  journal = {Phys. Rev. C},
  volume = {95},
  pages = {044909},
  year = {2017},
  eprint = {1701.07951},
  archivePrefix = {arXiv},
  primaryClass = {nucl-th},
  doi = {10.1103/PhysRevC.95.044909}
}
